\documentclass[12pt]{article}

\usepackage{latexsym,amsmath,amscd,amssymb,graphics}
\usepackage{enumerate}

\usepackage{graphicx}

\usepackage[colorlinks]{hyperref}
\usepackage{url}
\usepackage{framed}

\usepackage[all]{xy}

\makeatletter

\@addtoreset{figure}{section}
\def\thefigure{\thesection.\@arabic\c@figure}
\def\fps@figure{h, t}
\@addtoreset{table}{bsection}
\def\thetable{\thesection.\@arabic\c@table}
\def\fps@table{h, t}
\@addtoreset{equation}{section}

\makeatother

\begin{document}

\newtheorem{theorem}{Theorem}[section]
\newtheorem{definition}[theorem]{Definition}
\newtheorem{lemma}[theorem]{Lemma}
\newtheorem{remark}[theorem]{Remark}
\newtheorem{proposition}[theorem]{Proposition}
\newtheorem{corollary}[theorem]{Corollary}
\newtheorem{example}[theorem]{Example}

\def\below#1#2{\mathrel{\mathop{#1}\limits_{#2}}}

%%%%%%%%%%%%%%%%%%%%%%%%%%%%%%%%%%%%%%%%%%%%%%%%%%%%%%%%%%%%%%%%%%%%%%%%%%%%%%%
%%%%%%%%

%%%%%%%%%%%%%%%%%%%%%%%%%%%%%%%%%%%%%%%%%%%%%%%%%%%%%%%%%%%%%%%%%%%%%%%%%%%%%%%
%%%%%%%%

\title{Clebsch variational principles in field theories and singular solutions of covariant EPDiff equations}

\author{Fran\c{c}ois Gay-Balmaz$^{1}$}
\date{}

\addtocounter{footnote}{1}
\footnotetext{CNRS, LMD, \'Ecole Normale Sup\'erieure de Paris, France.
\texttt{francois.gay-balmaz@lmd.ens.fr}
}
\maketitle

\makeatother

\maketitle

\begin{abstract}
This paper introduces and studies a field theoretic analogue of the Clebsch variational principle of classical mechanics.  This principle yields an alternative derivation of the covariant Euler-Poincar\'e equations that naturally includes covariant Clebsch variables via multisymplectic momentum maps. In the case of diffeomorphism groups, this approach gives a new interpretation of recently derived singular peakon solutions of $ \operatorname{Diff}( \mathbb{R}  )$-strand equations, and allows for the construction of singular solutions (such as filaments or sheets) for a more general class of equations, called covariant EPDiff equations.
The relation between the covariant Clebsch principle and other variational principles arising in mechanics and field theories, such as Hamilton-Pontryagin principles, is explained through the introduction of a class of covariant Pontryagin variational principles in field theories.

\end{abstract}
\tableofcontents

%|||-------------------text width----------------------|||

\section{Introduction}

Variational principles are of central importance in mechanics and field theories, as illustrated by \textit{Hamilton's principle of critical action}, stating that the Euler-Lagrange equations are obtained by computing the critical curves, or the critical sections, of the action functional. On the Hamiltonian side, the canonical Hamilton equations of classical mechanics can also be obtained via a variational principle, the so called \textit{Hamilton phase space principle}. This principle extends to the field theoretic context in a natural way since the multisymplectic Hamilton equations can be characterized as  the critical sections of an action functional defined with the help of the Hamiltonian density and the Cartan form.

Variational principles also arise naturally in optimal control theory, via the Pontryagin maximum principle. 
Under some assumptions, one can characterize the extremals of optimal control problems as the critical points of a functional obtained from the Lagrange multipliers theorem, e.g. \cite{Bl2003}. Such a functional is obtained by adding to the cost function of the problem, the control equation constraint multiplied by a costate variable. This variational principle generalizes naturally to the case of nonlinear optimal control problems whose control function takes value in a fiber bundle, e.g. \cite{IbPeSa2010}. It will be referred to as the \textit{Pontryagin variational principle}.

Based on earlier works on geometric optimal control theory such as, e.g., \cite{Br1994}, \cite{BlCr1996}, \cite{BlBrCr1997}, \cite{BlCrMaRa1998}, \cite{BlCrSa2006}, \cite{Ho2009}, \cite{CoHo2009}, a particular class of optimal control problems and variational principles associated to group actions was introduced and studied in \cite{GBRa2011}, under the name of \textit{Clebsch optimal control problem}. The main feature of this class of problems is that the control function takes values in the Lie algebra of the acting Lie group and that the constraint ODE consists in imposing the state of the system to evolve according to the infinitesimal action of the control function.
The variational principle associated to this class of problems, called \textit{Clebsch variational principle}, provides an
alternative derivation of the Euler-Poincar\'e equations of classical mechanics based on Clebsch variables (as defined by \cite{MaWe1983}). Moreover, 
this approach unifies and generalizes a wide range of examples appearing in the literature such as the symmetric formulation of $N$-dimensional rigid body, optimal control for ideal flow using the back-to-labels map, double bracket equations associated to symmetric spaces, and optimal control problems in imaging science.

In a similar way with the Pontryagin variational principle arising in optimal control problems, the \textit{Hamilton-Pontryagin principle} of classical mechanics is a modification of Hamilton's principle that includes the Legendre transform as a constraint in the action functional, see \cite{YoMa2006}. This principle combines variables from both the Lagrangian and Hamiltonian formulations and is especially appropriate to deal with degenerate Lagrangians. Using similar ideas, a field theoretic version of this principle has been recently formulated in \cite{VaYoMa2012}.

\medskip

The first goal of the present paper is to introduce and study a new class of variational principles for field theories, called \textit{covariant Clebsch variational principle} and to illustrate the applicability of this principle to a wide range of examples both in the finite or infinite dimensional situation. A main example, inspired by the recent work \cite{HoIvPe2012}, being given by the construction of singular solutions for the covariant Euler-Poincar\'e equations on diffeomorphism groups. The second goal of the paper is to present the unifying geometric framework underlying all the variational principles mentioned above (Hamilton phase space principle, Hamilton-Pontryagin principle, Pontryagin principle, Clebsch principle) both in the context of classical mechanics and for field theories.

\paragraph{Plan of the paper.} In Section \ref{Sec_2} we present some background material needed for the rest of the paper. We first review the definition and the main properties of the Clebsch variational principle in mechanics. In particular, we explain how the Euler-Poincar\'e equations are obtained from the stationarity conditions and mention the occurrence of Clebsch variables and collective Hamiltonian. Then we present basic facts about classical field theory, such as jet bundles, covariant Euler-Lagrange and Hamilton equations, and covariant momentum maps. Finally, we recall the process of covariant Euler-Poincar\'e reduction on a trivial principal bundle and relate the spacetime decomposition of these equations with the $G$-strand equations of \cite{HoIvPe2012}. In Section \ref{Sec_3} we introduce the covariant Clebsch variational principle and study its main properties. In particular, we present the stationarity conditions, explain their covariant Hamiltonian formulation through the introduction of a collective Hamiltonian via the covariant momentum map, and show that they yield the covariant Euler-Poincar\'e equations. In Section \ref{Sec_4} we apply the covariant Clebsch variational principle to various particular cases of special interests such as the case of a trivial fiber bundle, the case of group translations, and the case of a linear action.
In the case of diffeomorphism groups, this approach gives a new interpretation of recently derived singular peakon solutions of $ \operatorname{Diff}( \mathbb{R}  )$-strand equations, and allows for the construction of singular solutions for a more general class of equations, called covariant EPDiff equations. We also consider the covariant version of the Euler equations for perfect fluids, called covariant EPDiff$_{vol}$ equations and provide covariant Clebsch variables for these equations. Finally, in Section \ref{Sec_5} we present a class of covariant variational principles for field theories that underlies the covariant Clebsch variational principle and other known principles such as the Hamilton phase space principle and the Hamilton-Pontryagin principle, both in their classical and covariant form.

\section{Preliminaries}\label{Sec_2}

In this section we review some needed facts concerning the Clebsch variational principle in mechanics and the multisymplectic geometry of classical fields.

\subsection{The Clebsch variational principle in mechanics}\label{subsec_Clebsch} 

Recently, a particular class of geometric optimal control problems associated to Lie group actions has been considered and studied in \cite{GBRa2011}, see also \cite{CoHo2009}, in relation with variational principles and Clebsch variables in mechanics. The main feature of this class of problems is that the control function takes values in the Lie algebra of the acting Lie group and that the constraint ODE consists in imposing the state of the system to evolve according to the infinitesimal action of the control function. It was shown that the necessary condition for an optimal solution, obtained via the Pontryagin maximum principle, naturally encodes a Clebsch representation of the momentum variable conjugate to the control. The associated variational principle is therefore naturally referred to as the \textit{Clebsch variational principle}. It is reminiscent of the \textit{Hamilton-Pontryagin principle} which incorporates, instead of a Clebsch representation, the Legendre transformation, and enables one to treat degenerate Lagrangian systems, see \cite{YoMa2006} and references therein. Both variational principles can be seen as particular instances of the variational principle for optimal control problems provided in \cite{IbPeSa2010} in connection with Pontryagin maximum principle. This latter variational approach will be adapted to the covariant field theoretic context in the last section of the present paper.

Various optimal control formulations of mechanical systems were shown in \cite{GBRa2011} to arise as particular instances of Clebsch optimal control problems, obtained by an appropriate choice of the Lie group action on the state manifold. Such examples include various double bracket equations and systems on Lie algebras and symmetric spaces (\cite{Br1994}, \cite{BlCr1996}, \cite{BlBrCr1997}); optimal control problems for fluids (\cite{BlCrHoMa2000}, \cite{Ho2009}, \cite{CoHo2009}); the symmetric representation of the rigid body and its generalizations (\cite{BlCr1996}, \cite{BlCrMaRa1998}, \cite{BlCrSa2006}, \cite{BlCrMaSa2008}).
A particularly relevant Clebsch representation for the present paper is the one provided by the singular solution of the Camassa-Holm equations and its higher dimensional extensions (\cite{HoMa2004}, \cite{CoHo2009}).
We now recall the definition of the Clebsch optimal control problem.

\paragraph{The Clebsch optimal control problem.} Let $\Phi: G \times Q \rightarrow Q$, $(g,q) \mapsto \Phi _g (q)$ be a left action of a Lie group $G$ on a manifold $Q$. Let us denote by $ \mathfrak{g}  $ the Lie algebra of the Lie group $G$ and by $ \mathfrak{X}  (Q)$ the space of all smooth vector fields on the manifold $Q$. Given a Lie algebra element $ \xi \in \mathfrak{g}  $, the associated infinitesimal generator $ \xi _Q \in \mathfrak{X}  (Q)$ is defined by
\[
\left.\xi _Q(q):=\frac{d}{dt}\right|_{t=0}\Phi_{\operatorname{exp}(t \xi )}(q),\quad q \in Q,
\]
where $ \operatorname{exp}: \mathfrak{g}  \rightarrow G$ is the exponential map of $G$. Given a cost function $\ell:\mathfrak{g}\rightarrow\mathbb{R}$, a real number $T>0$, and two elements $ q _0 , q_T \in Q$ belonging to the same $G$-orbit, the \textit{Clebsch optimal control problem} associated to the action $ \Phi $ consists in minimizing the integral
\begin{equation}\label{Clebsch_optimal_control}
\int_0^T\ell(\xi (t))dt
\end{equation}
subject to the following conditions:
\begin{itemize}
\item[\rm (A)] $\dot q(t)=\xi (t)_Q(q(t))$;
\item[\rm (B)] $q(0)=q_0$ and $q(T)=q_T$.
\end{itemize}

From the Pontryagin maximum principle, it follows that an extremal solution $(q(t), \xi (t))$ of \eqref{Clebsch_optimal_control} is necessarily a solution of the \textit{Clebsch variational principle} 
\begin{equation}\label{CVP} 
\delta \int_0^T\left(  \left\langle \alpha  , \dot q- \xi  _Q(q) \right\rangle +\ell( \xi ) \right) dt=0
\end{equation} 
over curves $ \xi (t) \in \mathfrak{g}  $ and $ \alpha (t) \in T^*Q$, where $ \alpha (t)$ is a curve covering $q(t) \in Q$ with fixed endpoints $q _0 $ and $ q _T $. This principle yields the stationarity conditions 
\begin{equation}\label{Pontryagin_solution}\frac{\delta \ell}{\delta \xi }=\mathbf{J}(\alpha),\quad \frac{d}{dt} \alpha=\xi _{T^*Q}(\alpha)
\end{equation}
(see \cite{GBRa2011}), where $ \mathbf{J} :T^*Q \rightarrow \mathfrak{g}  ^\ast $ is the momentum map associated to the cotangent lifted action of $G$ on $T^*Q$, and $ \xi _{T^*Q}\in \mathfrak{X}  (T^*Q)$ is the infinitesimal generator of this cotangent lifted action. Recall that $ \mathbf{J} $ is given by $\left\langle  \mathbf{J} ( \alpha _q ) , \xi \right\rangle := \left\langle \alpha _q , \xi _Q (q) \right\rangle $, for all $ \xi \in \mathfrak{g}, \alpha _q \in T^*Q$. Recall also that the cotangent lifted action of $ \Phi $, denoted by $\Phi^{T^*}$, is defined by
\[
\Phi ^{T^*}: G \times T^*Q \rightarrow T^*Q, \quad 
\Phi^{T^*}_g( \alpha _q ):=T^*\Phi_{g^{-1}}( \alpha _q ),\quad \text{for all $ \alpha _q \in T^*Q$}, 
\]
where $T^* \Phi _{ g ^{-1} }: T^*Q \rightarrow T^*Q$ is the dual of the tangent map $T \Phi _{ g ^{-1} }: TQ \rightarrow TQ$.

\paragraph{Recovering the Euler-Poincar\'e equations.} A solution of the second equation in \eqref{Pontryagin_solution} is necessarily of the form $\alpha(t)=\Phi^{T^*}_{g(t)}(\alpha(0))$, where $\dot g(t)g(t)^{-1}=\xi (t)$. Using the first equation in \eqref{Pontryagin_solution}, we obtain that  the control $ \xi (t)$ satisfies the \textit{Euler-Poincar\'e equations}
\[
\frac{d}{dt}\frac{\delta \ell}{\delta \xi }+\operatorname{ad}^*_\xi \frac{\delta \ell}{\delta \xi }=0,
\]
where $\delta\ell/\delta \xi \in \mathfrak{g}  ^\ast $ denotes the functional derivative of $\ell$ and $\left\langle \operatorname{ad}^*_ \xi \mu , \eta \right\rangle = \left\langle \mu , [ \xi ,\eta ] \right\rangle $, with $\xi , \eta \in \mathfrak{g}$ and $\mu \in \mathfrak{g}  ^\ast $, is the infinitesimal coadjoint action. 
Recall that the Euler-Poincar\'e equations can be obtained by Lagrangian reduction of the Euler-Lagrange equations associated to a Lagrangian $L:TG\rightarrow\mathbb{R}$ defined on the tangent bundle of $G$ and $G$-invariant under the tangent lift of right translations. A curve $g(t)\in G$ is a solution of the Euler-Lagrange equations for $L=L(g, \dot g)$ if and only if the curve $\xi(t)=\dot g(t)g(t)^{-1}\in \mathfrak{g}  $ is a solution of the Euler-Poincar\'e equations for the reduced Lagrangian $\ell:=L|_{\mathfrak{g}}:\mathfrak{g}\rightarrow\mathbb{R}$. The same procedure on the Hamiltonian side leads to the so called \textit{Lie-Poisson equations}
\begin{equation}\label{LP_equ} 
\displaystyle\dot\mu+\operatorname{ad}^\ast _{\frac{\delta h}{\delta\mu}}\mu=0
\end{equation} 
associated to a given Hamiltonian $h: \mathfrak{g}  ^\ast \rightarrow \mathbb{R}  $. These equations can be obtained by Poisson reduction of canonical Hamilton's equations for the right invariant Hamiltonian $H$ on $T^*G$ with $h=H|_ \mathfrak{g}  $. If $\xi \mapsto \frac{\delta\ell}{\delta \xi }$ is a diffeomorphism, then Euler-Poincar\'e and Lie-Poisson equations are equivalent, since one can pass from one to the other by the Legendre transformation
\begin{equation}\label{Leg_trsf} 
h(\mu)=\langle\mu,\xi\rangle-\ell(\xi),\qquad \mu=\frac{\delta\ell}{\delta \xi}.
\end{equation} 
Let us recall that the equations \eqref{LP_equ} are Hamiltonian relative to the \textit{Lie-Poisson bracket}
\begin{equation}\label{LP_bracket} 
\{f,g\}( \mu ) = \left\langle \mu , \left[\frac{\delta f}{\delta \mu }, \frac{\delta g}{\delta \mu }\right] \right\rangle ,\quad \mu \in \mathfrak{g}  ^\ast
\end{equation} 
on functions $f,g: \mathfrak{g}  ^\ast \rightarrow \mathbb{R}  $. We refer to \cite{MaRa1999} for a detailed account about Euler-Poincar\'e and Lie-Poisson reductions.

\paragraph{Collective Hamiltonian and Clebsch variables.} The \textit{Pontryagin Hamiltonian} associated to the Clebsch optimal control problem \eqref{Clebsch_optimal_control} reads
\begin{equation}\label{Pontryagin_Hamiltonian} 
\hat{H}( \alpha _q , \xi )= \left\langle \alpha ,\xi _Q (q) \right\rangle -\ell( \xi ).
\end{equation} 
Assume that $\xi \mapsto \frac{\delta\ell}{\delta \xi }$ is a diffeomorphism and denote by $\xi ^\star(\alpha_q )$ the optimal control uniquely determined by the condition $\delta\ell/\delta \xi =\mathbf{J}(\alpha_q )$. The \textit{optimal Hamiltonian} of the problem, defined by $H(\alpha_q ):=\hat H(\alpha_q ,\xi ^\star(\alpha))$, is thus given by
\begin{equation}\label{coll_ham}
H(\alpha_q )=\left\langle\frac{\delta \ell}{\delta \xi },\xi  \right\rangle-\ell(\xi )=h\left(\frac{\delta \ell}{\delta \xi }\right)=h(\mathbf{J}(\alpha_q )),
\end{equation}
where $h:\mathfrak{g}^*\rightarrow\mathbb{R}$ is the Hamiltonian
associated to $\ell$ via the Legendre transformation $\xi \mapsto
\delta\ell/\delta \xi $, see \eqref{Leg_trsf}. Therefore, in the case of Clebsch optimal control problems, the optimal Hamiltonian coincides with the so called \textit{collective Hamiltonian} associated to Lie-Poisson Hamiltonian $h$ and the momentum map $ \mathbf{J}  :T^*Q \rightarrow \mathfrak{g}  ^\ast $, (see \cite{GuSt1980}, \cite{MaRa1999}). 

We now recall from \cite{MaWe1983} the abstract notion of Clebsch variables. Let $P$ be a Poisson manifold, thought of as the phase space  of a physical system. By \textit{Clebsch variables} (or \textit{symplectic variables}) for this system, we mean a symplectic manifold $R$ and a Poisson map
\begin{equation}\label{Clebsch_variable}
\Psi:R\rightarrow P.
\end{equation}
Classical examples of such Clebsch variables are the \textit{Cayley-Klein} parameters for the free rigid body, the \textit{Kustaanheimo-Stiefel} coordinates in quantum mechanics, and, of course, the \textit{classical Clebsch variables} in various fluid dynamical models (homogeneous incompressible, isentropic, MHD, etc), see \cite{HoKu1983}. Since the momentum map $ \mathbf{J} :T^*Q \rightarrow \mathfrak{g}  ^\ast $ is a Poisson map with respect to the canonical symplectic structure on $T^*Q$ and the (noncanonical) Lie-Poisson structure \eqref{LP_bracket} on $\mathfrak{g}^{\ast}  $, the optimal control condition $\delta\ell/\delta \xi =\mathbf{J}(\alpha_q )$, naturally defines a Clebsch representation for the momentum $\delta\ell/\delta \xi$. More precisely, 
the noncanonical Hamiltonian motion of the control is found to arise from canonical Hamilton's equations for the collective Hamiltonian, via a momentum map. Such an observation is particularly relevant for the dynamics of the singular solutions of the Camassa-Holm equations and their generalizations (called EPDiff), which arise from the momentum map
\[
( \mathbf{Q} , \mathbf{P} ) \in T^* \operatorname{Emb}(S,M) \mapsto  \mathbf{m} =\int_S \mathbf{P} (s) \delta (x- \mathbf{Q} (s))ds \in \mathfrak{X}  (M) ^\ast,
\]
(\cite{HoMa2004}) as we will recall later. This expression of $ \mathbf{m} $ in terms of $(\mathbf{Q},\mathbf{P})$  is a Clebsch representation which yields a singular solution of the EPDiff
equations in terms of canonical variables that obey canonical Hamilton's equations on the cotangent bundle of the space $ \operatorname{Emb}(S,M)$ of all embeddings $ \mathbf{Q} :S \rightarrow M$.

\subsection{The geometry of jet bundles and classical field theory}\label{subsec_geometryofjetbundles} 

In this section, we recall some basic facts about the geometry of classical field theory. This includes the notion of jet bundles, dual jet bundles, canonical multisymplectic forms, covariant momentum maps, covariant Euler-Lagrange and Hamilton equations. For detailed informations, the reader can consult e.g. \cite{GiMaSa1997}, \cite{GiMmsy}, \cite{EEMLRR2000}, \cite{CaGuMa2012} and references therein.

\paragraph{Jet bundles.} Let $ \pi _{X,Y} : Y \rightarrow  X$ be a fiber bundle
which we call the \textit{covariant configuration bundle}. The fiber at $x \in X$ is denoted $Y_x= \pi _{X,Y} ^{-1} (x)$. The physical fields are sections
of this bundle, that is, smooth maps $ \phi : X \rightarrow Y$ such that $ \pi \circ \phi = id_X$. We suppose that $ \operatorname{dim}X=n+1$ and $ \operatorname{dim}Y=n+N+1$. Coordinates on $X$ are denoted $x ^\mu $, $ \mu =0,...,n$, and fiber coordinates on $Y$ are denoted $y ^A $, $A=1,...,N$, so that a section reads locally $ \phi (x ^\mu )=( x ^\mu , y ^A ( x ^\mu ))$.

The \textit{first jet first bundle} $ \pi _{Y, J ^1 Y}: J ^1 Y \rightarrow Y$ is the affine bundle over $Y$ whose fiber at $y\in Y_x$ is
\[
J ^1 _yY:=\left\{  \gamma _y \in L(T_x X, T_y Y)\mid T \pi _{X,Y} \circ \gamma_y  = id_{T_xX} \right\},
\]
where $L(T_x X, T_y Y)$ denotes the space of linear maps $T_xX \rightarrow T_yY$.
It is the covariant analogue of the tangent bundle in classical mechanics.
Given a section $ \phi $ of $ \pi _{X,Y}: Y \rightarrow X$, the map $x\mapsto j ^1 \phi (x):=T_x\phi $ defines a section of the fiber bundle $\pi _{X,J^1Y}: J ^1 Y\rightarrow X$. This section is called the first jet prolongation of $ \phi $ and reads locally
\[
x ^\mu \mapsto ( x ^\mu , y ^A ( x ^\mu ), \partial _\nu y ^A (x ^\mu ) ).
\]

\paragraph{Dual jet bundles.} The field-theoretic analogue of the cotangent bundle or phase space of classical mechanics, is given by the \textit{dual jet bundle} $\pi _{Y,J^1Y^\star}:J ^1 Y ^\star\rightarrow Y$, defined as the vector bundle over $Y$ whose fiber at $y \in Y _x $ is the space
\begin{equation}\label{def_djb} 
J^1_yY^\star= \operatorname{Aff}(J^1_yY, \Lambda _x ^{n+1}X) 
\end{equation} 
of all affine maps $J^1_yY \rightarrow \Lambda _x ^{n+1}X$, where $ \Lambda ^{n+1}X$ is vector bundle of $(n+1)$ forms on $X$, supposed to be an orientable manifold.  Any affine map from $J^1_yY$ to $\Lambda _x ^{n+1}X$ can be locally written as
\[
v _\mu ^A \mapsto ( \pi + p _A ^\mu v _\mu ^A ) d^{n+1}x,
\]
where $d^{n+1}x= dx ^0 \wedge d x ^1 \wedge ... \wedge dx^n$, so that coordinates on $J^1Y ^\star$ are given by $( x ^\mu , y ^A , p _A ^\mu, \pi  )$.

The dual jet bundle $\pi _{Y,J^1Y^\star}:J^1Y^\star \rightarrow Y$ is canonically isomorphic to the vector bundle $ \pi _{Y,Z} : Z \rightarrow Y$, whose fiber at $y \in Y$ is given by
\begin{equation}\label{def_Z} 
Z_y=\left\{ z _y \in \Lambda _y ^{n+1}Y\mid \mathbf{i} _v \mathbf{i} _w z _y =0, \;\; \text{for all $v,w \in V_yY$}  \right\},
\end{equation} 
where $ \mathbf{i} _v $ denotes the interior multiplication by $v$ and $V_yY:= \operatorname{ker}(T_y \pi _{X,Y} ) \subset T_yY$ is the vertical subspace. The canonical isomorphism is provided by the vector bundle map
\[
\Phi : Z \rightarrow J^1Y^\star, \quad \left\langle \Phi (z_y), \gamma _y \right\rangle := \gamma _y ^\ast z,\;\;\text{for all $z _y \in Z_y$ and $ \gamma _y \in J^1_yY$}. 
\]
Locally the inverse of this isomorphism reads 
\[
\left( \pi + p _A ^\mu \partial _\mu  \otimes dy ^A \right) d^{n+1}x \mapsto  \pi d^{n+1}+ p _A ^\mu dy ^A \wedge d ^n x _\mu, 
\]
where $d ^n x _\mu:= \mathbf{i} _{ \partial _\mu } d ^{n+1}x$.

The \textit{restricted dual jet bundle} $\pi _{Y \Pi }: \Pi \rightarrow Y$ is, by definition, the vector bundle over $Y$ whose fiber at $y$ is
\begin{equation}\label{def_rdjb} 
\Pi _y:=L(L(T_xX, V_yY), \Lambda _x ^{n+1}X).
\end{equation} 
The total space $ \Pi $ can be seen as the quotient of $J^1_y Y^\star$ by constant affine transformations along the fibers of $ \pi _{Y,J^1Y}$, so that coordinates on $ \Pi $ are given by $ ( x ^\mu , y ^A , p_A^\mu)$ and the quotient map $ \mu : J^1Y^\star \rightarrow \Pi $ reads locally
\[
\mu ( x ^\mu , y ^A , p _A ^\mu , \pi )= ( x ^\mu , y ^A , p _A ^\mu).
\]
Note that $ \mu : J^1Y^\star \rightarrow \Pi $ is a line bundle, the fiber being isomorphic to $ \Lambda ^{n+1}_x X$.

\paragraph{Canonical multisymplectic Forms.} Analogous to the canonical
forms $\Theta _{can}= p _Ad q ^A  $ and $\Omega _{can}=d q ^A \wedge d p _A $ on a cotangent bundle, there are canonical forms on the dual jet bundle $J^1Y^\star\simeq Z$. The \textit{canonical $(n+1)$-form} $ \Theta \in \Omega ^{n+1}(Z)$ is defined by
\[
\Theta (z) ( u _0 , ..., u_{n}):= z (T \pi _{Y,J^1Y^\star}(u _0 ),..., T \pi _{Y,J^1Y^\star}(u _{n} ))= (\pi^\ast  _{Y,J^1Y^\star}z)(u _0 ,..., u _{n})
\]
and the \textit{canonical multisymplectic form} $ \Theta \in \Omega ^{n+2}(Z)$ is defined by
\[
\Omega := - \mathbf{d} \Theta .
\]
These forms are locally given by
\begin{equation}\label{local_expr_forms} 
\Theta = p_A  ^\mu d y ^A \wedge d ^n x _\mu + \pi d^{n+1}x \quad \text{and} \quad \Omega =  d y ^A \wedge dp_A^\mu \wedge d ^n x _\mu - d \pi \wedge d^{n+1}x.
\end{equation}

\paragraph{Lagrangian formalism.} Assume that $X$ is an oriented manifold. A \textit{Lagrangian density} is a smooth map $ \mathcal{L} : J ^1 Y \rightarrow \Lambda ^{n+1}X$ covering the identity on $X$. In local coordinates, we can write
\[
\mathcal{L} ( x ^\mu , y ^A , v _\mu ^A )=L(x ^\mu , y ^A , v _\mu ^A ) d^{n+1}x.
\]
Let $U \subset X$ be an open subset whose closure $\bar U$ is compact. Throughout the paper, we will assume that the boundary $ \partial U$ is regular enough to enable the use of Stokes' theorem.
Recall that a local section $ \phi :\bar U \subset X\rightarrow Y$ of $ \pi _{X,Y}$ is, by definition, smooth if for every point $x \in \bar U$ there is an open neighborhood $U_x$ of $x$ and a smooth section $ \phi _x :U_x \rightarrow Y$ extending $ \phi $.
A critical section of the action functional defined by $ \mathcal{L} $ is a smooth local section $ \phi :\bar U \subset X\rightarrow Y$ that satisfies
\begin{equation}\label{Ham_Principle} 
\left.\frac{d}{d\varepsilon}\right|_{\varepsilon=0} \int_{U} \mathcal{L} ( j^1 \phi _\varepsilon )=0,
\end{equation} 
for all smooth variations $ \phi_ \varepsilon : \bar U \subset X\rightarrow Y$ such that $\phi _0 = \phi $ and $ \phi _\varepsilon |_{ \partial U}= \phi |_{ \partial U}$. 

It is well-known that a section is critical if and only if it verifies the \textit{covariant Euler-Lagrange equations}
\[
\frac{\partial L}{\partial y ^A }( j ^1 \phi )- \frac{\partial }{\partial x ^\mu }\left(\frac{\partial L}{\partial v ^A _\mu }( j ^1 \phi ) \right) =0
\]
in local coordinates.

\paragraph{Hamiltonian formalism.} There are two equivalent field-theoretic analogues of the Hamiltonian function of classical mechanics: the Hamiltonian density and the Hamiltonian section.

For the approach using the Hamiltonian density, we follow the formulation given in \cite{CaGuMa2012} (see in particular the Appendix), which is independent of the choice of an Ehresmann connection on $ \pi _{X,Y}: Y \rightarrow X$ or a volume form on $X$.  By definition, a \textit{Hamiltonian density} is a smooth map $ \mathcal{H} :J^1Y^\star \rightarrow \Lambda ^{n+1}X$ covering the identity on $X$ such that
\begin{equation}\label{cond_Ham_dens} 
\mathbf{i} _X (\mathbf{d} \mathcal{H} + \Omega)=0 , \;\;\text{for all $X \in \mathfrak{X}^V  (J^1Y^\star) $} ,
\end{equation} 
where $ \mathfrak{X}^V  (J^1Y^\star)= \{X \in \mathfrak{X}  (J^1Y^\star)\mid T \mu \circ X=0\}$ is the space of all $ \mu $-vertical vector fields on the dual jet bundle. From the condition \eqref{cond_Ham_dens} we obtain that a Hamiltonian density is locally given by
\[
\mathcal{H} (x ^\mu , y ^A , p _A ^\mu , \pi )= (\pi + h_{loc}(x ^\mu , y ^A , p _A ^\mu))d^{n+1}x.
\]

A \textit{Hamiltonian section} is, by definition (see \cite{GiMaSa1997}), a smooth section $\mathsf{h}: \Pi \rightarrow J^1Y ^\star$ of the line bundle $ \mu : J ^1 Y^\star \rightarrow \Pi $. It is therefore locally written as
\[
\mathsf{h}( x ^\mu , y ^A , p_A ^\mu )=( x ^\mu , y ^A , p_A ^\mu, -h_{loc}(x ^\mu , y ^A , p_A ^\mu)).
\]

As the local expressions suggest, there is a bijective correspondence between Hamiltonian densities and Hamiltonian sections. This correspondence being given by
\[
\operatorname{im} \mathsf{h}= \mathcal{H} ^{-1} (0). 
\]

The Cartan $(n+1)$-form and $(n+2)$ forms associated to a Hamiltonian section $\mathsf{h}$ are defined by
\[
\Theta _\mathsf{h}:= \mathsf{h}^* \Theta \in \Omega ^{n+1}( \Pi ) \quad \text{and} \quad  \Omega  _\mathsf{h}:= \mathsf{h}^* \Omega \in  \Omega ^{n+2}( \Pi )
\]
and we have the relations
\[
\mu ^\ast \Theta _\mathsf{h}= \Theta - \mathcal{H} \quad \text{and} \quad  \mu ^\ast\Omega  _\mathsf{h}=\Omega + \mathbf{d} \mathcal{H}.
\]

On the Hamiltonian side, the physical fields of the theory are represented by local smooth sections $ \omega  : \bar U \subset X \rightarrow \Pi $ of the restricted dual jet bundle $ \pi _{ X ,\Pi }: \Pi \rightarrow X$. 
The covariant analogue of the classical Hamilton's phase space variational principle reads
\begin{equation}\label{covariant_phase_space}
\left.\frac{d}{d\varepsilon}\right|_{\varepsilon=0} \int_U \omega _\varepsilon  ^\ast\Theta _\mathsf{h}=0,
\end{equation} 
for all smooth variations $\omega  _ \varepsilon : \bar U \subset X\rightarrow \Pi $ of a given section $ \omega : \bar U \rightarrow \Pi $, such that $\omega  _0 = \omega  $ and $ \phi _\varepsilon |_{ \partial U}= \phi |_{ \partial U}$, where $ \phi_\varepsilon  := \pi _{Y, \Pi } \circ \omega  _\varepsilon: \bar U \subset X \rightarrow Y$.
A section $ \omega  $ is critical if and only if it verifies the \textit{covariant Hamilton equations}
\[
\omega  ^\ast (\mathbf{i} _V \Omega_\mathsf{h}) =0, \quad \text{for every $ \pi _{X, \Pi }$-vertical vector fields $V$ on $\Pi $.}
\]
In local coordinates, these equations take the standard form
\begin{equation}\label{covHE} 
\frac{\partial \omega  ^A }{\partial x ^\mu }= \frac{\partial h_{loc}}{\partial \omega  _A ^\mu }( \omega   ), \quad   \frac{\partial \omega  ^\mu _A  }{\partial x ^\mu }=- \frac{\partial h_{loc}}{\partial \omega  ^A }( \omega   ).
\end{equation} 

\paragraph{The Legendre transformations.} Given a Lagrangian density $ \mathcal{L} :J^1Y \rightarrow \Lambda ^{n+1}X$, the associated \textit{covariant Legendre transformation} is defined by
\[
\mathbb{F}  \mathcal{L} : J^1Y \rightarrow J^1Y^\star, \quad \left\langle \mathbb{F}  \mathcal{L} ( \gamma ), \gamma ' \right\rangle := \mathcal{L} ( \gamma )+ \left.\frac{d}{d\varepsilon}\right|_{\varepsilon=0} \mathcal{L} ( \gamma + \varepsilon ( \gamma - \gamma ')),
\]
where $ \gamma , \gamma '\in J^1_y Y$ for some $ y \in Y$. The \textit{reduced covariant Legendre transformation} is, by definition,
\[
\widehat{ \mathbb{F}  \mathcal{L} }: J^1Y \rightarrow \Pi , \quad \widehat{ \mathbb{F}  \mathcal{L} }:= \mu \circ \mathbb{F}  \mathcal{L}.
\]
In local coordinates, they read
\[
\mathbb{F}  \mathcal{L}( x ^\mu , y ^A , v_ \mu ^A )= \left(  x ^\mu , y ^A , \frac{\partial L}{\partial v_ \mu ^A } , L- \frac{\partial L}{\partial v_ \mu ^A } v_ \mu ^A \right) \quad \text{and} \quad  \widehat{ \mathbb{F}  \mathcal{L} }( x ^\mu , y ^A , v_ \mu ^A )= \left(  x ^\mu , y ^A , \frac{\partial L}{\partial v_ \mu ^A } \right).
\]

The Lagrangian density is said to be \textit{hyperregular} when $ \widehat{ \mathbb{F}  \mathcal{L} }: J^1Y \rightarrow \Pi $ is a diffeomorphism. In this case, it is possible to define the Hamiltonian section associated to $ \mathcal{L} $ as
\[
\mathsf{h}_ \mathcal{L} := \mathbb{F}  \mathcal{L} \circ \widehat{ \mathbb{F}  \mathcal{L} }^{-1} : \Pi \rightarrow J^1Y^\star.
\]
It is locally given by
\[
\mathsf{h}_ \mathcal{L} ( x ^\mu , y ^A , p_A ^\mu )= ( x ^\mu , y ^A , p_A ^\mu , L( x ^\mu , y ^A , v _\mu ^A )-p_A ^\mu  v_ \mu ^A),
\]
where, by the hyperregularity hypothesis, $v _\mu ^A $ is uniquely determined from the other variables by the condition $\frac{\partial L}{\partial v_ \mu ^A }= p_A ^\mu $. The associated Hamiltonian density $ \mathcal{H} _ \mathcal{L} : J^1Y^\star \rightarrow \Lambda ^{n+1}X$ is given locally by the expression
\[
\mathcal{H} _ \mathcal{L} (x ^\mu , y ^A , p _A ^\mu , \pi )= (\pi + p_A ^\mu  v_ \mu ^A-L( x ^\mu , y ^A , v _\mu ^A ))d^{n+1}x.
\]
In the hyperregular case, one checks that the covariant Euler-Lagrange and covariant Hamilton equations are equivalent.

\paragraph{Covariant momentum maps.} Let $ \mathcal{G} $ be a Lie group acting by bundle automorphisms on the left on the configuration bundle $ \pi _{X,Y}: Y \rightarrow X$. Let us denote by $ \psi : \mathcal{G} \times Y \rightarrow Y$ this action and by $ \psi ^X :\mathcal{G} \times X \rightarrow X$ the induced action on $X$. This action naturally induces $ \mathcal{G} $-actions on $J^1Y$ and $J^1Y^\star$. They are respectively given by
\[
\psi _g ^{J^1Y}: J^1Y \rightarrow J^1Y, \quad \psi _g ^{J^1Y}( \gamma _y ):= T \psi _g \circ \gamma _y \circ \psi ^X _{ g ^{-1} }
\]
and
\[
\psi _g ^{J^1Y^\star}: J^1Y^\star \rightarrow J^1Y^\star, \quad \psi _g ^{J^1Y^\star}( z_y ):= (\psi _{g^{-1} } ) ^\ast z _y ,
\]
where we identified $J^1Y^\star$ with $Z$. The lifted action on the dual jet bundle is a special covariant canonical transformation in the sense that its preserves the canonical $(n+1)$-form $ \Theta $ and, therefore, the multisymplectic form $ \Omega $:
\[
(\psi _g ^{J^1Y^\star}) ^\ast \Theta = \Theta \quad \text{and} \quad  (\psi _g ^{J^1Y^\star}) ^\ast\Omega  = \Omega, \quad \text{for all $g \in \mathcal{G} $.} 
\]

Recall (\cite{GiMmsy}) that the covariant momentum map associated to the lifted action $ \psi _g ^{J^1Y^\star}$ is given by
\begin{equation}\label{cov_momap_global} 
J: J^1Y^\star \rightarrow L( \mathfrak{g}  , \Lambda ^n J^1Y^\star), \quad J(z)( \xi )= \pi _{Y,Z} ^\ast \mathbf{i} _{ \xi _Y } z,
\end{equation} 
where $ \xi _Y \in \mathfrak{X}  (Y)$ is the infinitesimal generator associated to the action of $ \mathcal{G} $. It is locally given by
\begin{equation}\label{cov_momap} 
J(z) ( \xi )= (p_A^\mu \xi ^A + \pi \xi ^\mu )d^n x _\mu -p_A^\mu \xi ^\nu dy ^A \wedge d^{n-1}x_{ \mu \nu },
\end{equation} 
where $d^{n-1}x_{ \mu \nu }:= \mathbf{i} _{ \partial _\nu } \mathbf{i} _{ \partial _\mu } d^{n+1}x$ and $( \xi ^\mu , \xi ^A )$ is the local expression of $ \xi _Y $.
The map $J$ verifies the momentum map condition
\[
\mathbf{d} J_ \xi = \mathbf{i} _{ \xi _{J^1 Y^\star}} \Omega ,\quad \text{for all $ \xi \in \mathfrak{g}  $}, 
\]
where $ \xi _{J^1Y^\star} $ is the infinitesimal generator of the lifted action on $J^1Y^\star$ and $J_ \xi \in \Omega ^{n+1}(J^1Y^\star)$ is the $(n+1)$-form defined by $J_\xi (z):= J(z)(\xi) $.

\subsection{Covariant Euler-Poincar\'e reduction and $G$-strands}\label{covEP_Gstrands}

In this paragraph, we first recall the process of covariant Euler-Poincar\'e reduction on a trivial principal bundle. It extends the Euler-Poincar\'e theory recalled above in \S\ref{subsec_Clebsch}, to the field theoretic context. Covariant Euler-Poincar\'e reduction for principal bundle field theories has been developed by \cite{CaRaSh2000}, and was then extended to more general situations in \cite{CaRa2003} and \cite{ElGBHoRa2011}. In the second part of this paragraph, we 
relate the spacetime decomposition of these equations with the continuum spin chain or $G$-strand equations considered in \cite{Ho2011}, \cite{HoIvPe2012}.

\paragraph{Covariant Euler-Poincar\'e reduction.}
Suppose that the covariant configuration bundle is a trivial right principal bundle $\pi _{X,P}:P= X \times G \rightarrow X$. Since the bundle is trivial, the fiber $J^1P_{(x,g)}$ of the first jet bundle can be canonically identified with the vector space $L(T_x X , T_g G)$. Sections $ \phi $ of this bundle are identified with smooth maps $g:X \rightarrow G$, via the relation $ \phi (x)=(x, g (x))$.

Let $ \mathcal{L} :J ^1 P \rightarrow \Lambda ^{n+1}X$ be a Lagrangian density supposed to be right $G$-invariant, that is $ \mathcal{L} (TR_h \circ v_{(x,g)})= \mathcal{L} (v_{(x,g)})$, for all $h \in G$, $v_{(x,g)} \in L(T_x X , T_g G)$, where $TR_h: TG \rightarrow TG$ denotes the tangent map to right translation $R_h$ by $h$ on $G$. Since the quotient vector bundle $J^1P/G\rightarrow X$ can be identified with the vector bundle $L(TX, \mathfrak{g}  )\rightarrow X$, the map  $ \mathcal{L} $ induces a reduced Lagrangian density $\ell: L(TX, \mathfrak{g}  ) \rightarrow \Lambda ^{n+1}X$ defined by $\ell(TR_{g ^{-1} } \circ v_{(g,x)}  )= \mathcal{L} (v_{(x,g)})$. The relation between the field $g:X \rightarrow G$ of the theory and its reduced expression $ \sigma $ is
$\sigma = TR_{g ^{-1} } \circ \mathbf{d} g$, where $ \mathbf{d} g:TX \rightarrow TG$ is the tangent map to $g$. We will use the simpler notation $\sigma = \mathbf{d} g g ^{-1} $.
Hamilton's variational principle \eqref{Ham_Principle} for $ \mathcal{L} $ induces the following constrained variational principle for $\ell$:
\begin{equation}\label{cov_EP_VP} 
\delta \int_ {U} \ell( \sigma ) =0, \quad \text{for all $ \delta \sigma = \mathbf{d} \zeta + \operatorname{ad}_ \zeta \sigma $},
\end{equation} 
where $ \sigma :\bar U \subset X \rightarrow L(TX, \mathfrak{g}  )$ is a local section and $ \zeta :\bar U \subset X \rightarrow \mathfrak{g}  $ is an arbitrary smooth map such that $ \zeta |_ {\partial \bar U}=0$. The stationarity condition
yields the \textit{covariant Euler-Poincar\'e equations}
\begin{equation}\label{cove_EP} 
\operatorname{div} \frac{\delta\ell}{\delta \sigma}+ \operatorname{ad}^*_ \sigma \frac{\delta\ell }{\delta \sigma }=0,
\end{equation} 
where $ \frac{\delta \ell}{\delta \sigma }\in L(L(TX, \mathfrak{g}  ), \Lambda ^{n+1}X)$ is the functional derivative of $\ell$. In addition to the covariant Euler-Poincar\'e equation, the reduced field $ \sigma $ also satisfies the \textit{zero curvature condition}
\begin{equation}\label{compatibility_condition} 
\mathbf{d} \sigma = [ \sigma , \sigma ].
\end{equation}

Conversely, in order to reconstruct from \eqref{cove_EP} a solution $g:\bar U \subset X \rightarrow G$ of the covariant Euler-Lagrange equations for $ \mathcal{L} $, one has to assume the zero curvature condition \eqref{compatibility_condition}. 

We refer to \cite{CaRaSh2000} for the detailed derivation of these equations together with the full treatment on arbitrary principal bundles.

\paragraph{Spacetime decomposition and $G$-strand equations.} In the special case when the base manifold decomposes as $X= \mathbb{R}  \times M\ni (t,m)$, we can write the reduced field as $ \sigma (t,m)= \nu(t,m) dt+ \gamma (t,m)$, where $\nu (t,m) \in \mathfrak{g}  $ and $ \gamma(t,m) \in L(T_mM, \mathfrak{g}  )$. Note that the relation $ \sigma = \mathbf{d} gg ^{-1} $ between the initial field $g$ and its reduced expression yields the equalities
\[
\nu= \partial _t g g ^{-1} \quad \text{and} \quad \gamma = \mathbf{d}  _M  g g ^{-1},
\]
where $ \mathbf{d}  _M g:TM \rightarrow TG$ denotes the partial derivative with respect to the variable $m \in M$.
By writing explicitly the covariant Euler-Poincar\'e variational principle \eqref{cov_EP_VP} in this case, we get
\begin{equation}\label{VP_decomposed} 
\delta \int_U \ell( \nu , \gamma )=0, \quad \text{for all $ \delta \nu = \partial _t \zeta + \operatorname{ad}_ \zeta \nu $ and $ \delta \gamma = \mathbf{d} _M   \zeta + \operatorname{ad}_ \xi \gamma $}, 
\end{equation} 
where we choose $U=]t _1 , t _2 [ \times V$ with $V$ an open subset in $M$ and $ \zeta : \bar U \rightarrow \mathfrak{g}  $ is an arbitrary map with $ \zeta |_{\bar U}=0$. The covariant Euler-Poincar\'e \eqref{cove_EP} can thus be equivalently written as
\begin{equation}\label{Cov_EP_split} 
\frac{\partial }{\partial t}\frac{\delta \ell}{\delta \nu }+ \operatorname{ad}^*_ \nu  \frac{\delta\ell }{\delta \nu  }+\operatorname{div} _M\frac{\delta\ell}{\delta \gamma }+ \operatorname{ad}^*_ \gamma  \frac{\delta\ell }{\delta \gamma  }=0,
\end{equation} 
and the zero curvature condition becomes
\begin{equation}\label{ZCC}
\partial_t  \gamma - \mathbf{d} _M \nu = [ \nu , \gamma ] \quad\text{and} \quad    \mathbf{d} _M \gamma = [ \gamma , \gamma ],
\end{equation} 
see \cite{GBRa2010}.
\medskip

In the case $M= \mathbb{R}  $, the equations \eqref{Cov_EP_split}, together with the variational principle \eqref{VP_decomposed}, coincide with the ones considered in \cite{Ho2011} (see \S10) in connection with the dynamics of a continuum spin chain. These equations, also called \textit{$G$-strands}, were further studied in \cite{HoIvPe2012} for several remarkable choices for the group $G$, in connection with integrable classical chiral models. The case $G= \operatorname{Diff}( \mathbb{R}  )$, where $\operatorname{Diff}( \mathbb{R}  )$ denotes the diffeomorphism group of the real line,  is especially important. Indeed, \cite{HoIvPe2012} show that when the Sobolev $H ^1 $ metric is used in the Lagrangian, the $\operatorname{Diff}( \mathbb{R}  )$-strands equations admit peakon solutions in a similar way as the Camassa-Holm equations. We will recall this fact in detail later (in \S\ref{sing_sol}) and extend the peakon solution to sheets and filaments. Moreover, we will provide a covariant momentum map interpretation of these singular solutions, and prove that they undergo canonical covariant Hamiltonian dynamics.

Note that in the case of $G$-strands, denoting $\mathbb{R}  \times M= \mathbb{R}  \times \mathbb{R}  \ni (t,s)$, equations \eqref{Cov_EP_split} and the first equation in \eqref{ZCC} reduce to
\begin{equation}\label{GStrands} 
\frac{\partial }{\partial t}\frac{\delta \ell}{\delta \nu }+ \operatorname{ad}^*_ \nu  \frac{\delta\ell }{\delta \nu  }+\frac{\partial }{\partial s} \frac{\delta\ell}{\delta \gamma }+ \operatorname{ad}^*_ \gamma  \frac{\delta\ell }{\delta \gamma  }=0, \quad \partial_t  \gamma - \partial _s  \nu = [ \nu , \gamma ],
\end{equation} 
whereas the second equation in \eqref{ZCC} vanishes.

\begin{remark}[A classical Lagrangian reduction approach]{\rm Remarkably, the covariant Euler-Poincar\'e equations \eqref{VP_decomposed} can also be obtained via a classical (as opposed to covariant) Lagrangian reduction approach in the sense that the field
\[
g: X= \mathbb{R}  \times M \rightarrow G, \quad (t,m) \mapsto g(t,m)
\]
is interpreted as a curve $g: t \in \mathbb{R}  \rightarrow g (t) \in \mathcal{F} (M, G)$ in the infinite dimensional Lie group $ \mathcal{F} (M, G)$ rather than a section $(t,m) \in \mathbb{R}  \times M \rightarrow (t,m,g(t,m)) \in \mathbb{R}  \times M \times G$ of the principal bundle. This approach (developed in \cite{GBRa2010}) makes use of the process of affine Euler-Poincar\'e reduction for the Lie group $\mathcal{F} (M, G)$ acting by affine representation on the space $ \Omega ^1 (M, \mathfrak{g}  ) \ni \gamma $. In this case, the first equation in \eqref{ZCC} becomes an advection equation, while the second equation is a consequence of the initial value (zero, for instance) chosen for the variable $ \gamma $ and can be generalized to include nontrivial dynamic of the curvature $B= \mathbf{d} \gamma -[ \gamma , \gamma ]$. We refer to \cite{GBRa2009} for a detailed treatment of affine Euler-Poincar\'e reduction, and to \cite{GBRa2010} for its application to covariant Euler-Poincar\'e reduction.}
\end{remark}

\section{The covariant Clebsch variational principle}\label{Sec_3} 

\subsection{Covariant formulation of the standard Clebsch principle}

In this section we formulate the \textit{classical} Clebsch variational principle reviewed in \S\ref{subsec_Clebsch}, by using the geometric tools of covariant field theories recalled in \S\ref{subsec_geometryofjetbundles}. This will be crucial to introduce the \textit{covariant} Clebsch variational principle in the next paragraph.

Recall from \S\ref{subsec_Clebsch} that, for a given Lie group action $ \Phi :G \times Q \rightarrow Q$ and a Lagrangian $ \ell: \mathfrak{g}  \rightarrow \mathbb{R}  $, the Clebsch variational principle is
\begin{equation}\label{Clebsch_var_principle} 
\delta \int_{0}^{T}\left( \ell( \xi )+ \left\langle \alpha , \dot q- \xi _Q (q) \right\rangle\right)  dt=0
\end{equation} 
over the space of smooth curves $(\xi,  \alpha) : [0,T ] \rightarrow \mathfrak{g}  \times T^*Q$, such that $q(0)=q _0 $ and $q(T)=q_T$, where $q(t) \in Q$ is the curve induced by $ \alpha (t) \in T^*Q$ on $Q$.
We now rewrite the integrand in a covariant form suitable for the extension from $ \mathbb{R}$ to an arbitrary base manifold $X$. Defining the function $e(\alpha _q , t,\xi ):= \left\langle \mathbf{J} ( \alpha _q ), \xi \right\rangle - \ell( \xi )$, we can write
\begin{align*} 
(\ell( \xi )+ \left\langle \alpha_q  , \dot q- \xi _Q (q) \right\rangle) dt &=\left( \ell( \xi )- \left\langle \mathbf{J} ( \alpha _q ), \xi \right\rangle + \left\langle \alpha_q  , \dot q \right\rangle \right) dt\\
 &=-e(\alpha _q , t, \xi )dt+\left\langle \alpha , \dot q \right\rangle dt\\
&=\psi ^\ast \mathsf{e} ^\ast \Theta (t),
\end{align*} 
where the map $ \psi $, defined by $ \psi (t):= (\alpha (t),t , \xi (t))$ is interpreted as a section of the trivial bundle
\[
T^*Q \times \mathbb{R}  \times \mathfrak{g}  \rightarrow \mathbb{R}, \quad ( \alpha _q , t, \xi ) \mapsto t;
\]
the function $ \mathsf{e}$, defined by $\mathsf{e}( \alpha _q,t , \xi ):=( \alpha _q,t , -e(t, \alpha _q , \xi ), \xi )$ is interpreted as a section of the trivial bundle
\[
T^*(Q \times \mathbb{R} ) \times \mathfrak{g} \rightarrow T^*Q \times \mathbb{R}  \times \mathfrak{g}, \quad ( \alpha _q ,t, \pi , \xi ) \mapsto ( \alpha _q ,t, \xi );
\]
and the one-form $ \Theta $ on $T^*(Q \times \mathbb{R}  ) \times \mathfrak{g}  $ is defined by $ \Theta( \alpha _q,t,\pi) = \pi dt+ \Theta _{can}( \alpha _q )$, where $ \Theta _{can} \in \Omega ^1 (T^*Q)$ is the canonical one-form on $T^*Q$.

At this point one recognizes that for the covariant configuration bundle of classical mechanics, namely
\[
Y= \mathbb{R}  \times Q \rightarrow \mathbb{R}, \quad (t,q) \rightarrow t,
\]
we have (from \eqref{def_djb} \eqref{def_rdjb})
\[
J^1Y^\star= T^*( Q \times \mathbb{R}  )\ni ( \alpha _q , t, \pi ) \quad\text{and} \quad   \Pi =T^*Q \times \mathbb{R}  \ni ( \alpha _q , t).
\]
We now incorporate the Lie algebra $ \mathfrak{g}  $ in each $Y$-fiber by considering the bundles
\begin{equation}\label{extended_mathfrakg}
J^1Y^\star_ \mathfrak{g}  :=J^1Y^\star\times _Y \mathfrak{g} \quad\text{and} \quad\Pi _ \mathfrak{g}  := \Pi \times _Y \mathfrak{g}  
\end{equation} 
over $Y= Q \times \mathbb{R}  $, so that $ \psi $ and $ \mathsf{e}$ can be seen, respectively, as sections of the bundles $\Pi _ \mathfrak{g}  \rightarrow \mathbb{R}  $ and $J^1Y^\star_ \mathfrak{g} \rightarrow \mathbb{R}  $. The variational principle \eqref{Clebsch_var_principle} can thus be rewritten as
\[
\delta \int_{ 0}^{T } \psi   ^\ast \mathsf{e}^\ast \Theta =0.
\]
It is this geometric formulation that we will use to define the covariant Clebsch variational principle in the next paragraph.

Note also that the action of $G$ on $Q$ naturally induces an action by bundle automorphisms on $Y= Q \times \mathbb{R}$ obtained by acting trivially on $ \mathbb{R}$. The covariant momentum map $J$ associated to this action recovers the usual momentum map $ \mathbf{J} $ on $T^*Q$ since, from formula \eqref{cov_momap}, we get
\[
J(\alpha _q , t , \pi )= \mathbf{J}  ( \alpha _q ).
\]
Therefore, the section $ \mathsf{e}$ can be expressed in terms of the covariant momentum map as follows
\begin{equation}\label{mathsfe_cassical_clebsch} 
\mathsf{e}( \alpha _q,t , \xi ):=( \alpha _q,t , -e(t, \alpha _q , \xi ), \xi ), \quad e(t, \alpha _q , \xi ) =\left\langle J(\alpha _q , t , \pi ), \xi \right\rangle - \ell( \xi ).
\end{equation}

\subsection{The covariant Clebsch variational principle and its properties}

\subsubsection{Definition of the covariant Clebsch variational principle}

In order to define the covariant Clebsch variational principle, we fix a fiber bundle $ \pi _{X,Y}:Y \rightarrow X$ and an action $ \psi _g :Y \rightarrow Y$  of a Lie group $ \mathcal{G} $. We suppose that $ \psi _g$ covers the identity on $X$, that is $ \pi _{X,Y} \circ \psi _g = \pi _{X,Y}$. The coordinates of a Lie algebra element $ \xi \in \mathfrak{g}  $ will be denoted by $ \xi ^\alpha $ and the coordinates of the associated infinitesimal generator $ \xi _Y $ by $ ( \xi ^\mu , \xi ^A )$.

Let $L(TX, \mathfrak{g}  )\rightarrow X $   be the vector bundle whose fiber at $x$ consists of linear maps $ \sigma_x  \in L(T_xX, \mathfrak{g}  )$, where $ \mathfrak{g}  $ is the Lie algebra of $ \mathcal{G} $, and fix a map $\ell: L(TX, \mathfrak{g}  ) \rightarrow \Lambda ^{n+1}X$ covering the identity on $X$.

Such a map can be obtained by reduction of a $ \mathcal{G} $-invariant Lagrangian density $ \mathcal{L} :J^1P\rightarrow \Lambda ^{n+1}X$, defined on the first jet bundle of the trivial principal bundle $P:= X \times \mathcal{G} \rightarrow X$, as recalled in the last paragraph of \S\ref{subsec_geometryofjetbundles}.

As in \eqref{extended_mathfrakg}, we need to extend the line bundle $ \mu : J^1Y^\star \rightarrow \Pi $ in order to incorporate the vector bundle $L(TX , \mathfrak{g}  ) \rightarrow X$. This is done by considering the line bundle 
\begin{align*} 
\mu _ \mathfrak{g}  :J^1Y^\star_ \mathfrak{g} \rightarrow \Pi _ \mathfrak{g}  , \quad \text{where}\quad   &J^1Y^\star_ \mathfrak{g}:=J^1Y^\star \times _Y \pi _{X,Y} ^\ast L(TX, \mathfrak{g}  )\\
&\Pi _ \mathfrak{g}  :=  \Pi \times _Y \pi _{X,Y} ^\ast L(TX, \mathfrak{g}  )\\
&\mu _ \mathfrak{g}  ( z, \sigma ):= ( \mu (z) , \sigma ).
\end{align*}
In order to formulate the variational principle, one needs to define a section $ \mathsf{e}$ of this line bundle
which appropriately generalizes the expression \eqref{mathsfe_cassical_clebsch}. We will use the following reformulation of the covariant momentum map of the $ \mathcal{G} $-action, namely, we define the map
\begin{equation}\label{mathcalJ} 
\mathcal{J} :J^1Y^\star \rightarrow L( \mathfrak{g}  , \Lambda ^n Y), \quad \mathcal{J} (z)( \xi ):= \mathbf{i} _{ \xi _Y } z.
\end{equation} 
We thus have the relation $J( z)( \xi )= \pi _{Y,J^1Y^\star}^\ast ( \mathcal{J} (z)( \xi ))$, where $J: J^1Y^\star \rightarrow L( \mathfrak{g}  , \Lambda ^n J^1Y^\star)$ is the covariant momentum map as defined in \eqref{cov_momap_global}.

The construction of the section $\mathsf{e}$ will use the following lemma.

\begin{lemma} Let $ \mathcal{G} $ be a Lie group acting on $\pi _{X,Y}:Y\rightarrow X$ by bundle automorphisms. Then, for all $ \sigma _x \in L(T_xX, \mathfrak{g}  )$ and $z_y  \in Z_y\simeq J^1_yY^\star$, where $ \pi _{X,Y}(y)=x$, we have
\[
\pi _{X,Y} ^\ast (\sigma _x) \wedge \mathcal{J} (z_y ) \in J^1_yY^\star,
\]
where, by definition, the wedge product involves a contraction of the Lie algebra indices. We thus obtain a well defined bilinear bundle map
\[
\tilde{j}:J^1Y^\star _ \mathfrak{g} = J^1Y^\star \times _Y\pi _{X,Y} ^\ast  L(TX, \mathfrak{g}  ) \rightarrow J^1Y^\star , \quad \tilde{j}(z_y , \sigma _x )= \pi _{X,Y} ^\ast (\sigma _x) \wedge \mathcal{J} (z)
\]
covering the identity on $Y$.

When the $ \mathcal{G} $-action covers the identity on $X$, this map induces a well defined bilinear bundle map
\begin{equation}\label{bundle_map_j} 
j:\Pi _ \mathfrak{g}=\Pi  \times _Y \pi _{X,Y} ^\ast L(TX,\mathfrak{g}  ) \rightarrow J^1Y^\star 
\end{equation} 
such that $j \circ \mu =\tilde{j}$.
\end{lemma} 
\textbf{Proof.} Since $ \pi _{X,Y} ^\ast (\sigma _x) \in L(T_y Y, \mathfrak{g}  )$ and $\mathcal{J} (z) \in L( \mathfrak{g}  , \Lambda ^n_y  Y)$, we have  $\pi _{X,Y} ^\ast (\sigma _x) \wedge \mathcal{J} (z)\in \Lambda ^{n+1}Y$. So, to prove that this is an element in the subspace $Z_y\subset \Lambda ^{n+1}_yY$, we need to check that it verifies the definition \eqref{def_Z}. This is trivially the case as the one-form $\pi _{X,Y} ^\ast (\sigma _x)$ annihilates vertical vectors.

If the action covers the identity on $X$, the infinitesimal vector field $ \xi _Y $ is vertical. Thus, from the local expression of the momentum map \eqref{cov_momap}, it follows that $ \mathcal{J} $ does not depend on the coordinates $ \pi $ in any local charts. This proves that the expression $ \mathcal{J}$ only depends on the projection $ \mu (z) \in \Pi $ of $z$. We thus obtain the desired result. $\qquad\blacksquare$ 

\bigskip

In order to define the section $ \mathsf{e}$, we need to fix an Ehresmann connection on the fiber bundle $ \pi _{X,Y}:Y \rightarrow X$, that is, a smooth subbundle $HY$ of $TY$ called the horizontal subbundle such that $TY=HY \oplus VY$. This is equivalent to the choice of a smooth $VY$-valued one-form $ \gamma \in \Omega ^1 (Y,VY)$ such that $ \gamma |_{VY}=id_{VY}$. In coordinates, the evaluation of  $\gamma$ on a tangent vector to $Y$, namely $( v ^\mu , v ^A )$ is written as $(0, v^A +\gamma _\mu ^A v ^\mu )$.
As shown in \cite{GiMaSa1997}, such a connection defines a linear section $\mathsf{s}^ \gamma $ of the line bundle $ \mu : J^1Y^\star \rightarrow \Pi$ given by
\[
\mathsf{s}^ \gamma : \Pi =L(L(TX, VY), \Lambda ^{n+1}X) \rightarrow J^1Y^\star, \quad \mathsf{s} ^\gamma ( v \otimes \alpha \otimes \mu ):= \gamma ^\ast \alpha \wedge \mathbf{i} _v \mu,
\]
where $ \alpha \in VY ^\ast $, $ \mu \in \Lambda ^{n+1}X$, and $v \in TX$. Locally, the section reads
\[
\mathsf{s} ^\gamma ( x ^\mu , y ^A , p _A ^\mu )=  (x ^\mu , y ^A , p _A ^\mu , p _A ^\mu \gamma _\mu ^A ).
\]

By assuming that  the $ \mathcal{G} $-action covers the identity on $X$, we can now define the map $\mathsf{e}$ as
\begin{equation}\label{def_section_e} 
\mathsf{e}( \omega , \sigma ):=\left( \mathsf{s} ^\gamma ( \omega )-j( \omega , \sigma )+ \pi _{X,Y} ^\ast \ell( \sigma ), \sigma \right) 
\end{equation} 
and we have the following result.

\begin{lemma} The map $\mathsf{e}$ is a section of the vector bundle
\[
\mu _ \mathfrak{g}  : J^1Y^\star_ \mathfrak{g}  \rightarrow \Pi _\mathfrak{g}.
\]
It is locally given by
\begin{equation}\label{local_e} 
\mathsf{e}( x ^\mu , y ^A , p _A ^\mu , \sigma^\alpha _\mu  )=( x ^\mu , y ^A , p _A ^\mu , -p _A ^\mu ( \sigma _\mu ^A - \gamma _\mu ^A )+  l( \sigma _\mu ^\alpha ), \sigma^\alpha _\mu ),
\end{equation} 
where $l$ is the local expression of $\ell$ defined by $\ell( \sigma _\mu ^\alpha )=l( \sigma _\mu ^\alpha ) d^{n+1}x$.
\end{lemma} 
\textbf{Proof.} It suffices to show that $ \mu _ \mathfrak{g}  \circ \mathsf{e}= id_{ \Pi _ \mathfrak{g}  }$. We have
\begin{align*} 
\mu _ \mathfrak{g}  ( \mathsf{e}( \omega , \sigma ))&= (\mu (\mathsf{s}^ \gamma ( \omega )-j( \omega , \sigma )+ \pi _{X,Y} ^\ast \ell( \sigma )), \sigma )\\
&=(\mu (\mathsf{s}^ \gamma ( \omega )) -\mu (j( \omega , \sigma ))+ \mu (\pi _{X,Y} ^\ast \ell( \sigma )), \sigma )\\
&=( \omega , \sigma ),
\end{align*} 
where we used $\mu (\mathsf{s}^ \gamma ( \omega ))= \omega $ which follows from the fact that $ \mathsf{s}^ \gamma $ is a section of the line bundle $ \mu : J^1Y^\star \rightarrow \Pi $, and the equalities $\mu (j( \omega , \sigma ))= 0$, $\mu (\pi _{X,Y} ^\ast \ell( \sigma )))=0$ which both follow from the property $ \mu |_{ \pi _{X,Y} ^\ast \Lambda ^{n+1}X}=0$. To see that $j( \omega , \sigma )$ is the pull back by $ \pi _{X,Y}$ of a $(n+1)$-form on $X$, we use local coordinates. We have, locally,
\begin{align*} 
j( \omega , \sigma )&= \sigma ^\alpha \wedge \mathcal{J} (z)_\alpha = \sigma ^\alpha _\mu dx ^\mu \wedge \mathcal{J} (z)_\alpha= dx ^\mu \wedge \mathcal{J} (z)( \sigma _\mu )\\
&= dx ^\mu \wedge p_A ^\nu \sigma _\mu ^Ad^nx _\nu  = p_A ^\mu\sigma _\mu ^Ad^{n+1}x,
\end{align*} 
which yields the desired result.

The local expression \eqref{local_e} follows from the local expression of each term. $\qquad\blacksquare$

\medskip
Making use of the above considerations we can now state the main definition of this section.

\begin{definition}[Covariant Clebsch variational principle]\label{Def_CCVP} Let $ \pi _{X,Y}: Y \rightarrow X$ be a fiber bundle endowed with an Ehresmann connection $ \gamma $, let $ \mathcal{G} $ be a Lie group acting on $Y$ and covering the identity on $X$, and let $\ell:L(TX, \mathfrak{g}  ) \rightarrow \Lambda ^{n+1}X$ be a Lagrangian density. Define the $(n+1)$ form
\[
\Theta _\mathsf{e}:= \mathsf{e}^\ast  \pi _{J^1Y^\star,J^1Y^\star_ \mathfrak{g} } ^\ast \Theta \in \Omega ^{n+1}(J^1Y^\star_ \mathfrak{g} ),
\]
where $\mathsf{e}$ is the section associate to $\ell$ and $\gamma$ defined in \eqref{def_section_e}.
Let $U \subset X$ be an open subset whose closure $\bar U$ is compact, and let $ \psi : \bar U \subset X \rightarrow \Pi _ \mathfrak{g} $ be a local smooth section of $ \pi _{X ,\Pi _ \mathfrak{g}  }: \Pi _ \mathfrak{g}  \rightarrow X$. The covariant Clebsch variational principle is
\begin{equation}\label{def_cov_Clebsch} 
\delta \int_{ U} \psi ^\ast   \Theta _ {\mathsf{e}}   =0,
\end{equation} 
for all variations $\psi _ \varepsilon : \bar U \subset X\rightarrow  \Pi \times _Y L(TX, \mathfrak{g}  )$ of $ \psi $ (among smooth sections) such that $\psi _0 = \psi  $ and $ \phi _\varepsilon |_{ \partial U}= \phi |_{ \partial U}$, where $ \phi _\varepsilon$ is the section of $ \pi _{X,Y}: Y \rightarrow X$ induced from $ \psi _\varepsilon$.
\end{definition}

\subsubsection{Stationarity conditions}

In this paragraph we compute the equations characterizing the critical points of the variational principle \eqref{def_cov_Clebsch}.
We present the computations both in global formulation and in local coordinates.

\begin{proposition} A smooth section $ \psi :\bar U \subset X \rightarrow \Pi _ \mathfrak{g}  $ is a critical point of the covariant Clebsch variational principle if and only if it satisfies
\begin{equation}\label{critical_Clebsch} 
\psi ^\ast (\mathbf{i} _V \Omega _\mathsf{e})=0, \quad \text{for any $ \pi _{X, \Pi _ \mathfrak{g}  }$-vertical vector fields $V$ on $ \Pi _ \mathfrak{g}  $,}
\end{equation} 
where $ \Omega _ \mathsf{e}:= - \mathbf{d} \Theta _\mathsf{e}$.
\end{proposition} 
\textbf{Proof.} Since
\[
\delta \psi (x):= \left.\frac{d}{d\varepsilon}\right|_{\varepsilon=0} \psi _\varepsilon (x)\in V_{ \psi (x)} \Pi _ \mathfrak{g}  := \operatorname{ker}(T_{ \psi (x)} \pi _{ X, \Pi _ \mathfrak{g}  } ) \quad \text{and} \quad \delta \psi |_ { \partial U}=0,  
\]
one may assume without loss of generality that $ \psi _\varepsilon = \eta_\varepsilon \circ \psi $, where $ \eta _\varepsilon $ is the flow of a  $ \pi _{ X, \Pi _ \mathfrak{g}  }$-vertical vector field $V \in \mathfrak{X}  ( J ^1 Y ^\star_ \mathfrak{g}  )$ such that $V( \psi (x))=0$ for all $ x \in \partial U$. Let us denote by $ \pounds _V\alpha $ the Lie derivative of a $k$-form $ \alpha $. Recall that it verifies the Cartan's magic formula $ \pounds _V \alpha = \mathbf{i} _V \mathbf{d} \alpha + \mathbf{d} \mathbf{i} _V \alpha $. It follows that
\begin{align*} 
\delta \int_{ U} \psi ^\ast \Theta _ { \mathsf{e}} &=\left.\frac{d}{d\varepsilon}\right|_{\varepsilon=0} \int_{ U} \psi_\varepsilon ^\ast \Theta _ { \mathsf{e} } = \int_ U \psi ^\ast \pounds _V \Theta _ \mathsf{e}\\
&= \int_ U \psi ^\ast \left( -\mathbf{i} _ V \Omega _ \mathsf{e}+ \mathbf{d} ( \mathbf{i} _ V \Theta _ \mathsf{e})\right) \\
&=- \int_U \psi ^\ast \mathbf{i} _V \Omega _\mathsf{e}+\int_{ \partial U}\psi \mathbf{i} _V^\ast \Theta _\mathsf{e}=- \int_U \psi ^\ast \mathbf{i} _V \Omega _\mathsf{e},
\end{align*} 
for all $V$, where we utilized Stokes' theorem and $V( \psi (x))=0$ for all $ x \in \partial U$. Thus $ \psi $ is a critical point if and only if \eqref{critical_Clebsch} holds.$\qquad\blacksquare$ 

\medskip

We now compute the stationary condition in local coordinates.

\begin{lemma}\label{local_express}  The stationary condition \eqref{critical_Clebsch} reads locally
\begin{equation}\label{stationary_cond_local} 
\frac{\partial \psi _A^\mu  }{\partial x ^\mu } =\psi _B ^\nu( \gamma ^B _\nu - \sigma ^B_\nu)_{,A}, \quad \frac{\partial \psi ^A}{\partial x ^\mu }=\sigma _\mu ^A-\gamma _\mu ^A , \quad \frac{\delta \ell}{\delta \sigma _\mu ^\alpha }=d x ^\mu \wedge \mathcal{J} ( z) _\alpha,
\end{equation} 
where we denoted by $ \psi ( x ^\mu ) =( x ^\mu , \psi ^A ( x ^\mu ), \psi _A ^\nu ( x ^\mu ), \sigma ^\alpha _\nu ( x ^\mu ))$ the section $ \psi $ in local coordinates.
\end{lemma} 
\textbf{Proof.}  Using the local  expression of the multisymplectic form $ \Omega $, see \eqref{local_expr_forms}, we get
\[
\Omega _\mathsf{e}= d y ^A \wedge d p _A ^\mu \wedge d ^n x _\mu + \left( \frac{\partial e_{loc}}{\partial y ^A }d y ^A + \frac{\partial e_{loc}}{\partial p_A ^\mu }d p _A ^\mu + \frac{\partial e_{loc}}{\partial \sigma _\mu ^\alpha }d \sigma _\mu ^\alpha    \right)  \wedge d^{n+1}x  ,
\]
where $e_{loc}(x ^\mu , y ^A , p _A ^\mu , \sigma _\mu ^\alpha )= p _A ^\mu ( \sigma _\mu ^A - \gamma _\mu ^A )-  l( \sigma _\mu ^\alpha )$. Note that a $ \pi _{X, \Pi _{ \mathfrak{g}}}$-vertical vector field on $ \Pi _ \mathfrak{g}  $ reads locally $V=(V ^\mu =0, V ^A , V_A ^\mu , V _\mu ^\alpha )$. We thus have
\[
\mathbf{i} _V \Omega _\mathsf{e}=V ^A dp_A ^\mu \wedge d ^n x _\mu -V_A ^\mu dy ^A \wedge d ^n x _\mu +\left(  \frac{\partial e_{loc}}{\partial y ^A }V^A+\frac{\partial e_{loc}}{\partial p_A ^\mu }V _A ^\mu + \frac{\partial e_{loc}}{\partial \sigma _\mu ^\alpha }V_\mu ^\alpha    \right) d^{n+1}x,
\]
and, therefore,
\begin{align*}
\psi ^\ast ( \mathbf{i} _V \Omega )&= V ^A \left( \frac{\partial \psi _A ^\mu }{\partial x ^\mu }+ \frac{\partial e_{loc}}{\partial y ^A }  \right) +V _A ^\mu \left(- \frac{\partial \psi _A }{\partial x ^\mu }+ \frac{\partial e_{loc}}{\partial p_A ^\mu }  \right)+ V _\mu ^\alpha \frac{\partial e_{loc}}{\partial \sigma _\mu ^\alpha }=0. 
\end{align*} 
Since the vector field $V$ is arbitrary, we get the conditions
\begin{equation}\label{Hamiltonian_like_conditions} 
\frac{\partial \psi _A ^\mu }{\partial x ^\mu }=- \frac{\partial e_{loc}}{\partial y ^A }, \quad \frac{\partial \psi _A }{\partial x ^\mu }=\frac{\partial e_{loc}}{\partial p_A ^\mu } , \quad \frac{\partial e_{loc}}{\partial \sigma _\mu ^\alpha }=0.
\end{equation} 
Using the expression of $e_{loc}$ and the equality $ p _A ^\mu \sigma _\mu ^A d^{n+1} x= dx ^\nu \wedge \mathcal{J} (z) _\alpha \sigma _\nu ^\alpha $, the previous conditions read
\[
 \frac{\partial \psi _A^\mu  }{\partial x ^\mu } =\psi _B ^\nu( \gamma ^B _\nu - \sigma ^B_\nu)_{,A}, \quad \frac{\partial \psi ^A}{\partial x ^\mu }=\sigma _\mu ^A-\gamma _\mu ^A , \quad \frac{\delta \ell}{\delta \sigma _\mu ^\alpha }=d x ^\mu \wedge \mathcal{J} ( z) _\alpha.
\]
This proves the lemma. $\qquad\blacksquare$

\paragraph{Derivation in local coordinates.} It is also instructive to carry out the variational principle directly in local coordinates. Using the local expressions of $ \Theta $ and $\mathsf{e}$, see \eqref{local_expr_forms} and \eqref{local_e}, we obtain
\[
 \Theta _ {\mathsf{e}} = (l( \sigma _\mu ^\alpha )-p_A ^\mu ( \sigma _\mu ^A - \gamma _\mu ^A )) d^{n+1}x+p_A ^\mu dy ^A \wedge d^n x _\mu,
\]
so that, writing locally $\psi ( x ^\mu )=( x ^\mu , \psi ^A ( x ^\mu ), \psi ^A _\nu ( x ^\mu ), \sigma ^\alpha _\nu ( x ^\mu ))$, we get
\begin{align*} 
\psi ^\ast   \Theta _ {\mathsf{e}}& =(l( \sigma _\mu ^\alpha )-\psi _A ^\mu ( \sigma _\mu ^A - \gamma _\mu ^A ))d^{n+1}x+ \psi _A ^\mu \frac{\partial \psi ^A}{\partial x ^\nu }dx ^\nu \wedge d^n x _\mu \\
&= \left( \psi _A ^\mu\left( \frac{\partial \psi ^A}{\partial x ^\mu }+\gamma _\mu ^A - \sigma _\mu ^A \right) +l( \sigma _\mu ^\alpha )\right)  d^{n+1}x.
\end{align*} 
Applying the variational principle in local coordinates yields
\begin{align*} 
&\delta \int_U \left( \psi _A ^\mu\left( \frac{\partial \psi ^A}{\partial x ^\mu }+\gamma _\mu ^A - \sigma _\mu ^A \right) +l( \sigma _\mu ^\alpha )\right)  d^{n+1}x\\
&=\int_U \delta \psi _A ^\mu \left(  \frac{\partial \psi ^A}{\partial x ^\mu }+\gamma _\mu ^A - \sigma _\mu ^A \right)d^{n+1}x + \delta \sigma _\mu ^\alpha \left( -d x ^\mu \wedge \mathcal{J} ( z) _\alpha +\frac{\delta l}{\delta \sigma _\mu ^\alpha } d^{n+1}x\right) \\
& \qquad \qquad + \psi _A ^\mu \left( \frac{\partial \delta \psi ^A}{\partial x ^\mu } +( \gamma ^A _\mu - \sigma ^A _\mu)_{,B} \delta \psi ^B   \right) \\
&=\int_U \delta \psi _A ^\mu \left(  \frac{\partial \psi ^A}{\partial x ^\mu }+\gamma _\mu ^A - \sigma _\mu ^A \right)d^{n+1}x + \delta \sigma _\mu ^\alpha \left( -d x ^\mu \wedge \mathcal{J} ( z) _\alpha +\frac{\delta l}{\delta \sigma _\mu ^\alpha } d^{n+1}x\right) \\
& \qquad \qquad + \delta \psi ^A \left( ( \gamma ^B _\mu - \sigma ^B_\mu)_{,A}\psi _B ^\mu - \frac{\partial \psi _A^\mu  }{\partial x ^\mu } \right) d^{n+1}x+\int_U \frac{\partial }{\partial x ^\mu }( \psi _A ^\mu \delta \psi ^A ) d^{n+1}x.
\end{align*} 
Therefore, since the last term vanishes, we get the conditions
\[
\frac{\delta \ell}{\delta \sigma _\mu ^\alpha }=d x ^\mu \wedge \mathcal{J} ( z) _\alpha, \quad \frac{\partial \psi ^A}{\partial x ^\mu }=\sigma _\mu ^A-\gamma _\mu ^A , \quad \frac{\partial \psi _A^\mu  }{\partial x ^\mu } =\psi _B ^\nu( \gamma ^B _\nu - \sigma ^B_\nu)_{,A},
\]
which consistently recover the conditions \eqref{stationary_cond_local} obtained by writing the condition \eqref{critical_Clebsch} in local coordinates.

\bigskip

The following Theorem summarizes the results obtained in this paragraph.

\begin{theorem}\label{3_statements}   The following statements for a smooth section $ \psi : \bar U \subset X \rightarrow \Pi _ \mathfrak{g}  $ are equivalent:
\begin{itemize}
\item[\rm (1)] $ \psi $ is a critical point of the covariant Clebsch variational principle \eqref{def_cov_Clebsch}.
\item[\rm (2)] $ \psi ^\ast ( \mathbf{i} _V \Omega _\mathsf{e})=0$, for all $ \pi _{X, \Pi _ \mathfrak{g}  }$-vertical vector fields on $ \Pi _ \mathfrak{g}  $.
\item[\rm (3)] $ \psi $ satisfies the equations \eqref{stationary_cond_local}.
\end{itemize}
\end{theorem}

\subsubsection{Hamiltonian formulation}

Recall that, under the hypothesis that $\ell$ is hyperregular,  the critical curve $q(t)$ of the classical (as opposed to covariant) Clebsch variational principle \eqref{CVP}, is necessarily the projection to $Q$ of the canonical Hamiltonian flow on $T^*Q$ associated to the optimal Hamiltonian $H( \alpha _q)=\hat H( \alpha _q , \xi ^\star( \alpha _q ))$, see \eqref{Pontryagin_Hamiltonian}.
This is a particular instance of a well-known result in optimal control theory based on the Pontryagin maximum principle, see e.g. \cite{AgSa2004}. As already mentioned earlier, for the Clebsch optimal control problem, the optimal Hamiltonian coincides with the collective Hamiltonian $h \circ \mathbf{J} $ associated to the Lie-Poisson Hamiltonian $h$ obtained from the cost function Lagrangian $\ell$ by Legendre transformation.

We shall show that the same situation arises in the covariant case, namely, that the stationarity condition obtained from the covariant Clebsch variational principle admits, in the hyperregular case, a canonical covariant Hamiltonian formulation with respect to a Hamiltonian section $\mathsf{h}: \Pi \rightarrow J^1Y^\star$. This section is obtained from the section $\mathsf{e}$ by inserting the momentum map relation obtained from one of the stationarity condition.
This Hamiltonian section naturally involves the Hamiltonian density $h:L(L(TX, \mathfrak{g}  ), \Lambda ^{n+1}X) \rightarrow \Lambda ^{n+1}X$ associated to $\ell$ by Legendre transformation, i.e.
\begin{equation}\label{def_covLP_Ham} 
h( \nu _\alpha  ^\mu )= \nu _\alpha ^\mu \sigma _\alpha ^\mu d^{n+1}x- \ell( \sigma _\alpha ^\mu ), \quad \nu _\alpha ^\mu = \frac{\delta \ell}{\delta \sigma _\alpha ^\mu}. 
\end{equation} 
To make clear the distinction with the Hamiltonian density $\mathsf{h}$, we call $h$  the \textit{covariant Lie-Poisson Hamiltonian}. This terminology is justified by the fact that, from \eqref{def_covLP_Ham}, the covariant Euler-Poincar\'e equations \eqref{cove_EP} for $\ell$ are equivalent to the \textit{covariant Lie-Poisson equations} for $h$ given by
\begin{equation}\label{cov_LP} 
\operatorname{div}\nu + \operatorname{ad}^*_{ \frac{\delta h}{\delta \nu }} \nu =0.   
\end{equation}

\medskip

It will be convenient to define the following map, which is yet another formulation of the covariant momentum map. We define
\begin{equation}\label{cov_momap_reformulation} 
\mathfrak{j}: \Pi \rightarrow L(L(TX, \mathfrak{g}  ), \Lambda ^{n+1}X), \quad\mathfrak{j} ( \omega )_\alpha ^\mu := dx ^\mu \wedge \mathcal{J} (z) _\alpha,
\end{equation} 
where $ z \in J^1Y^\star$ is such that $ \mu (z) = \omega$. Note that $ \mathfrak{j} $ verifies
\[
\pi _{X,Y} ^\ast \left\langle \mathfrak{j} (\omega) , \sigma \right\rangle = j( \omega , \sigma ), \quad \text{for all $ \sigma \in L(TX, \mathfrak{g}  )$}, 
\]
where $j$ is the bilinear bundle map defined in \eqref{bundle_map_j}.

\begin{theorem}\label{omega_cov_Ham}  Suppose that the Lagrangian density $\ell$ is hyperregular, and define the associated Hamiltonian density $h:L(L(TX, \mathfrak{g}  ), \Lambda ^{n+1}) \rightarrow \Lambda ^{n+1}X$ by Legendre transformation.

Then, a section $\psi =( \omega , \sigma ):\bar U \subset X \rightarrow \Pi _ \mathfrak{g}  $ is a critical point of the Clebsch variational principle if and only if the section $ \omega $ is a solution of covariant Hamilton's equations
\[
\omega ^\ast (\mathbf{i} _V \Omega _\mathsf{h})=0, \quad \text{for all $V \in  \mathfrak{X}  ^V(\Pi  )$},
\]
i.e., in local coordinates
\[
\frac{\partial \omega  ^A }{\partial x ^\mu }= \frac{\partial h_{loc}}{\partial \omega  _A ^\mu }( \omega   ), \quad   \frac{\partial \omega  ^\mu _A  }{\partial x ^\mu }=- \frac{\partial h_{loc}}{\partial \omega  ^A }( \omega   ),
\]
relative to the Hamiltonian section $\mathsf{h}: \Pi \rightarrow J^1Y^\star$ defined by
\begin{equation}\label{cov_coll_Ham} 
\mathsf{h}( \omega )= \mathsf{s}^ \gamma (\omega) - \pi _{X,Y} ^\ast h( \mathfrak{j} (\omega) ).
\end{equation} 
\end{theorem} 
\textbf{Proof.}  We first check that inserting one of the stationary condition, namely $ \frac{\delta \ell}{\delta \sigma _\mu ^\alpha }= d x ^\mu \wedge \mathcal{J} (z)_\alpha = \mathfrak{j} ( \omega ) _\alpha ^\mu $, in the section $\mathsf{e}$ defined in \eqref{def_section_e} yields the desired expression $\mathsf{h}$ in \eqref{cov_coll_Ham}. This is the covariant analogue of the passage from the Pontryagin Hamiltonian $\hat H$ to the optimal Hamiltonian $H$, recalled in \eqref{Pontryagin_Hamiltonian}-\eqref{coll_ham}. We have
\begin{align*} 
\mathsf{e}( \omega , \sigma )&= ( \mathsf{s} ^\gamma ( \omega )-j( \omega , \sigma )+ \pi _{X,Y} ^\ast \ell( \sigma ), \sigma )\\
&= ( \mathsf{s}^ \gamma ( \omega )- \pi _{X,Y} ^\ast (\left\langle  \mathfrak{j} ( \omega ), \sigma \right\rangle -\ell( \sigma )), \sigma )\\
&=\left(  \mathsf{s}^ \gamma ( \omega )- \pi _{X,Y} ^\ast \left( \left\langle \frac{\delta \ell}{\delta \sigma } , \sigma \right\rangle -\ell( \sigma )\right) , \sigma \right) \\
&=\mathsf{s}^ \gamma (\omega) - \pi _{X,Y} ^\ast h( \mathfrak{j} (\omega) ),
\end{align*} 
which recovers the Hamiltonian section \eqref{cov_coll_Ham}. Locally, it reads
\[
\mathsf{h}( x ^\mu , y ^A , p _A ^\mu )=(( x ^\mu , y ^A , p _A ^\mu,-h_{loc}( x ^\mu , y ^A , p _A ^\mu ) )
\]
with $h_{loc}( x ^\mu , y ^A , p _A ^\mu ) =h( \mathfrak{j} ( x ^\mu , y ^A , p _A ^\mu)_\alpha ^\nu )- p _A ^\mu \gamma _\mu ^A = h( p _A ^\mu ( \partial _\alpha ) ^A )-  p _A ^\mu \gamma _\mu ^A$, where $ \partial _\alpha $ denote a basis of the Lie algebra $ \mathfrak{g}  $. We now compute the covariant Hamilton equations. Denoting by $ \nu _\alpha ^\mu $ the variable of $h$, and using the notation $\sigma_\mu ^\alpha := \frac{\partial h}{\partial \nu _\alpha ^\mu }$ for the functional derivative of $h$, we have  
\begin{align*}
\frac{\partial h_{loc}}{\partial p _A ^\mu }&= \frac{\partial h}{\partial \nu _\alpha ^\mu } ( \partial _\alpha ) ^A- \gamma _\mu ^A = \sigma _\mu ^\alpha ( \partial _\alpha ) ^A- \gamma _\mu ^A = \sigma _\mu ^A - \gamma _\mu ^A\\
\frac{\partial h_{loc}}{\partial y ^B}&= \frac{\partial h}{\partial \nu _\alpha ^\mu } p_A ^\mu ( \partial _\alpha ) ^A_{,B}- p_A  ^\mu( \gamma _\mu ^A) _{,B}= p _A ^\mu (\sigma _\mu ^\alpha ( \partial _\alpha ) ^A_{,B}-( \gamma _\mu ^A) _{,B})=p _A ^\mu (\sigma _\mu ^A-\gamma _\mu ^A) _{,B}.
\end{align*} 
From this we obtain that the canonical covariant Hamilton equations
\[
\frac{\partial \omega  ^A }{\partial x ^\mu }= \frac{\partial h_{loc}}{\partial \omega  _A ^\mu }( \omega   ), \quad   \frac{\partial \omega  ^\mu _A  }{\partial x ^\mu }=- \frac{\partial h_{loc}}{\partial \omega  ^A }( \omega   )
\]
coincide with the first two equations in \eqref{stationary_cond_local}. To check complete equivalence with the stationarity condition \eqref{stationary_cond_local}, it remains to check that the definition we took for $ \sigma _\mu ^\alpha $ in the present proof coincides with the third equation for $\sigma _\mu ^\alpha $ in \eqref{stationary_cond_local}. This follows by using the Legendre transformation relation between $h$ and $\ell$:
\[
h( \nu _\alpha ^\mu )=  \nu _\alpha ^\mu  \sigma _\alpha ^\mu  d^{n+1}x -\ell( \sigma _\mu ^\alpha ), \quad \text{with} \quad \nu_\alpha ^\mu =\frac{\partial \ell}{\partial \sigma _\mu ^\alpha }.  \qquad\blacksquare
\]

The Hamiltonian section $\mathsf{h}$ in \eqref{cov_coll_Ham} will be referred to as the \textit{covariant collective Hamiltonian}, since the relation \eqref{cov_coll_Ham} is the extension of the relation $H= h \circ \mathbf{J} $ to the field theoretic context. Note that, in general, an Ehresmann connection is needed to obtain a globally defined Hamiltonian section $\mathsf{h}$ from the covariant Lie-Poisson Hamiltonian $h$.

\subsubsection{Clebsch variables and the covariant Euler-Poincar\'e equations}

Recall from \S\ref{subsec_Clebsch} that the classical Clebsch variational principle reproduces the Euler-Poincar\'e equations for the cost function Lagrangian $\ell( \xi )$. We now prove a similar statement in the covariant case. To obtain this result, we need to assume that the open subset $U$ on which the sections $ \psi $ are defined is sufficiently small in order to allow the choice of the local trivial connection on $U$, which is simply the natural projection whose action in coordinates is $(0, v_A)$, i.e., the components $ \gamma_\mu ^A  = 0$.

\begin{theorem}\label{sigma_cov_EP}  Let $\psi= ( \omega , \sigma ): \bar U \subset X \rightarrow \Pi _ \mathfrak{g}  $ be a critical point of the covariant Clebsch variational principle
\[
\delta \int_U \psi ^\ast \Theta _\mathsf{e}=0.
\]
Then the section $ \sigma :\bar U \subset X \rightarrow L(TX, \mathfrak{g}  )$ is a solution of the covariant Euler-Poincar\'e equations
\begin{equation}\label{COV_EP} 
\operatorname{div} \frac{\delta \ell}{\delta \sigma }+ \operatorname{ad}^*_ \sigma \frac{\delta \ell}{\delta \sigma }=0.
\end{equation} 
\end{theorem} 
\textbf{Proof.} The proof follows from a computation in local coordinates. Let $ \xi : \bar U \subset X \rightarrow \mathfrak{g}  $ be an arbitrary smooth function. Using the stationarity conditions \eqref{stationary_cond_local} and the fact that we chose the locally trivial connection $\gamma _\mu ^A =0$, we have
\begin{align*}
\left\langle \partial _\mu  \frac{\delta \ell}{\delta \sigma _\mu  } , \xi \right\rangle & = \partial _\mu \left(\frac{\delta \ell}{\delta \sigma _\mu ^\alpha }  \xi ^\alpha \right) - \frac{\delta \ell}{\delta \sigma _\mu ^\alpha } \partial _\mu  \xi ^\alpha= \partial _\mu (\psi _A ^\mu \xi ^A) d^{n+1}x- \psi _A ^\mu \xi _{, \mu } ^A d^{n+1}x\\
&=\left( \partial _\mu \psi _A ^\mu \xi ^A + \psi _A ^\mu \left( \xi _{, \mu } ^A + \xi _{, B} ^A \frac{\partial \psi ^B }{\partial x ^\mu }\right) - \psi _A ^\mu \xi _{, \mu } ^A  \right) d ^{n+1}x\\
&= \psi _A ^\mu ( \xi ^A _{,B} \sigma _\mu ^B - \sigma _{ \mu , A} ^B \xi ^A )d ^{n+1}x\\
&= \psi _A ^\mu [ \sigma _\mu , \xi ]_{JL} ^A d ^{n+1}x= - dx ^\mu \wedge \mathcal{J} (z)_\alpha [ \sigma _\mu , \xi ] ^\alpha \\
&= -  \frac{\delta \ell}{\delta \sigma _\mu ^\alpha }  [ \sigma _\mu , \xi ] ^\alpha =- \left\langle \operatorname{ad}^*_{ \sigma _\mu } \frac{\delta \ell}{\delta \sigma _\mu  }, \xi \right\rangle,
\end{align*} 
which proves the result. In the fifth equality we used the local expression of the Jacobi-Lie bracket of vertical vector fields on $Y$. In the sixth equality, we used the formula $[ \xi _Y , \eta _Y ]_{JL}=-[ \xi , \eta ]_Y$ for a left Lie algebra representation (see e.g. \S9.3 in \cite{MaRa1999}) together with the definition of the mapping $ \mathcal{J}$. $\qquad\blacksquare$

\paragraph{Canonical VS noncanonical structures.} Recall that, in classical mechanics, Hamilton's equations are sometimes called \textit{noncanonical} when their underlying Poisson structure is not associated to a symplectic form. A lot of effort has been devoted to recast noncanonical Hamilton's equations into a canonical (i.e. symplectic) form by introducing auxiliary variables, like the well-known Clebsch variables in fluid mechanics, \cite{Cl1857}, \cite{Cl1859}, \cite{MaWe1983}, \cite{HoKu1983}. They represent a first attempt to understand these equations as Hamiltonian system, before the introduction of Poisson manifolds.

In the covariant case, and in order to emphasize the analogy with classical mechanics, we will use the same terminology to characterize the covariant Hamilton's equations arising in this paper: the covariant Lie-Poisson equations \eqref{cov_LP} will be referred to as \textit{noncanonical} as opposed to the \textit{canonical} covariant Hamilton equations \eqref{covHE} associated to the canonical multisymplectic form.

\paragraph{Curvature condition.} It is important to mention that the covariant Clebsch variational principle only yields the covariant Euler-Poincar\'e equations and that the zero curvature condition \eqref{compatibility_condition} may not be satisfied in general. In the Lemma below we make this statement more precise.

\begin{lemma}\label{lemma_CC}  Assume that $Y$ is a trivial bundle, $Y=X \times Q$ and let $ \psi = ( \omega , \sigma )$ be a critical point of the covariant Clebsch variational principle. Then the following relation is satisfied
\begin{equation}\label{curvature_condition_Q} 
(\mathbf{d} \sigma -[ \sigma , \sigma ])_Q(q)=0
\end{equation} 
as a skew-symmetric linear map $T_xX \times T_xX \rightarrow T_qQ$.

This condition is weaker than the zero curvature condition \eqref{compatibility_condition}. In the particular situation when the $G$-action on $Q$ is free, then \eqref{curvature_condition_Q} and \eqref{compatibility_condition} are equivalent.

When $ \operatorname{dim}X=1$, then both conditions are always satisfied. 
\end{lemma} 
\textbf{Proof.} Let $ \nabla $ be a torsion free connection on $Q$ and let $(t,s) \in \mathbb{R}  ^2 \rightarrow x(s,t)\in X$ be a smooth function. We have the relation
\begin{equation}\label{torsion_free_relation} 
\left.\frac{D^ \nabla }{Ds}\right|_{s=0} \left.\frac{d}{dt}\right|_{t=0} q(x(s,t))- \left.\frac{D^ \nabla }{Dt}\right|_{t=0} \left.\frac{d}{ds}\right|_{s=0} q(x(s,t))=0,
\end{equation}
where $D ^ \nabla /Ds$ denotes the covariant derivative associated to $ \nabla $.
Using the notations $x(s)=x(s,0)$, $x(t)=x(0,t)$, $\delta x(t):= \left.\frac{d}{ds}\right|_{s=0} x(s,t)$, $ \Delta x(s)=\left.\frac{d}{dt}\right|_{t=0} x(s,t)$, and the relation $ \mathbf{d} q(x)= \sigma _Q(q(x))$, we compute
\begin{align*} 
\left.\frac{D^ \nabla }{Ds}\right|_{s=0} \left.\frac{d}{dt}\right|_{t=0} q(x(s,t))&=\left.\frac{D^ \nabla }{Ds}\right|_{s=0}  \mathbf{d} q(x(s)) \cdot \delta x(s)= \left.\frac{D^ \nabla }{Ds}\right|_{s=0}  \big( \sigma(x(s)) \cdot \delta x(s)\big) _Q(q(x(s))\\
&= \nabla _{ \mathbf{d} q(x) \cdot \Delta  x} \big(\sigma (x)( \delta x) \big)_Q(q(x))+ \left( \left.\frac{d}{ds}\right|_{s=0} \sigma (x(s))\cdot \delta x (s) \right) _Q(q(x))
\end{align*} 
so that the relation \eqref{torsion_free_relation} reads
\begin{align*} 
0&=\nabla _{ \big(\sigma (x) \cdot  \Delta  x\big)_Q(q(x))} \big(\sigma (x)( \delta x) \big)_Q(q(x))- \nabla _{ \big(\sigma (x) \cdot  \delta   x\big)_Q(q(x))} \big(\sigma (x)( \Delta  x) \big)_Q(q(x))\\
&\qquad + \left( \left.\frac{d}{ds}\right|_{s=0} \sigma (x(s))\cdot \delta x (s) - \left.\frac{d}{dt}\right|_{t=0} \sigma (x(t))\cdot \Delta  x (t) \right) _Q(q(x))\\
&= \left[\big(\sigma (x) \cdot  \Delta    x\big)_Q, \big(\sigma (x) \cdot  \delta   x\big)_Q\right] _{JL}(q(x))+ \big(\mathbf{d} \sigma (x) \cdot ( \Delta x, \delta x)\big)_Q(q(x))\\
&=(- [ \sigma , \sigma ]+ \mathbf{d} \sigma )_Q(q(x)) \cdot ( \Delta x, \delta x),
\end{align*} 
where we used the property $[ \xi _Q , \eta _Q ]=-[ \xi ,\eta ]_Q$ for the infinitesimal generators of left action associated to Lie algebra elements $ \xi , \eta \in \mathfrak{g}  $, and a property of the exterior differential of a one-form. Since $ \delta  x, \Delta x \in T_xX$ are arbitrary, the result follows. $\qquad\blacksquare$

\paragraph{Covariant Clebsch variables.} From Theorem \ref{omega_cov_Ham}  we know that if $ ( \omega , \sigma )$ is a critical point of the covariant Clebsch variational principle, then $ \omega $ is necessarily a solution of the canonical covariant Hamilton equations for the collective Hamiltonian section. Furthermore, from Theorem \ref{sigma_cov_EP} we know that $ \sigma $ verifies the covariant Euler-Poincar\'e equations and therefore, by Legendre transformation, $\nu = \mathfrak{j} (\omega )$ is solution of the noncanonical covariant Lie-Poisson equations \eqref{cov_LP}.

This means that $ \omega $ is a Clebsch variable for the covariant Lie-Poisson equation system, the relation between the canonical and noncanonical equations being given by the map
\[
\omega \in \Pi \mapsto \mathfrak{j} ( \omega ) \in L(L(TX, \mathfrak{g}  ), \Lambda ^{n+1}X).
\]
Recall that this map is nothing else than a reformulation of the covariant momentum map associated to an action of the Lie group $ \mathcal{G} $ on the total space $Y$ of the fiber bundle.

Note finally that given a Lie group $ \mathcal{G} $ and a manifold $X$, any fiber bundle $ \pi _{X,Y}: Y \rightarrow X$ on which $ \mathcal{G} $ acts (with trivial action on $X$) provide Clebsch variables for the covariant Lie-Poisson equations on $L(TX, \mathfrak{g}  )$, by means of the covariant momentum map. 

The situation is summarized in the next theorem.

\begin{theorem}[Clebsch variables for the covariant Lie-Poisson equations] $\;\;\;$\\Consider a manifold $X$ and a Lie group $ \mathcal{G} $. Let $\ell:L(TX, \mathfrak{g}  ) \rightarrow \Lambda ^{n+1}X$ be an hyperregular Lagrangian density and denote by $h$ the associated Hamiltonian density.

Let $\pi _{X,Y}: Y \rightarrow X$ be a fiber bundle over $X$ and suppose that $ \mathcal{G} $ acts on $Y$ and induces the trivial action on $X$. Let $\mathsf{h}$ be the collective Hamiltonian section \eqref{cov_coll_Ham} defined locally with the help of the trivial connection, that is, we have
\[
\mathsf{h}( x ^\mu , y ^A , p _A ^\mu )=(x ^\mu , y ^A , p _A ^\mu, -h( \mathfrak{j} (x ^\mu , y ^A , p _A ^\mu))).
\]

If $\omega:\bar U \subset X \rightarrow \Pi $ is a solution of the (canonical) covariant Hamilton's equations for $\mathsf{h}$, then $ \mathfrak{j}( \omega ) \in L(L(TX, \mathfrak{g}  ), \Lambda ^{n+1}X)$ is solution of the covariant (noncanonical) Lie-Poisson equations for $h$. 
\end{theorem}

\section{Examples}\label{Sec_4} 

In this section we apply the covariant Clebsch variational principle (VP) to various particular cases of special interest. First we consider the case given by a trivial fiber bundle $Y= X \times Q$ and a Lie group action of $G$ on $Q$. Then we present the special cases when $Q$ is a vector space on which $G$ acts by representation and the case when $Q$ is a group on which $G$ acts by translation. This approach allows us to develop the covariant analog of several Clebsch variational formulations of the Euler-Poincar\'e equations in mechanics (\cite{GBRa2011}), such as the \textit{symmetric formulation of rigid bodies}, the \textit{coupled double bracket equations}, the \textit{Clebsch variables for perfect fluids}, and the \textit{singular solutions of EPDiff equations}. The approach developed here applies to any equations admitting a covariant Euler-Poincar\'e formulation, such as the $G$-strand equations for various choice of Lie groups $G$ and Lagrangian $\ell$, \cite{HoIvPe2012}. We consider in details the cases $G= \operatorname{Diff}(M)$ and $G=\operatorname{Diff}_{vol}(M)$, the associated covariant momentum maps and covariant Euler-Poincar\'e equations (called \textit{covariant EPDiff} and \textit{covariant EPDiff$_{vol}$}). In the first case, our approach allows for the construction of singular solutions (filaments, sheets, etc.) of the equations.

\subsection{Covariant Clebsch VP on trivial fiber bundles} We now consider in detail the case when the bundle $\pi_{X,Y} :Y\rightarrow X$ is trivial. We will denote by $Q$ the fiber so that we have $Y= X \times Q$. Note that a smooth section $ \phi :\bar U \subset X \rightarrow Y$ is of the form $\phi (x)=(x,q(x))$, where $ q: \bar U \subset X \rightarrow Q$.

Recall that the fiber of the first jet bundle is an \textit{affine space}. In the case of a trivial bundle it reads $J^1(X \times Q)_{(x,q)}=id_{T_xX}\times L(T_xX,T_qQ)$ so it can be canonically identified with the \textit{vector space} $L(T_xX, T_qQ)$.
This corresponds to the isomorphism $j^1 \phi (x) \mapsto \mathbf{d} q(x)$ for a section $ \phi (x)=(x, q(x))$.

The dual jet bundle and restricted dual jet bundle are given by
\[
\begin{aligned}
J^1(X \times Q)^\star &&=L(L(TX,TQ), \Lambda ^{n+1}X) \times \Lambda ^{n+1}X &&\;\;\text{and}\;\;\;   \Pi &&=L(L(TX,TQ), \Lambda ^{n+1}X)\\
&&= (TX \otimes T^*Q \otimes \Lambda ^{n+1}X) \times \Lambda ^{n+1}X&&  &&= (TX \otimes T^*Q \otimes \Lambda ^{n+1}X)
\end{aligned} 
\]
so that the line bundle $ \mu  : J^1(X \times Q)^\star \rightarrow \Pi  $ is trivial, and
therefore Hamiltonian sections can be identified with maps $\mathsf{h}: \Pi \rightarrow \Lambda ^{n+1}X$ covering the identity on $X$.

\paragraph{Covariant momentum maps.} Recall that in the field theoretic context, covariant momentum maps are defined on the dual jet bundle and take values in  the bundle $L( \mathfrak{g}  , \Lambda ^n J^1(X \times Q)^\star )$.

In the special case of lifted actions we have seen in  \eqref{cov_momap_global} that the value of the covariant momentum map is the pull-back of an expression in $L( \mathfrak{g}  , \Lambda ^n (X \times Q) )$. This was the motivation for the definition of the map $ \mathcal{J} :J^1(X \times Q)^\star  \rightarrow L( \mathfrak{g}  , \Lambda ^n (X \times Q))$ in \eqref{mathcalJ}. In the more special case of a group action on $Y$ covering the identity on $X$, further simplifications arise in the expression of $ \mathcal{J} $. In view of the its occurrence in the stationarity condition of the Clebsch variational principle, it was natural to reformulate the covariant momentum map in this case as a map
\[
\mathfrak{j} : L(L(TX,TQ), \Lambda ^{n+1}X) \rightarrow L(L(TX,\mathfrak{g}  ), \Lambda ^{n+1}X),
\]
see \eqref{cov_momap_reformulation}. In the case of a trivial bundle, the $\mathcal{G} $-action on $Y= X \times Q$ is the one naturally induced by a given $ \mathcal{G} $-action $\Phi$ on $Q$. We know check that the map $ \mathfrak{j} $ has a very simple expression in terms of the momentum map $ \mathbf{J} :T^*Q \rightarrow \mathfrak{g}  ^\ast $ of the cotangent lifted action $ \Phi ^{T^*}: \mathcal{G} \times T^*Q \rightarrow T^*Q$ of $ \Phi $. Indeed, using \eqref{mathcalJ} and \eqref{cov_momap_reformulation}, we have
\begin{align*}
\left\langle \mathfrak{j} ( \omega ), \sigma \right\rangle =\mathfrak{j}( \omega ) _\alpha ^\mu \sigma _\mu ^\alpha = dx ^\mu \wedge \mathcal{J} ( z) (\sigma _\mu )= dx ^\mu \wedge P _A ^\nu \sigma _\mu ^A d ^n x _\nu = P _A ^\mu \sigma _\mu ^A d^{n+1}x.
\end{align*} 
So, for $\omega =v_x  \otimes \alpha_q  \otimes \mu_x  $ and $ \sigma_x  \in L(TX , \mathfrak{g}  )$, we get
\begin{align*} 
\left\langle \mathfrak{j} ( v_x  \otimes \alpha_q  \otimes \mu_x  ) , \sigma_x  \right\rangle = \left\langle \alpha , \sigma_x  (v_x ) _Q \right\rangle  \mu = \left\langle \mathbf{J} ( \alpha_q  ), \sigma_x  (v_x )\right\rangle \mu_x  = \left\langle v_x  \otimes \mathbf{J} ( \alpha _q ) \otimes \mu_x  , \sigma_x  \right\rangle.
\end{align*} 
Therefore, we have the formula
\[
\mathfrak{j} ( v_x  \otimes \alpha_q  \otimes \mu_x  )= v_x  \otimes \mathbf{J} ( \alpha_q  ) \otimes \mu_x .
\]
It will be also useful to write $ \mathfrak{j}$ in coordinates as follows
\begin{align}\label{rewriting_J}  
\mathfrak{j} ( \omega )&= \mathfrak{j} ( \omega _A ^\mu \partial _\mu \otimes d y ^A \otimes d ^{n+1}x)= \omega _A ^\mu \partial _\mu \otimes \mathbf{J} (d y ^A ) \otimes d^{n+1}x\nonumber\\
& =  \partial _\mu \otimes \mathbf{J} (\omega _A ^\mu d y ^A ) \otimes d^{n+1}x=   \partial _\mu \otimes \mathbf{J} (\omega ^\mu  ) \otimes d^{n+1}x.
\end{align}

\paragraph{The covariant Clebsch variational principle.} In the trivial bundle case, the vector bundle $ \Pi _ \mathfrak{g}  $ reads
\[
\Pi _ \mathfrak{g}  = L(L(TX, TQ), \Lambda ^{n+1}X) \times_{X \times Q} L(TX , \mathfrak{g}  ),
\]
and we will denote by $ \psi = ( \omega , \sigma ):\bar U \subset X \rightarrow \Pi _ \mathfrak{g}  $ the smooth sections of this bundle. Note that such section naturally encodes a smooth map $q: \bar U \subset X \rightarrow Q$, given by the projection of $ \omega $ on $Q$. In other words, we have $ \omega (x) \in L(L(T_xX,T_{q(x)}Q), \Lambda _x^{n+1}X)$.

Using these notations, one easily verifies that the covariant Clebsch variational principle \eqref{def_cov_Clebsch} reads
\[
\delta \int_U \ell( \sigma )+ \left\langle \omega , \mathbf{d} q - \sigma _Q (q) \right\rangle =0,
\]
over the space of smooth sections $\psi =(\omega , \sigma ): \bar U \subset X \rightarrow \Pi _\mathfrak{g}$. 

A detailed explanation of the formula appearing under the integral is in order. The symbol $ \mathbf{d} q$ denotes the differential of the map $q: \bar U \subset X \rightarrow Q$ induced by $ \omega $, therefore $ \mathbf{d} q(x) \in L(T_xX, T_{q(x)}Q)$. We have $ \sigma (x) \in L(T_xX, \mathfrak{g}  )$ and the notation $ \sigma _Q (q)$ denotes the function $\bar U \subset X \rightarrow L(TX, TQ)$ sending $x\in \bar U$ to the linear map $v _x\in T_xX \mapsto ( \sigma (v _x ))_Q(q) \in T_{q(x)}Q$. The duality pairing denotes the natural duality between the spaces $L(T_xX, T_qQ)$ and $L(L(T_xX, T_qQ), \Lambda _x ^{n+1}X )$. In local coordinates, we thus have
\[
\delta \int_U \ell( \sigma _\mu ^\alpha )+  \omega _A ^\mu (\partial _\mu  q^A  - \sigma_\mu ^A  )d^{n+1}x =0.
\]
A direct computation or an application of Lemma \ref{local_express} yields the stationarity conditions
\begin{equation} \label{local_coord_SC} 
\frac{\partial \omega _A^\mu  }{\partial x ^\mu } =-\omega  _B ^\nu (\sigma ^B_\nu)_{,A}, \quad \frac{\partial q ^A}{\partial x ^\mu }=\sigma _\mu ^A, \quad \frac{\delta \ell}{\delta \sigma }=\mathfrak{j} ( \omega ).
\end{equation} 
Thanks to the expression \eqref{rewriting_J}, the last condition can be rewritten as
\[
\frac{\delta \ell}{\delta \sigma_\mu  }= \mathbf{J} ( \omega ^\mu ) d ^{n+1}x.
\]

We now derive these conditions by using a global geometric formulation. To achieve this goal we need to fix a covariant derivative $ \nabla $ on $Q$. For simplicity, we assume that $ \nabla$ has no torsion. This covariant derivative induces a differential operator, denoted $ \tilde{\nabla }$, acting on smooth functions $ f: X \rightarrow TQ$ and defined by
\[
\tilde{ \nabla }_{ v _x }f (x):= \left.\frac{D^ \nabla }{D\varepsilon}\right|_{\varepsilon=0} f(c( \varepsilon )),
\]
where $c( \varepsilon ) \in X$ is a curve such that $\left.\frac{d}{d\varepsilon}\right|_{\varepsilon=0} c( \varepsilon )=v _x $ and $D^ \nabla /D \varepsilon $ is the covariant derivative associated to $ \nabla $ of the curve $ \varepsilon \mapsto f(c( \varepsilon ))$ in $TQ$. Note that $\tilde{ \nabla }_{ v _x }f (x) \in T_{q(x)}Q$, where $q:X \rightarrow Q$ is the map induced by $f$. Given a variation $\psi _\varepsilon =( \omega _\varepsilon , \sigma _\varepsilon )$ of the section $ \psi =( \omega , \sigma )$, we compute
\begin{align}\label{glob_comput_VP} 
&\left.\frac{d}{d\varepsilon}\right|_{\varepsilon=0} \int_U \ell( \sigma )+ \left\langle \omega , \mathbf{d} q - \sigma _Q (q) \right\rangle \nonumber\\
&= \int_U \left\langle \frac{\delta \ell}{\delta \sigma }, \delta \sigma \right\rangle + \left.\frac{d}{d\varepsilon}\right|_{\varepsilon=0}  \left\langle \omega_\varepsilon  , \mathbf{d} q_\varepsilon  - (\sigma _\varepsilon )_Q (q_\varepsilon ) \right\rangle \nonumber\\
&= \int_U \left\langle \frac{\delta \ell}{\delta \sigma }, \delta \sigma \right\rangle +   \left\langle \left.\frac{D^ \nabla }{D\varepsilon}\right|_{\varepsilon=0}\omega_\varepsilon  , \mathbf{d} q  - \sigma _Q (q)  \right\rangle+\left\langle \omega  ,\left. \frac{D^ \nabla }{D\varepsilon}\right|_{\varepsilon=0}\mathbf{d} q_\varepsilon  - (\sigma _\varepsilon )_Q (q_\varepsilon ) \right\rangle\nonumber\\
&= \int_U \left\langle \frac{\delta \ell}{\delta \sigma }, \delta \sigma \right\rangle +   \left\langle \delta ^ \nabla \omega , \mathbf{d} q  - \sigma _Q (q) \right\rangle+\left\langle \omega  ,\tilde{ \nabla } \delta  q- (\delta \sigma  )_Q (q )- \nabla _{ \delta q} \sigma _Q  \right\rangle\nonumber\\
&= \int_U \left\langle \frac{\delta \ell}{\delta \sigma }- \mathfrak{j} ( \omega ), \delta \sigma \right\rangle +   \left\langle \delta ^ \nabla \omega , \mathbf{d} q  - \sigma _Q (q) \right\rangle-\left\langle \tilde{\operatorname{div}}^ \nabla (\omega) + \left\langle \omega , \nabla \sigma _Q \right\rangle  ,\delta q \right\rangle.
\end{align}
In the second equality we used the general formula
\[
\frac{d}{d\varepsilon} \left\langle \sigma_\varepsilon  ,e _\varepsilon  \right\rangle = \left\langle \frac{D^{\nabla^{E^\ast }} }{D\varepsilon} \sigma _\varepsilon  , e _\varepsilon  \right\rangle+ \left\langle  \sigma _\varepsilon  ,\frac{D^{\nabla^E}}{D\varepsilon} e _\varepsilon   \right\rangle,
\]
where $e( \varepsilon ) \in E$ is a curve in the total space of a given vector bundle $E \rightarrow X$ endowed with the covariant derivative $ \nabla ^E $, and $\sigma ( \varepsilon ) \in E^*$ is a curve in the total space of the dual vector bundle $E ^\ast \rightarrow X$ endowed with the induced covariant derivative $\nabla^{E^\ast }$. The curves $e(\varepsilon) $ and $\sigma(\varepsilon) $ are assumed to cover the same curve $x( \varepsilon ) \in X$. In the present case, the vector bundle is $E=L(TX, TQ) \rightarrow X \times Q$ and the covariant derivative is the one induced from the covariant derivative $ \nabla $ on $Q$ and a covariant derivative $ \nabla ^X $ on $X$. One checks that our expressions do not depend on $ \nabla ^X $. For simplicity, we thus also denote by $ \nabla $ the covariant derivatives on $L(TX,TQ)$ and its dual.

 In the third equality we defined $ \delta ^ \nabla \omega := \left.\frac{D ^ \nabla }{D\varepsilon}\right|_{\varepsilon=0} \omega _\varepsilon $ and we use the following formulas.
\begin{align*}
\left.\frac{D ^ \nabla }{D\varepsilon}\right|_{\varepsilon=0}  \mathbf{d} q _\varepsilon (x)\cdot  v _x &=\left.\frac{D ^ \nabla }{D\varepsilon}\right|_{\varepsilon=0} \left.\frac{d}{dt}\right|_{t=0} q _\varepsilon (x(t) )=\left.\frac{D ^ \nabla }{Dt}\right|_{t=0} \left.\frac{d}{d\varepsilon }\right|_{\varepsilon =0} q _\varepsilon (x(t) )\\
&= \left.\frac{D ^ \nabla }{Dt}\right|_{t=0} \delta q(x(t))= \tilde{\nabla} _{v _x } \delta q(x)
\end{align*} 
\begin{align*} 
\left.\frac{D ^ \nabla }{D\varepsilon}\right|_{\varepsilon=0} ( \sigma _\varepsilon )_Q(q _\varepsilon )(x)\cdot v_x&= \left.\frac{D ^ \nabla }{D\varepsilon}\right|_{\varepsilon=0} ( \sigma _\varepsilon (x) \cdot v _x )_Q(q _\varepsilon (x) )\\
&=( \delta \sigma (x) \cdot v _x )_Q(q(x))+ \nabla _{ \delta q} ( \sigma (x) \cdot v _x )_Q(q(x))\\
&=\big( ( \delta \sigma )_Q(q)+ \nabla _{ \delta q} ( \sigma  )_Q(q)\big)  (x) \cdot v _x,
\end{align*} 
where the last equality shows how one applies the section $( \delta \sigma )_Q(q)+ \nabla _{ \delta q} ( \sigma  )_Q(q)$ to $ v _x $.

In the fourth equality, we used Stokes theorem together with the definition of the covariant divergence $ \tilde{\operatorname{div}}$ associated to $\tilde{ \nabla }$ and the fact that variations vanish at the boundary. We refer to \cite{ElGBHoRa2011} for more details about the operators $\tilde{ \nabla }$ and $\tilde{ \operatorname{div}}^ \nabla $. 

From \eqref{glob_comput_VP} we thus obtain the intrinsic formulation of the stationarity conditions \eqref{local_coord_SC}: 
\begin{equation}\label{intrinsic_stationarity_cond} 
\tilde{\operatorname{div}}^ \nabla (\omega) = -  \left\langle \omega , \nabla \sigma _Q \right\rangle , \quad \mathbf{d} q  = \sigma _Q (q), \quad \frac{\delta \ell}{\delta \sigma }= \mathfrak{j} ( \omega ).
\end{equation}

\paragraph{Clebsch variables and collective Hamiltonian.} 
We first compute the collective Hamiltonian of the covariant Clebsch principle. Since the bundle is trivial, we can choose the trivial connection, i.e. the one given by the horizontal subbundle $H(X \times Q)=TX\times\{0\} \subset T(X \times Q)$. Recall also that the line bundle $ \mu  : J^1(X \times Q)^\star \rightarrow \Pi  $ is trivial, and
therefore Hamiltonian section $ \mathsf{h}: \Pi \rightarrow J ^1 Y ^\star= \Pi \times _X \Lambda ^{n+1}X$ can be globally written as $\mathsf{h}(\omega )=( \omega ,-\mathsf{H}( \omega ) ) $, where $\mathsf{H}: \Pi \rightarrow \Lambda ^{n+1}X$ is a smooth map covering the identity on $X$. With this identification, we obtain that the relation \eqref{cov_coll_Ham} is equivalently written as
\[
\mathsf{H}( \omega )=h( \mathfrak{j} ( \omega ))\in \Lambda ^{n+1}X.
\]
We recognize clearly here the generalization of the collective Hamiltonian relation
\[
H( \alpha )= h( \mathbf{J} ( \alpha ))
\]
to the covariant case. 

From the expression of $ \mathfrak{j} $ in the case of a trivial bundle, we obtain that the covariant Clebsch variables for the noncanonical covariant Lie-Poisson equations for $h$ are provided by the variables $ \omega _A ^\mu $ through the mapping
\[
\omega =\omega _A ^\mu \partial _\mu \otimes d q ^A \otimes d^{n+1}x \mapsto  \mathfrak{j} ( \omega )= \partial _\mu \otimes \mathbf{J} (\omega ^\mu  ) \otimes d^{n+1}x.
\]
The variables $ \omega _A ^\mu $ evolve according to the canonical covariant Hamilton equations for $\mathsf{h}$. For completeness, we present here a global formulation of these equations. A convenient way is to use the covariant Hamilton's phase space principle
\[
\delta \int_U \omega  ^\ast \mathsf{h} ^\ast \Theta =0.
\]
In our case, since $\mathsf{h}( \omega )=( \omega , -\mathsf{H}( \omega ))$, we have $\omega ^\ast  \mathsf{h} ^\ast \Theta = \omega \cdot \mathbf{d} q- \mathsf{H}( \omega )$, so that the covariant Hamilton's phase space principle reads
\[
\delta \int_ U  \omega \cdot \mathbf{d} q- \mathsf{H}( \omega )=0.
\]
Similar computations as the ones used in \eqref{glob_comput_VP} yield the equations
\begin{equation}\label{cov_Ham_equ_global}
\mathbf{d} q= \frac{\partial \mathsf{H}}{\partial \omega }, \quad \tilde{\operatorname{div}}^\nabla \omega= - \frac{\partial ^ \nabla \mathsf{H}}{\partial q}, 
\end{equation}  
where a covariant derivative $\nabla $ has been fixed on $Q$. Here $ \partial \mathsf{H}/ \partial \omega $ is the fiber derivative of $ \mathsf{H}$ and $ \partial^ \nabla  \mathsf{H}/ \partial q$ is the partial derivative of $ \mathsf{H}$ relative to the connection $ \nabla $.

When $\mathsf{H}( \omega )= h( \mathfrak{j} ( \omega ))$, we get
\[
\frac{\partial \mathsf{H}}{\partial \omega} = \left( \frac{\delta h}{\delta \nu }\right) _Q \quad \frac{\partial ^ \nabla \mathsf{H}}{\partial q}=\left\langle \omega , \nabla \left( \frac{\delta h}{\delta \nu }\right) _Q \right\rangle ,
\]
where $ \nu = \mathfrak{j} (\omega) $. In this case \eqref{cov_Ham_equ_global} consistently recovers \eqref{intrinsic_stationarity_cond}.

The properties of the covariant Clebsch variational principle in the trivial bundle case are summarized in the following proposition.

\begin{proposition}[Covariant Clebsch variational principle on trivial bundles]\label{summary_trivial_case}  Let $G $ be a Lie group acting on a manifold $Q$ and let $X$ be a manifold of dimension $n+1$. Let $\ell: L(TX, \mathfrak{g}  ) \rightarrow \Lambda ^{n+1}X$ be a Lagrangian density. Then we have the following statements.
\begin{itemize}
\item[\bf(1)] The covariant Clebsch variational principle for a section $ \psi =( \omega, \sigma ): U \subset X \rightarrow \Pi _\mathfrak{g} $ is
\[
\delta \int_U\ell( \sigma )- \left\langle \omega , \mathbf{d} q- \sigma _Q (q) \right\rangle =0.
\]
\item[\bf(2)] The section $ \psi =(\omega, \sigma )$ is a critical section of the covariant Clebsch variational principle if and only if it verifies
\begin{equation}\label{intrinsic_stationarity_cond_prop} 
\tilde{\operatorname{div}}^ \nabla (\omega) = -  \left\langle \omega , \nabla \sigma _Q \right\rangle , \quad \mathbf{d} q  = \sigma _Q (q), \quad \frac{\delta \ell}{\delta \sigma }= \mathfrak{j} ( \omega ),
\end{equation} 
where $ \nabla $ is a covariant derivative on $Q$ with vanishing torsion.
\item[\bf(3)] The section $ \sigma :\bar U \subset X \rightarrow L(TX, \mathfrak{g}  )$ is a solution of the covariant Euler-Poincar\'e equations
\begin{equation}\label{COV_EP_prop} 
\operatorname{div} \frac{\delta \ell}{\delta \sigma }+ \operatorname{ad}^*_ \sigma \frac{\delta \ell}{\delta \sigma }=0
\end{equation}
and verifies the compatibility condition $( \mathbf{d} \sigma- [ \sigma , \sigma ])_Q(q)=0$.
\item[\bf(4)] Suppose that $\ell$ is hyperregular and define the associated Hamiltonian $h:L(TX, \mathfrak{g}  ) ^\ast \rightarrow \Lambda ^{n+1}X$. Let $ \mathsf{H}: \Pi \rightarrow \Lambda ^{n+1}X$ be the collective covariant Hamiltonian defined by $ \mathsf{H}=h \circ \mathfrak{j} $. Then $( \omega , \sigma )$ is a solution of \eqref{intrinsic_stationarity_cond_prop} if and only if $ \omega $ is a solution of the covariant canonical Hamilton's equations
\begin{equation}\label{cov_Ham_equ_global_prop}
\mathbf{d} q= \frac{\partial \mathsf{H}}{\partial \omega }, \quad \tilde{\operatorname{div}}^\nabla \omega= - \frac{\partial ^ \nabla \mathsf{H}}{\partial q}.
\end{equation}
\end{itemize}
\end{proposition}

\subsection{Covariant Clebsch VP for linear actions}

We now suppose that the Lie group $G$ acts by representation on a vector space $V$. We denote by $ v \mapsto gv$ this action. The infinitesimal generator is denoted by $ \xi _V (v)= \xi v$ and the cotangent bundle momentum $ \mathbf{J} :T^*V \rightarrow \mathfrak{g}  ^\ast $, is denoted by $\mathbf{J} (v,p)=v \diamond p$.

Let $X$ be a $(n+1)$-dimensional manifold. Since $T^\ast V= V \times V ^\ast $, a section of the restricted dual jet bundle $ \Pi \rightarrow X$ can be written as a couple $(v, \omega )$, where $v: X \rightarrow V$ and $ \omega : X \rightarrow TX \otimes V ^\ast \otimes \Lambda ^{n+1}X$. Writing $ \omega = \partial _\mu \otimes \omega ^\mu \otimes dx^{n+1}$, the covariant momentum map is given by $ \mathfrak{j} (v, \omega )= \partial _\mu \otimes v \diamond \omega ^\mu \otimes dx^{n+1}\in TX \otimes \mathfrak{g}  ^\ast \otimes \Lambda ^{n+1}X$.

The covariant Clebsch variational principle reads
\[
\delta \int_U \ell( \sigma )+ \left\langle \omega , \mathbf{d} v- \sigma v \right\rangle =0
\]
for maps $ \sigma : U \subset X\rightarrow T^\ast X \otimes \mathfrak{g} $, $v: U \subset X \rightarrow V$, and $ \omega : U \subset X \rightarrow TX \otimes V ^\ast \otimes \Lambda ^{n+1}X$. It yields the stationarity conditions
\[
\frac{\delta \ell}{\delta \sigma }= \mathfrak{j} (v, \omega ), \quad \mathbf{d} v- \sigma v=0, \quad \operatorname{div} \omega - \sigma_\mu  \omega ^\mu =0
\]
which imply the covariant Euler-Poincar\'e equations for $\ell$, together with the compatibility condition
\[
( \mathbf{d} \sigma -[ \sigma , \sigma ])v=0.
\]

Assuming that the Lagrangian $\ell$ is hyperregular, with associate Hamiltonian $h$, the collective Hamiltonian is $\mathsf{H}(v, \omega )=h(\partial _\mu \otimes  v \diamond \omega^\mu \otimes d^{n+1}x )$ and the covariant canonical Hamilton's equations are
\[
\mathbf{d} v= \frac{\partial \mathsf{H}}{\partial \omega }, \quad \operatorname{div} \omega =- \frac{\partial \mathsf{H}}{\partial v},
\]
or, in terms of $h$,
\[
\mathbf{d} v= \frac{\delta h}{\delta \nu}v, \quad \operatorname{div} \omega = \frac{\delta h}{\delta \nu^\mu  }\omega ^\mu , \quad \text{with} \quad  \nu = \mathfrak{j} (v, \omega ).
\]
\paragraph{The case of $G$-strands.} When $X= \mathbb{R}  ^2 \ni (t,s)$, we write $ \sigma = \xi  dt+ \gamma ds$, $ \omega =(\partial _t \otimes m+ \partial _s \otimes n ) dt \wedge ds$, $ \nu =(\partial _t \otimes  \gamma +\partial _s \otimes \lambda ) dt \wedge ds$, with $ \xi(t,s) , \gamma (t,s)\in \mathfrak{g}  $, $m(t,s), n(t,s) \in V^\ast $, $ \gamma (t,s), \lambda  (t,s) \in \mathfrak{g}  ^\ast$. The covariant Clebsch VP becomes
\[
\delta \int_U \ell( \xi  , \gamma  )+ \left\langle m , \partial _t  v- \xi   v \right\rangle dt ds+\left\langle n , \partial _s  v- \gamma   v \right\rangle dt ds=0,
\]
and yields the stationarity conditions
\begin{align*} 
&\frac{\delta \ell}{\delta \xi }= (v \diamond m)dt \wedge ds,\quad \frac{\delta \ell}{\delta \gamma  }  =  (v \diamond n) dt \wedge ds\\
&\partial _t v= \xi v, \quad \partial _s v= \gamma v, \quad \partial _t m+ \partial _s n=\xi m+ \gamma n,
\end{align*} 
so that $ \xi $, $ \gamma $ verify the $G$-strand equations
\[
\partial _t \frac{\delta \ell}{\delta \xi }+ \operatorname{ad}^*_ \xi    \frac{\delta \ell}{\delta \xi }+\partial _s\frac{\delta \ell}{\delta \gamma  }+ \operatorname{ad}^*_ \gamma     \frac{\delta \ell}{\delta \gamma  }= 0.
\]
The compatibility condition is $(\partial _t \gamma - \partial _s \xi -[ \xi , \gamma ])v=0$ and the Hamilton equations read
\[
\partial _t v = \frac{\delta h}{\delta \gamma  }v, \quad  \partial _s v  = \frac{\delta h}{\delta \lambda   }v, \quad \partial _t m+ \partial _s n= \frac{\delta h}{\delta \gamma }m+ \frac{\delta h}{\delta \lambda }n.  
\]

\paragraph{Adjoint action and covariant coupled double bracket equations.} A particularly interesting example is the case of the adjoint action, that is, we take $V= \mathfrak{g}  $ and $ gm= \operatorname{Ad}_gm$  for $ g \in G$ and $ m \in \mathfrak{g}$. Note that the infinitesimal generator is $\xi _ \mathfrak{g}  ( m)= \operatorname{ad}_ \xi m=[ \xi , m]$, and the cotangent bundle momentum map is $ \mathbf{J} ( m, p)= m \diamond p=- \operatorname{ad}^*_mp$. In this case, the classical Clebsch variational principle yields the coupled double bracket equations
\begin{equation}\label{CDBE} 
\dot m=[m,[p,m]], \quad \dot p= [p,[p,m]]
\end{equation} 
of \cite{BlCr1996}, as shown in \cite{GBRa2011}, for an appropriate choice of the Lagrangian. We now present the covariant analogue of this system.

Given a $(n+1)$-dimensional manifold $X$, the covariant Clebsch variational principle reads
\[
\delta \int_U \ell( \sigma )+ \left\langle \omega , \mathbf{d} m- [\sigma ,m] \right\rangle =0
\]
for maps $ \sigma : U \subset X\rightarrow T^\ast X \otimes \mathfrak{g} $, $m: U \subset X \rightarrow \mathfrak{g}  $, and $ \omega : U \subset X \rightarrow TX \otimes \mathfrak{g}  ^\ast \otimes \Lambda ^{n+1}X$, and where $[ \sigma , m] \in L(TX, \mathfrak{g}  )$ is defined by $[ \sigma , m]( v _x ):= [ \sigma (v _x ),m]$.  It yields the stationarity conditions
\begin{equation}\label{stat_cond_ad} 
\frac{\delta \ell}{\delta \sigma  }= \mathfrak{j} ( m , \omega ), \quad \partial _\mu m-[ \sigma _\mu ,m]=0, \quad \partial _\mu \omega ^\mu + \operatorname{ad}^*_{ \sigma _\mu } \omega ^\mu =0,
\end{equation} 
where $ \mathfrak{j} (m, \omega )=   - \partial _\mu  \otimes \operatorname{ad}^*_ m \omega ^\mu    \otimes d^{n+1}x$, together with the compatibility condition
\[
[\mathbf{d} \sigma -[ \sigma , \sigma ], m]=0.
\]
Remarkably, the last two equations in \eqref{stat_cond_ad} can be rewritten as
\[
\mathbf{d} ^\sigma m=0, \quad \operatorname{div} ^\sigma \omega =0, 
\]
where we interpret $ \sigma $ as a connection on the (left) trivial principal bundle $X \times G \rightarrow G$, and where $ \mathbf{d} ^ \sigma $ and $ \operatorname{div}^ \sigma $ denote, respectively, the covariant differential and covariant divergence associated to $ \sigma $.

In the hyperregular case, these equations can be written with the help of the Hamiltonian $h$ as
\[
\mathbf{d} ^{ \delta h/\delta \nu } m=0, \quad \operatorname{div} ^{ \delta h/\delta \nu } \omega =0, \quad \text{where} \quad \nu = \mathfrak{j} (m, \omega ).
\]

Suppose that $ \mathfrak{g}  $ is endowed with an $ \operatorname{Ad}$-invariant inner product $ \gamma $ and fix a Riemannian metric $g$ on $X$. In this case, one can naturally identify the dual $L(L(TX, \mathfrak{g}  ), \Lambda ^{n+1}X)$ of $L(TX, \mathfrak{g} )$ with  $L(TX, \mathfrak{g}  )$ by using the duality pairing $ \left\langle \! \left\langle \sigma , \eta \right\rangle \! \right\rangle := g ^{ \mu \nu } \gamma _{ab} \sigma _\mu ^a \eta _\nu ^b $. The stationarity conditions are
\[
\frac{\delta \ell}{\delta \sigma } =[m, \omega ]   , \quad \mathbf{d} m-[ \sigma ,m]=0, \quad \operatorname{div}_{g \gamma } \omega  - \operatorname{Tr}_g [ \sigma , \omega ] =0,
\]
where the functional derivative $ \delta \ell/ \delta \sigma \in L(TX, \mathfrak{g}  )$ is associated to the pairing $ \left\langle \! \left\langle \,, \right\rangle 
\! \right\rangle $, $\operatorname{div}_{g \gamma }: \Gamma (L(TX, \mathfrak{g}  )) \rightarrow \mathcal{F} (X, \mathfrak{g}  )$ is the divergence associated to $g$ and $ \gamma $, and $ \operatorname{Tr}_g$ is the trace relative to $g$. 
If, moreover, the reduced Lagrangian is $\ell( \sigma )= \frac{1}{2} \| \sigma \| ^2 $, where $\|\,\|$ is the vector bundle norm associated to $g$ and $ \gamma $ on $L(TX, \mathfrak{g}  )$, then we have
\[
[m, \omega _\mu ]= \sigma _\mu, 
\]
so that, inserting this in the equations above, we get
\[
\mathbf{d}  m-[ [m, \omega ] ,m]=0, \quad \operatorname{div}_{g \gamma } \omega  -\operatorname{Tr}_g [[m, \omega ] , \omega].
\]
These are the \textit{covariant coupled double bracket equations} which extend \eqref{CDBE} to the covariant case. They can also be written using covariant differential operators as
\[
\mathbf{d}^{\,[m, \omega ]\,}m=0, \quad \operatorname{div}_{\,g \gamma } ^{\,[m, \omega ]\,}\omega =0.
\]

Note that the associated covariant Euler-Poincar\'e equations read
\[
\operatorname{div}_{g \gamma }  \frac{\delta \ell}{\delta \sigma }- \operatorname{Tr}_g \left[ \sigma , \frac{\delta \ell}{\delta \sigma }\right]=0.
\]
For the choice $\ell( \sigma )= \frac{1}{2} \| \sigma \| ^2 $, the second term vanishes and therefore the covariant coupled double bracket equations imply 
\[
\operatorname{div}_{g \gamma }  \sigma =0.
\]

The results obtained in this paragraph, which extend those of \cite{BlCr1996} and \cite{GBRa2011} (see esp. equation (31)) to the covariant case, are summarized in the next proposition.

\begin{proposition}[Coupled double bracket formulation of covariant EP equations]\label{coupled_DB}  Let $\ell: L(TX, \mathfrak{g}  ) \rightarrow \Lambda ^{n+1}X$ be a Lagrangian density assumed to be hyperregular, and consider the associated Hamiltonian density $h$. Assume that $m: U \subset X \rightarrow \mathfrak{g}  $, and $ \omega : U \subset X \rightarrow TX \otimes \mathfrak{g}  ^\ast \otimes \Lambda ^{n+1}X$ verify the canonical covariant Hamilton equations
\[
\mathbf{d}  m-\left[  \frac{\delta h}{\delta [m, \omega ]} ,m\right] =0, \quad \operatorname{div}_{g \gamma } \omega  -\operatorname{Tr}_g \left[ \frac{\delta h}{\delta [m, \omega ]}, \omega\right].
\]
Then $ \sigma $, given by $ \frac{\delta \ell}{\delta \sigma }= \mathfrak{j} (m, \omega )$, verifies the covariant Euler-Poincar\'e equations, together with the compatibility condition $[ \mathbf{d} \sigma -[ \sigma , \sigma ],m]=0$. 
\end{proposition}

\subsection{Covariant Clebsch VP for Lie group translations}

Recall that in the special case when a Lie group $G$ acts on itself by \textit{left} translation, the cotangent bundle momentum map is given by \textit{right} translation at the identity:  $ \mathbf{J} : T^*G \rightarrow \mathfrak{g}  ^\ast $, $ \mathbf{J} ( \alpha _g )= \alpha _g g ^{-1} $, \cite{MaRa1999}.

The covariant analogue is obtained by taking $Y= X \times G$ and letting $G$ act on itself by left translation. We have $ \xi _G (g)=TR_g( \xi  )= \xi  g$ and $ \mathfrak{j}: \Pi \rightarrow L(L(TX, \mathfrak{g}  ) , \Lambda ^{n+1})$ reads
\[
\mathfrak{j} ( v _x \otimes \alpha _g \otimes \mu _x )=  v _x \otimes \alpha _gg ^{-1}  \otimes \mu _x.
\]

A section $ \omega $ of the restricted dual jet bundle $ \Pi \rightarrow X$ induces a map $g:X \rightarrow G$ such that $ \omega (x) \in L(L(T_xX, T_{g(x)}G), \Lambda ^{n+1}X)$.

The covariant Clebsch variational principle reads
\begin{equation}\label{CC_G}
\delta \int_ U \ell( \sigma )+ \left\langle \omega , \mathbf{d}g - \sigma g \right\rangle =0,
\end{equation} 
for sections $ \omega :U \subset X \rightarrow L(L(TX, TG), \Lambda ^{n+1}X)$ and  $ \sigma : U \subset X \rightarrow L(TX, \mathfrak{g}  )$, and where $ \mathbf{d}g(x) \in L(T_xX, T_{g(x)}G)$ denotes the tangent map to $g:X \rightarrow G$ and $ (\sigma g)(x)\in L(T_xX, T_{g(x)}G)$ is defined by $( \sigma g)(v _x )= ( \sigma (v _x ))g(x)=TR_{g(x)}( \sigma (v _x ))$.

One can obtain the stationary conditions together with the Hamiltonian formulation by applying Proposition \ref{summary_trivial_case} and fixing a covariant derivative on $G$. We will however take the advantage of having a Lie group to trivialize the variational principle, by using the trivialized section $ \nu := \omega g ^{-1} :U \subset X \rightarrow L(L(TX, \mathfrak{g}  ), \Lambda ^{n+1}X)$. Using this section, the variational principle \eqref{CC_G} can be rewritten as
\[
\delta \int_ U \ell( \sigma )+ \left\langle \nu  , (\mathbf{d}g)g ^{-1}  - \sigma\right\rangle =0.
\]
A direct computation yields
\begin{align*} 
&\delta \int_ U \ell( \sigma )+ \left\langle \nu  , (\mathbf{d}g)g ^{-1}  - \sigma\right\rangle \\
&\quad  = \int_U \left\langle \frac{\delta \ell}{\delta \sigma }, \delta \sigma \right\rangle + \left\langle \delta \mu , ( \mathbf{d}  g ) g ^{-1} - \sigma \right\rangle + \left\langle \nu  , (\mathbf{d} \delta g) g ^{-1}   - \mathbf{d} g g ^{-1} \delta g g ^{-1} - \delta \sigma \right\rangle \\
&\quad  = \int_U \left\langle \frac{\delta \ell}{\delta \sigma }- \nu , \delta \sigma \right\rangle + \left\langle \delta \nu , ( \mathbf{d}  g ) g ^{-1} - \sigma \right\rangle + \left\langle \nu  , \mathbf{d} (\delta g g ^{-1})   +[ \delta g g ^{-1} ,\mathbf{d} g g ^{-1} ] \right\rangle
\end{align*} 
and we get the stationarity conditions
\[
\frac{\delta \ell}{\delta \sigma }= \nu   , \quad \mathbf{d} g - \sigma g =0 , \quad \operatorname{div}( \nu  )+ \operatorname{ad}^*_{ \mathbf{d} g g ^{-1} } \nu =0,
\]
where $ \nu = \omega g ^{-1} $. Note that from this expression of the stationarity conditions, it is clear that they imply the covariant Euler-Poincar\'e equations, by simply inserting in the third equation, the expression for $ \sigma $ and $ \nu $ given from the first two equations.

Since the action is transitive, the compatibility condition coincides with the zero curvature condition \eqref{compatibility_condition}, see Lemma \ref{lemma_CC}. 

In the hyperregular case, the collective Hamiltonian is $\mathsf{H}(\omega )= h(\omega g ^{-1} )$ and the process of passing from the covariant canonical Hamilton's equation for $\mathsf{H}$ to the covariant Lie-Poisson equations for $h$, coincides with the covariant Lie-Poisson reduction process, see \cite{CaMa2003}.

The next proposition summarizes the properties obtained so far.

\begin{proposition} Let $\ell: L(TX, \mathfrak{g}  )\rightarrow \Lambda ^{n+1}X$ be a Lagrangian density and suppose that $G$ acts on $Q=G$ by left translation. Then the associated covariant Clebsch variational principle can be written as
\begin{equation}\label{CCVP_G} 
\delta \int_ U \ell( \sigma )+ \left\langle \nu  , (\mathbf{d}g)g ^{-1}  - \sigma\right\rangle =0,
\end{equation} 
where $ \sigma : U \subset X \rightarrow L(TX, \mathfrak{g}  )$, $g: U \subset X \rightarrow G$, and $ \nu :U \subset X \rightarrow L(L(TX, \mathfrak{g}  ), \Lambda ^{n+1}X) $. The stationarity conditions are
\[
\frac{\delta \ell}{\delta \sigma }= \nu   , \quad \mathbf{d} g - \sigma g =0 , \quad \operatorname{div}( \nu  )+ \operatorname{ad}^*_{ \mathbf{d} g g ^{-1} } \nu =0,
\]
and imply the covariant Euler-Poincar\'e equations for $\ell$ together with the zero curvature condition \eqref{compatibility_condition}. 
\end{proposition}

\paragraph{Example: convective strand dynamics.}
When the base manifold is $X= \mathbb{R}^2   \ni (t,s)$ and the Lie group is the special Euclidean group $G=SE(3)\ni ( \Lambda , \mathbf{r} )$, we can write $ \sigma = ( \omega , \gamma )dt+( \Omega , \Gamma )ds$ and $ \nu =(\partial _t \otimes (m, n)+ \partial _s \otimes (M,N) ) dt \wedge ds$, where
\[
( \omega , \gamma ), ( \Omega , \Gamma ): \mathbb{R}  ^2 \rightarrow \mathfrak{se}(3)\quad \text{and} \quad  ( m , n ), ( M , N ): \mathbb{R}  ^2 \rightarrow \mathfrak{se}(3)^\ast.
\]
In this case the covariant Clebsch variational principle \eqref{CCVP_G} reads
\begin{align*} 
&\delta \int \ell( \omega, \gamma , \Omega , \Gamma )+ \left\langle (m,n), ( \Lambda , \mathbf{r} ) ^{-1} ( \partial _t \Lambda , \partial _t \mathbf{r} )- ( \omega , \gamma ) \right\rangle dtds\\
& \qquad \qquad \qquad \qquad \qquad + \left\langle (M,N), ( \Lambda , \mathbf{r} ) ^{-1} ( \partial _s \Lambda , \partial _s \mathbf{r} )- ( \Omega , \Gamma ) \right\rangle dtds=0.
\end{align*}
For an appropriate choice of $\ell$, this principle yields the equations of motion for strand dynamics in convective representation (\cite{SiMaKr1988}), and recovers the variational principle developed in \cite{ElGBHoRa2011}. The variable $ \sigma =  ( \omega , \gamma )dt+( \Omega , \Gamma )ds$ involves the convective angular and linear velocities $( \omega , \gamma )$ and the convective angular and linear strains $( \Omega , \Gamma )$, whereas $ \nu =(\partial _t \otimes (m, n)+ \partial _s \otimes (M,N) ) dt \wedge ds$ involves the corresponding convective momenta $(m,n)$ and $(M,N)$.
We refer to \cite{ElGBHoRa2011} for the detailed derivation and alternative variational formulations of convective strand dynamics.

\paragraph{Example: the covariant version of the symmetric rigid body equations.} The so called \textit{symmetric rigid body equations}
\begin{equation}\label{symm_Nbody} 
\dot Q= QU, \quad \dot P= PU, \quad U \in \mathfrak{so}(N),
\end{equation} 
are an alternative formulation of $N$-dimensional rigid body dynamics, associated to an optimal control problem \cite{BlCr1996}, \cite{BlCrMaRa1998}, \cite{GBRa2011}. Given an hyperregular Lagrangian $\ell: \mathfrak{so}(N) \rightarrow \mathbb{R}  $, for example the $N$-dimensional rigid body Lagrangian, one observes that if $U$ is defined by $ \delta \ell / \delta U= \frac{1}{2} (Q^TP-P^TQ)$, then $U$ is a solution of the Euler-Poincar\'e equations for $\ell$ on $ \mathfrak{so}(N)$. We shall now present the covariant analogue of this statement.

Consider the right action of $SO(N)$ on the trivial bundle $ X \times GL(N) \rightarrow X$ by translation on $GL(N)$. The covariant Clebsch variational principle is
\[
\delta \int_U \ell(\upsilon )+ \left\langle \omega , \mathbf{d} Q- Q\upsilon  \right\rangle =0,
\]
for sections $ \omega :U \subset X \rightarrow L(L(TX, TGL(N)), \Lambda ^{n+1}X)$ and $\upsilon :U \subset X \rightarrow L(TX, \mathfrak{so}(N))$, and yields the stationarity conditions
\[
\frac{\delta \ell}{\delta \upsilon _\mu }= \mathfrak{j} (Q, \omega )^\mu = \frac{1}{2} (Q^T\omega^\mu  -(\omega^\mu) ^TQ), \quad \mathbf{d} Q=Q \upsilon , \quad \operatorname{div} \omega =\omega\upsilon.
\]
The last two equations are the covariant analogues of the symmetric equations \eqref{symm_Nbody} and imply the covariant Euler-Poincar\'e equations on $ \mathfrak{so}(N)$
\[
\operatorname{div} \frac{\delta \ell}{\delta \upsilon }+ \left[ \upsilon_\mu  , \frac{\delta \ell}{\delta \upsilon _\mu } \right]=0.
\]

This applies in particular to the $SO(N)$-strand equations if $X= \mathbb{R} ^2   \ni (t,s)$. In this case, using the notations $\upsilon  = U  dt+ V ds$, and $ \omega =(\partial _t \otimes M+ \partial _s \otimes N ) dt \wedge ds$, with $U(t,s), V(t,s) \in \mathfrak{so}(N)$ and $ M(t,s), N(t,s) \in T_{ Q(t,s)}GL(N)$, the stationarity conditions are
\[
\frac{\delta \ell}{\delta U}= \frac{1}{2} (Q^TM  -M^TQ)dt \wedge ds, \quad \frac{\delta \ell}{\delta V  }= \frac{1}{2} (Q^TN  -N^TQ)dt \wedge ds
\]
\[
\partial _t  Q=Q U , \quad \partial _s Q= Q V , \quad \partial _t M+ \partial _s N =M U + N V
\]
and imply the $SO(N)$-strand equations
\[
\partial _t \frac{\delta \ell}{\delta U }+ \partial _s \frac{\delta \ell}{\delta V}+ \left[ U  , \frac{\delta \ell}{\delta U} \right]+\left[ V  , \frac{\delta \ell}{\delta V} \right]=0.
\]
The particular case $N=3$ is relevant for chiral models, in which case the Lagrangian density is $\ell(U,V)= (\frac{1}{2} | U | ^2 - \frac{1}{2} | V | ^2 ) dt \wedge ds$, see \cite{HoIvPe2012} and references therein.

\subsection{The covariant EPDiff equations and singular solutions}\label{sing_sol}

\paragraph{The covariant EPDiff equations.} These equations are, by definition, the covariant Euler-Poincar\'e equations in the special case when the Lie group is the diffeomorphism group $ \operatorname{Diff}(M)$ of some manifold $ M$. The terminology that we use comes from the same denomination (EPDiff) for the Euler-Poincar\'e equations on the diffeomorphism group, \cite{HoMa2004}. The introduction of this class of partial differential equations, the covariant EPDiff equations, was motivated by the recent paper \cite{HoIvPe2012}.

Given a $(n+1)$-dimensional manifold $X$ we get from the general expression \eqref{cove_EP} of the covariant Euler-Poincar\'e equations, the covariant EPDiff equations
\begin{equation}\label{cov_EPDiff} 
\operatorname{div}_X \frac{\delta \ell}{\delta \boldsymbol{\sigma} }  + \operatorname{ad}^*_{\boldsymbol{\sigma} _\mu } \frac{\delta \ell}{\delta \boldsymbol{\sigma} _\mu }=0 ,
\end{equation} 
where $ \operatorname{ad}^*_\mathbf{u}  \mathbf{m} =  \nabla \mathbf{m} \cdot \mathbf{u} + \nabla \mathbf{u} ^\mathsf{T}\cdot \mathbf{m} + \mathbf{m} \operatorname{div} \mathbf{u} $ is the infinitesimal coadjoint operator associated to the Lie algebra $ \mathfrak{g}  = \mathfrak{X}  ( M )$ of vector fields on $M$, where we identified $ \mathfrak{g}  ^\ast $ with $ \mathfrak{g}  $ using the $L^2$-duality pairing. In \eqref{cov_EPDiff}, $ \operatorname{div}_X$ denotes the divergence of the $ \mathfrak{X} (M) ^\ast $-valued vector field density $ \frac{\delta \ell}{\delta \boldsymbol{\sigma} }\in TX \otimes \mathfrak{X}  (M) ^\ast \otimes \Lambda ^{n+1}X$.  Note that in these equations, $ \boldsymbol{\sigma}  $ is a smooth section $ \boldsymbol{\sigma} : U \subset X \rightarrow L(TX, \mathfrak{X} (M))$. This implies that, given $x \in X$ and $ m \in M$, we have $ \boldsymbol{\sigma} (x,m):= \boldsymbol{\sigma} (x)(m)\in L(T_xX, T_mM)$. The zero curvature conditions read
\begin{equation}\label{ZCC_EPDiff} 
\mathbf{d}_X  \boldsymbol{\sigma}  =[ \boldsymbol{\sigma}  , \boldsymbol{\sigma}  ],
\end{equation} 
where we used the notation $ \mathbf{d} _X $ to make clear that it means the exterior derivative relative to the variable $x$, that is, the exterior derivative of the one-form $ \boldsymbol{\sigma}  \in \Omega ^1(X, \mathfrak{X}  (M))$. Note that \eqref{ZCC_EPDiff} is an equality in $ \Omega ^2 (X, \mathfrak{X}  (M))$.

In the case $X= \mathbb{R}  ^2\ni (t,s) $ and $M= \mathbb{R}$, the covariant EPDiff equations recover the $ \operatorname{Diff}( \mathbb{R}  )$-strand equations studied in \cite{HoIvPe2012}.
Using the notation $ \boldsymbol{\sigma} (t,s)= \boldsymbol{\nu}  (t,s) dt+ \boldsymbol{\gamma}  (s,t) ds\in \mathfrak{X}  (M)$, we get from \eqref{cov_EPDiff} the $ \operatorname{Diff}( \mathbb{R}  )$-strand equations
\[
\partial _t \frac{\delta \ell}{\delta \boldsymbol{\nu}  }+\partial _s \frac{\delta \ell}{\delta \boldsymbol{\gamma}   } + \operatorname{ad}^*_{ \boldsymbol{\nu} }   \frac{\delta \ell}{\delta \boldsymbol{\nu}    }+ \operatorname{ad}^*_ {\boldsymbol{\gamma} }  \frac{\delta \ell}{\delta \boldsymbol{\gamma}   }=0,
\]
where $ \operatorname{ad}^*_uv= v' u+ 2 vu' $ is the infinitesimal coadjoint operator on $ \mathfrak{X}  ( \mathbb{R}  ) ^\ast $. The zero curvature condition \eqref{ZCC_EPDiff} reduces to 
\[
\partial_t  \boldsymbol{\gamma}  - \partial _s  \boldsymbol{\nu}  = [ \boldsymbol{\nu}  , \boldsymbol{\gamma}  ]= \boldsymbol{\gamma} \partial _m \boldsymbol{\nu}  -\boldsymbol{\nu}  \partial _m \boldsymbol{\gamma} . 
\]

In \cite{HoMa2004} a pair of momentum maps has been defined in the context of the (classical) EPDiff equations. These momentum maps are associated to the two natural diffeomorphism actions on the space $ \operatorname{Emb}(S,M)$ of embeddings of a $k$-dimensional manifold $S$ into $M$. The left action of the diffeomorphism group $ \operatorname{Diff}(M)$ yields the cotangent bundle momentum map $ \mathbf{J}_L :T^* \operatorname{Emb}(S, M) \rightarrow \mathfrak{X}  (M) ^\ast $ given by
\[
\mathbf{J}_L ( \mathbf{Q} , \mathbf{P} )=\int_S \mathbf{P} (s) \delta (m- \mathbf{Q} (s)) d^k s.
\]
This momentum map provides the expression for singular solutions of the EPDiff equations. They may, for example, be supported on sets of points (vector peakons, $ \operatorname{dim}S=0$), one-dimensional filaments (strings, $ \operatorname{dim}S=1$), or two-dimensional surfaces (sheets, $ \operatorname{dim}S=2$) when $ \operatorname{dim}M=3$. The right action of the diffeomorphism group $ \operatorname{Diff}(S)$ on $\operatorname{Emb}(S,M)$  yields the cotangent momentum map
\[
\mathbf{J} _R( \mathbf{Q} , \mathbf{P} )= \mathbf{P} \cdot \nabla \mathbf{Q}\in \mathfrak{X}  (S) ^\ast 
\]
which provides Clebsch variables for the EPDiff equations on $\mathfrak{X}  (S)$. Two Clebsch variational principles can be naturally associated to these actions, \cite{GBRa2011}.

We shall now present the covariant analogue of this setting and obtain singular solutions of the covariant EPDiff equations via covariant momentum maps.

\paragraph{Covariant Clebsch variational principle I: singular solutions.} We take the fiber bundle
$Y= X\times \operatorname{Emb}(S,M)$ and the principal bundle $P= X \times \operatorname{Diff}(M)$, and consider the left action of $ \operatorname{Diff}(M)$ on $  \operatorname{Emb}(S,M)$ by composition. The restricted dual jet bundle is $ \Pi = TX \otimes T^*\operatorname{Emb}(S,M)  \otimes \Lambda ^{n+1}X $. Using our general formula \eqref{rewriting_J}, we obtain the covariant momentum map $ \mathfrak{j}_L  : \Pi \rightarrow TX \otimes \mathfrak{X}  (M)  ^\ast \otimes \Lambda ^{n+1}X$,
\[
\mathfrak{j}_L  ( \mathbf{Q} ,\boldsymbol{\omega} )= \partial _\mu \otimes \mathbf{J} ( \mathbf{Q}, \boldsymbol{\omega} ^\mu ) \otimes d ^{n+1}x= \partial _\mu \otimes \int_S \boldsymbol{\omega} ^\mu (s) \delta (m- \mathbf{Q} (s)) d^k s \otimes d ^{n+1}x.
\]
From Proposition \ref{summary_trivial_case}, the covariant Clebsch variational principle reads
\begin{equation}\label{Cov_Clebsch_I} 
\delta \int_ U \ell( \boldsymbol{\sigma}  )+ \left\langle \boldsymbol{\omega}  , \mathbf{d}_X\mathbf{Q}  - \boldsymbol{\sigma}_ {\operatorname{Emb}}(  \mathbf{Q})  \right\rangle =0,
\end{equation} 
over sections $ \boldsymbol{\sigma} : U \subset X \rightarrow L(TX, \mathfrak{X}  (M))$ and $\boldsymbol{\omega} :U \subset U \rightarrow \Pi $. As before, $\mathbf{Q} : U \subset X \rightarrow \operatorname{Emb}(S,M)$ is the map induced by $ \boldsymbol{\omega} $, that is, we have $ \boldsymbol{\omega} (x) \in L(L(T_xX, T_{\mathbf{Q} (x)} \operatorname{Emb}(S,M)), \Lambda ^{n+1}X)$. We will use the notations $ \mathbf{Q} (x,s):= \mathbf{Q} (x)(s)$ and $ \boldsymbol{\omega} (x,s):= \boldsymbol{\omega} (x)(s)$. In the variational principle \eqref{Cov_Clebsch_I}, $ \mathbf{d}_X \mathbf{Q} $ denotes the derivative of the map $\mathbf{Q} : U \subset X \rightarrow \operatorname{Emb}(S,M)$, and the expression $\boldsymbol{\sigma}_ {\operatorname{Emb}}(  \mathbf{Q})\in L(TX, T \operatorname{Emb}(S,M))$ is given by $\boldsymbol{\sigma}_ {\operatorname{Emb}}(  \mathbf{Q})(v _x )(s)= (\boldsymbol{\sigma} (v_x))( \mathbf{Q} (x,s))\in T_{ \mathbf{Q} (x,s)}M$. The variational principle can be written as 
\[
\delta \int_ U \ell( \boldsymbol{\sigma} (x) )+ \left\langle \boldsymbol{\omega} ^\mu(x)  , \partial _\mu \mathbf{Q} (x) - (\boldsymbol{\sigma}_\mu(x) )_ {\operatorname{Emb}}(  \mathbf{Q}(x))  \right\rangle =0
\]
or, writing explicitly the duality pairing between $T \operatorname{Emb}(S,M)$ and its dual, 
\begin{equation}\label{explicit_covEPDiff_Clebsch} 
\delta \int_ U \left[ \ell( \boldsymbol{\sigma} (x)  )+ \int_S\left\langle  \boldsymbol{\omega} ^\mu (x,s) , \partial _\mu \mathbf{Q} (x,s) - \boldsymbol{\sigma}_\mu(x,   \mathbf{Q}(x,s))\right\rangle  d ^k s \right] =0.
\end{equation} 
The stationarity conditions read
\begin{equation*} 
\begin{aligned} 
\frac{\delta \ell}{\delta \boldsymbol{\sigma} _\mu }= \int_S \boldsymbol{\omega} ^\mu (s) \delta (m- \mathbf{Q} (s)) d^k s\qquad &  \partial _\mu \mathbf{Q} (x,s) = \boldsymbol{\sigma}_\mu(x,   \mathbf{Q}(x,s))\\
& \operatorname{div}_X \boldsymbol{\omega} =- (( \nabla \boldsymbol{\sigma} _\mu )^\mathsf{T} \circ \mathbf{Q} ) \cdot \boldsymbol{\omega} ^\mu.
\end{aligned} 
\end{equation*}
In the hyperregular case, eliminating the variable $ \boldsymbol{\sigma} $, these equations are equivalent to canonical covariant Hamilton's equations for the collective Hamiltonian $\mathsf{H}( \mathbf{Q} , \boldsymbol{\omega} )=h( \mathfrak{j}_L  ( \mathbf{Q} , \boldsymbol{\omega} ))$ and read
\[
\mathbf{d} \mathbf{Q} = \frac{\partial \mathsf{H}}{\partial \boldsymbol{\omega} }, \quad\operatorname{div}_X \boldsymbol{\omega} = -  \frac{\partial \mathsf{H}}{\partial \mathbf{Q} }
\]
on the restricted dual jet bundle $\Pi  $. Of course, one has to assume that the Hamiltonian $h$ converges on the value of the covariant momentum map $ \mathfrak{j}_L  $, that is, the collective Hamiltonian $\mathsf{H}=h \circ  \mathfrak{j} _L $ is well-defined.

In this case, it follows from Proposition \ref{summary_trivial_case} that  $ \boldsymbol{\sigma} $ is a solution of
\eqref{ZCC_EPDiff} and satisfy the compatibility conditions $ \left( \mathbf{d} _X \boldsymbol{\sigma} -[ \boldsymbol{\sigma} , \boldsymbol{\sigma} ]\right) _{ \operatorname{Emb}}( \mathbf{Q} ) =0$, which can be written more explicitly
\begin{equation}\label{CC_Left} 
\left( \mathbf{d} _X \boldsymbol{\sigma} (u _x , v _x )+ [ \boldsymbol{\sigma} ( u _x ), \boldsymbol{\sigma} ( v _x )] \right) \circ \mathbf{Q} (x)=0, \quad \text{for all $u _x , v _x \in T_xX$},
\end{equation} 
as an equality in $T_ { \mathbf{Q} (x)} \operatorname{Emb}(S,M)$. 

The covariant Clebsch variational principle generalizes easily to the case when the bundle is
\[
Y=X \times  \operatorname{Emb}(S,M)^N \ni ( x, \mathbf{Q} ^a ),\;\; a=1,...,N,
\] 
in which case the covariant momentum map is
\begin{equation}\label{CMM_covEPDiff} 
\mathfrak{j} _L(\mathbf{Q} ^a , \boldsymbol{\omega} _a )=\sum_{a=1}^N\partial _\mu \otimes \int_S \boldsymbol{\omega} ^\mu_a  (s) \delta (m- \mathbf{Q} ^a (s)) d^k s \otimes d ^{n+1}x.
\end{equation} 
In the Proposition below, we consider the case of a $H ^1 $ Lagrangian density, which is the natural generalization of the higher dimensional Camassa-Holm equations to the covariant case.

\begin{proposition}[Singular momentum map solutions of $H ^1 $ covariant EPDiff] Consider the Lagrangian density $\ell:L(TX, \mathfrak{X}(M)) \rightarrow \Lambda ^{n+1}X$ given by
\[
\ell( \boldsymbol{\sigma} )=\left( \frac{1}{2} \| \boldsymbol{\sigma} _0 \|^2 _ {H ^1 }-\frac{1}{2}\sum_{ i=1}^n \| \boldsymbol{\sigma} _i \|^2 _ {H ^1 }\right) d^{n+1}x,
\]
where $\| \mathbf{u} \|_{ H ^1 }^2 = \| \mathbf{u} \|_{L^2}^2 + \alpha^2  \| \nabla \mathbf{u} \|_{L^2}^2$, $\alpha >0$.
Then the corresponding covariant EPDiff equation is
\[
\partial _\mu \boldsymbol{\nu} ^\mu +\nabla \boldsymbol{\nu} ^\mu  \cdot \boldsymbol{\sigma} _\mu  + \nabla \boldsymbol{\sigma} _\mu  ^\mathsf{T}\cdot \boldsymbol{\nu} ^\mu  + \boldsymbol{\nu} ^\mu   \operatorname{div}\boldsymbol{\sigma} _\mu =0,
\]
with $\boldsymbol{\nu} ^0 =(1- \alpha ^2 \Delta ) \boldsymbol{\sigma} _0 $ and $\boldsymbol{\nu} ^i=-(1- \alpha ^2 \Delta ) \boldsymbol{\sigma} _i$, $i=1,...,n$.

This equation admits the singular solutions $ \boldsymbol{\sigma} = \boldsymbol{\sigma} _\mu dx ^\mu $ given by
\begin{align*} 
\boldsymbol{\sigma} _0 (x,m)&=\sum_{a=1}^N\int_S \boldsymbol{\omega} _a ^0 (x,s)G(m, \mathbf{Q} ^a (x,s))d ^k s\\
\boldsymbol{\sigma} _i (x,m)&=-\sum_{a=1}^N\int_S \boldsymbol{\omega} _a ^i (x,s)G(m, \mathbf{Q} ^a (x,s))d ^k s, \;\;i=1,...,n,
\end{align*} 
where $G$ is the Green's function for the linear operator $(1- \alpha ^2 \Delta )$ and where $( \mathbf{Q} ^a , \boldsymbol{\omega}_a  ), a=1,...,N$ verify the canonical covariant Hamilton's equations
\begin{equation}\label{cov_Ham_sing} 
\mathbf{d} \mathbf{Q}^a  = \frac{\partial \mathsf{H}}{\partial \boldsymbol{\omega} _a }, \quad\operatorname{div}_X \boldsymbol{\omega} _a = -  \frac{\partial \mathsf{H}}{\partial \mathbf{Q} ^a }, \;\; a=1,...,N,
\end{equation} 
for the collective Hamiltonian
\begin{align*}  
\mathsf{H}( (\mathbf{Q}^a   , \boldsymbol{\omega}_a)_{a=1}^N  )&=\sum_{ \mu =0} ^N \sum_{a,b=1}^N\frac{1}{2} \iint_S \boldsymbol{\omega} ^\mu  _a(x,s) G\left(  \mathbf{Q} _a (x,s),\mathbf{Q} _b (x,s')\right)  \boldsymbol{\omega} ^\mu _b(x,s') d^k s\;d ^k s'
\end{align*} 
on the restricted dual jet bundle $ \Pi = TX \otimes T^* \operatorname{Emb}(S,M) ^N \otimes \Lambda ^{n+1}X$.
\end{proposition} 
\textbf{Proof.} These results follows from the discussion before the proposition. In particular, the expression for $ \boldsymbol{\sigma} _0 $ and $ \boldsymbol{\sigma} _i $ follows by inverting the operator $(1- \alpha ^2 \Delta )$ and using the associated Green's function. The expression for the collective Hamiltonian is obtained by using the formula $ \mathsf{H}= h \circ \mathfrak{j} _L $ and noting that the covariant Lie-Poisson Hamiltonian $h$ associated to $\ell$ reads
\[
h ( \boldsymbol{\nu} )= \sum_{ \mu =0} ^n \frac{1}{2} \iint_M \boldsymbol{\nu}  ^\mu (m)G(m,n) \boldsymbol{\nu}  ^\mu (n)dm\;dn. \qquad \qquad\blacksquare
\]

\medskip

\begin{remark}\label{rmk_reconstruction} {\rm $ \textbf{(1)}$  Note that any covariant EPDiff equations whose Lagrangian density $\ell$ is such that the collective Hamiltonian $ \mathsf{H}= h \circ \mathfrak{j} _L $ is well-defined, admits these momentum map singular solutions.\\
$ \textbf{(2)}$ Note also that these singular solutions do not satisfy the zero curvature condition \eqref{ZCC_EPDiff} but only the compatibility conditions \eqref{CC_Left}, see Proposition \ref{summary_trivial_case}. Therefore, one cannot reconstruct the dynamics of the sections $x\in U \subset X \mapsto (x , \varphi (x))\in X \times \operatorname{Diff}(M) $ of the bundle $ X \times \operatorname{Diff}(M) \rightarrow X$. The dynamics of these singular solutions takes naturally place on the restricted dual jet bundle $ \Pi = TX \otimes T^*\operatorname{Emb}(S,M)^\nu \otimes \Lambda ^{n+1}X\ni ( \mathbf{Q}^a  , \boldsymbol{\omega} _a )$ and is given by the canonical covariant Hamilton equations for $ \mathsf{H}$. This remark holds in general for covariant Euler-Poincar\'e equations obtained through covariant Clebsch variational principles, see Proposition \ref{summary_trivial_case}.\\
$ \textbf{(3)}$ When $ \operatorname{dim}M=3$, for example, the components of the singular solutions may be supported on sets of points (vector peakons, $ \operatorname{dim}S=0$), one-dimensional filaments (strings, $ \operatorname{dim}S=1$), or two-dimensional surfaces (sheets, $ \operatorname{dim}S=2$).}
\end{remark}

\paragraph{The case of $ \operatorname{Diff}( \mathbb{R}  )$-strands.} When $X= \mathbb{R}  ^2 \ni (t,s)$ and $ M= \mathbb{R}  $, we recover the results of \cite{HoIvPe2012}. In this case, the manifold $ \operatorname{Emb}(S,M)$ is replaced by $ \mathbb{R}$ (the manifold $S$ is reduced to a point). We use the notations
\[
\boldsymbol{\sigma} = \boldsymbol{\nu} dt+ \boldsymbol{\gamma} ds\quad \text{and} \quad \omega_a = (\partial _t \otimes M_a + \partial _s \otimes N_a )dt \wedge ds,
\]
where $ \boldsymbol{\sigma} (t,s), \boldsymbol{\gamma} (t,s) \in \mathfrak{X}  ( \mathbb{R}  )$ and $M _a (t,s), N _a (t,s) \in T^*_{Q ^a (t,s)} \mathbb{R}  $. The covariant momentum map \eqref{CMM_covEPDiff} is therefore given by
\[
\mathfrak{j} _L ((Q ^a , \omega _a)_{a=1}^N )= \sum_{a=1}^N\left( \partial _t \otimes M _a \;\delta (m-Q ^a )+ \partial _s \otimes N _a \;\delta (m- Q ^a ) \right) dt \wedge ds
\]
and recovers the peakon solutions
\[
\boldsymbol{\nu}(t,s,m) =\sum_{a=1}^N M _a G(m, Q ^a (t,s)), \quad \boldsymbol{\gamma} (t,s,m)= -\sum_{a=1}^N N _a G(m, Q ^a (t,s))
\]
of the $ \operatorname{Diff}( \mathbb{R}  )$-strand equations found in \cite{HoIvPe2012}. In particular, our approach yields a natural interpretation of the equations governing the dynamics of these singular solutions. Indeed, from \eqref{cov_Ham_sing}, the fields $Q ^a, N _a , M _a $  obey the canonical covariant Hamilton equations for the collective Hamiltonian $\mathsf{H}$ given by
\begin{align*} 
\mathsf{H}( (Q ^a , N _a , M _a)_{a=1}^N )&= \frac{1}{2} \sum_{a,b=1}^N N _aG( Q_a , Q _b )N_b+\frac{1}{2} \sum_{a,b=1}^N M _a G( Q_a , Q _b )M_b
\end{align*} 
on the restricted dual jet bundle $ \Pi = T \mathbb{R}  ^2 \otimes T^* \mathbb{R}  \otimes \Lambda ^2 \mathbb{R} ^2 $. These equations are
\[
\partial _t Q _a (t,s)= \boldsymbol{\nu} (t,s, Q _a (t,s)), \quad \partial _s Q _a (t,s)= \boldsymbol{\gamma } (t,s, Q _a (t,s)),
\]
\[
\partial _t M _a + \partial _s N _a = - \sum_{b=1}^N(N _a N _b +M _a M _b ) \frac{\partial G}{\partial Q ^a }( Q ^a , Q ^b ). 
\]

They coincide with the equations derived in \cite{HoIvPe2012}, interpreted there as a constrained \textit{classical} Hamiltonian system, rather than an unconstrained \textit{covariant} Hamiltonian system.

The compatibility conditions \eqref{CC_Left} reduce to $\left( \partial_t  \boldsymbol{\gamma}  - \partial _s  \boldsymbol{\nu} -\boldsymbol{\gamma} \partial _m \boldsymbol{\nu}  +\boldsymbol{\nu}  \partial _m \boldsymbol{\gamma}\right) ( Q^a (t,s))=0$, for all $a=1,...,N$ and yield
\begin{align*} 
&\sum_{b=1}^N( \partial _t N _b + \partial _s M _b )G(Q _a , Q _b )\\
& \quad \quad +\sum_{b,c=1}^N(M _b N _c - N _b M _c )\left( \frac{\partial G}{\partial Q ^a }( Q _a , Q _b )G( Q _a , Q _c )+  \frac{\partial G}{\partial Q ^b }( Q _a , Q _b )G( Q _b , Q _c ) \right) =0,
\end{align*} 
for $a=1,...,N$. The same condition can be derived by writing $ \partial _{ts}Q _a = \partial _{st}Q _a $.

For completeness, we rewrite now the covariant Clebsch variational principle \eqref{explicit_covEPDiff_Clebsch} in the particular case of $ \operatorname{Diff}( \mathbb{R}  )$-strands. It reads
\begin{align*} 
&\delta \int_U \ell( \nu (t,s), \gamma (t,s))+  \sum_{a=1}^NM_a(t,s)(\partial _t Q ^a (t,s)- \nu (t,s,Q ^a (t,s))dtds\\
& \qquad\qquad \qquad \qquad \qquad \qquad +  \sum_{a=1}^NN_a(t,s)(\partial _s Q ^a (t,s)- \gamma  (t,s,Q ^a (t,s))dtds=0.
\end{align*}

\paragraph{Covariant Clebsch variational principle II: covariant Clebsch variables.} In order to implement the second covariant variational principle, we take the same fiber bundle
$Y= X\times \operatorname{Emb}(S,M)$ as before, and consider the principal bundle $P= X \times \operatorname{Diff}(S)$, where $ \operatorname{Diff}(S)$ acts on $  \operatorname{Emb}(S,M)$ by composition on the right. From our general formula \eqref{rewriting_J}, we obtain the covariant momentum map $ \mathfrak{j}_R : \Pi \rightarrow TX \otimes \mathfrak{X}  (S)  ^\ast \otimes \Lambda ^{n+1}X$,
\[
\mathfrak{j}_R ( \mathbf{Q} ,\boldsymbol{\omega} )= \partial _\mu \otimes \mathbf{J} ( \mathbf{Q}, \boldsymbol{\omega} ^\mu ) \otimes d ^{n+1}x= \partial _\mu \otimes \boldsymbol{\omega} ^\mu \cdot \nabla \mathbf{Q}  \otimes d ^{n+1}x.
\]

The covariant Clebsch variational principle reads
\begin{equation}\label{CCOC_right} 
\delta \int_ U \ell( \boldsymbol{\sigma}  )+ \left\langle \boldsymbol{\omega}  , \mathbf{d}_X \mathbf{Q}  + \nabla  \mathbf{Q} \cdot \boldsymbol{\sigma}   \right\rangle =0,
\end{equation} 
over sections $ \boldsymbol{\sigma} : U \subset X \rightarrow L(TX, \mathfrak{X}  (S))$ and $\boldsymbol{\omega} :U \subset X \rightarrow \Pi $, where $\mathbf{Q} : U \subset X \rightarrow \operatorname{Emb}(S,M)$ is the map induced by $ \boldsymbol{\omega} $. In the variational principle, $ \nabla \mathbf{Q} $ denote the derivative of $ \mathbf{Q}(x,s)$ relative to $s$, and $\nabla \mathbf{Q} \cdot \boldsymbol{\sigma} \in L(TX, T \operatorname{Emb}(S,M))$ is given by $(\nabla \mathbf{Q} \cdot \boldsymbol{\sigma}) (v _x )= \nabla \mathbf{Q} \cdot ( \boldsymbol{\sigma} (v _x ))\in T_{ \mathbf{Q} (x)} \operatorname{Emb}(S,M) $. The variational principle can be written as 
\[
\delta \int_ U \ell( \boldsymbol{\sigma} (x) )+ \left\langle \boldsymbol{\omega} ^\mu(x)  , \partial _\mu \mathbf{Q} (x) + \nabla \mathbf{Q} \cdot   \boldsymbol{\sigma} _\mu (x))  \right\rangle =0
\]
or, more explicitly,
\[
\delta \int_ U \left[ \ell( \boldsymbol{\sigma} (x)  )+ \int_S\left\langle  \boldsymbol{\omega} ^\mu (x,s) , \partial _\mu \mathbf{Q} (x,s) +\nabla \mathbf{Q} (x,s)\cdot   \boldsymbol{\sigma} _\mu (x,s))\right\rangle  d ^k s \right] =0.
\]
As above, in the hyperregular case the critical sections $( \mathbf{Q} , \boldsymbol{\omega} )$ of this principle satisfy the covariant Hamilton equations for the collective Hamiltonian $\mathsf{H}=h \circ \mathfrak{j} _R$ and $ \boldsymbol{\sigma} $ satisfies the covariant Euler-Poincar\'e equations for $\ell$ on $L(TX, \mathfrak{X}  (S))$, as well as the compatibility conditions \eqref{curvature_condition_Q} given here by 
\[
 \nabla \mathbf{Q} \cdot ( \mathbf{d} _X \boldsymbol{\sigma} - [ \boldsymbol{\sigma} , \boldsymbol{\sigma} ])=0.
\]
The detailed computations are left to the reader and are obtained by extending those of \cite[\S 4]{GBRa2011} to the covariant case. The sections $( \mathbf{Q} , \boldsymbol{\omega} )$ are interpreted here as covariant Clebsch variables for the covariant EPDiff equations, through the relation
\[
\frac{\delta \ell}{\delta \boldsymbol{\sigma} }= - \mathfrak{j} _R( \mathbf{Q} , \boldsymbol{\omega} )=- \partial _\mu \otimes \boldsymbol{\omega} ^\mu \cdot \nabla \mathbf{Q}  \otimes d ^{n+1}x. 
\]

Note that \eqref{CCOC_right} differs from the variational principle of Proposition \ref{summary_trivial_case} by a change of sign in front of the infinitesimal generator. This change of sign is due to the use of a \textit{right} action whereas Proposition \ref{summary_trivial_case} is formulated for \textit{left} actions.

\paragraph{A pair of covariant momentum maps.} The left and right actions of the diffeomorphism groups on $X \times \operatorname{Emb}(S,M)$ induce a pair of covariant momentum maps on the restricted dual jet bundle $ \Pi $:

\begin{picture}(150,100)(-70,0)%
\put(125,75){$\;\Pi $}
%top label

\put(90,50){$\mathfrak{j} _L$}
%left arrow label

\put(160,50){$\mathfrak{j} _R$}
%right arrow label

\put(0,15){$TX \otimes \mathfrak{X}(M)^{\ast}\otimes \Lambda ^{n+1}X$}
%left bottom label

\put(160,15){$TX \otimes \mathfrak{X}(S)^{\ast}\otimes \Lambda ^{n+1}X,$}
%right bottom label

\put(130,70){\vector(-1, -1){40}}
% left slanted arrow

\put(135,70){\vector(1,-1){40}}
% right slanted arrow

\end{picture}\\
similar to the dual pair of momentum map considered in \cite{HoMa2004} and further studied in \cite{GBVi2011}, the left leg being associated to singular solutions and the right leg to covariant Clebsch variables.

\subsection{The covariant EPDiff$_{vol}$ equations}

\paragraph{The covariant EPDiff$_{vol}$ equations.} We now consider the incompressible version of the covariant EPDiff equation studied above, that is, the case of the Lie group $ \operatorname{Diff}_{vol}(M)$ of volume preserving diffeomorphisms of a Riemannian manifold $M$ (assumed to be compact and without boundary, for simplicity). The Lie algebra of this group is given by the space $ \mathfrak{X}  _{vol}(M)$ of divergence free vector fields on $M$. Making the identification $ \mathfrak{X}  _{vol}(M)^\ast = \mathfrak{X}  _{vol}(M)$ and using the $L^2$ pairing associated to the Riemannian metric on $M$, we have the expression $\operatorname{ad}^*_ \mathbf{u} \mathbf{m} = \mathbb{P}  ( \nabla _ \mathbf{u} \mathbf{m} + \nabla \mathbf{u} ^\mathsf{T} \cdot \mathbf{m} )$, for all $ \mathbf{u} , \mathbf{m} \in \mathfrak{X}  _{vol}(M)$, where $ \mathbb{P}  : \mathfrak{X}  (M) \rightarrow \mathfrak{X}  _{vol}(M)$ denotes the orthogonal Hodge projector onto divergence free vector fields.

Given a $(n+1)$-dimensional manifold $X$ we get from \eqref{cove_EP}, the following covariant Euler-Poincar\'e equations on $ \operatorname{Diff}_{vol}(M)$:
\begin{equation}\label{cov_EPDiff_vol} 
\operatorname{div}_X \frac{\delta \ell}{\delta \boldsymbol{\sigma} }  + \nabla _{\boldsymbol{\sigma} _\mu } \frac{\delta \ell}{\delta \boldsymbol{\sigma}_\mu  }+ \nabla \boldsymbol{\sigma} _\mu ^\mathsf{T} \cdot \frac{\delta \ell}{\delta \boldsymbol{\sigma} _\mu }=\operatorname{grad}  p ,
\end{equation} 
where $ \boldsymbol{\sigma}  $ is a smooth section $ \boldsymbol{\sigma} : U \subset X \rightarrow L(TX, \mathfrak{X} (M))$ and $p: X \times M \rightarrow \mathbb{R}  $. The symbol $ \nabla $ and $ \operatorname{grad}$ denotes, respectively, the Levi-Civita covariant derivative and the gradient on $M$, both computed relative to the Riemannian metric on $M$. If the first cohomology $H ^1 (M)$ of $M$ is trivial, we can alternatively identify the dual Lie algebra with the space of exact two-forms on $M$, that is,
 $ \mathfrak{X}  _{vol}(M)^\ast = \Omega _{ex}^2(M)$. In this case, the equations take the vorticity form
\begin{equation}\label{cov_vol_vort} 
\operatorname{div}_X \frac{\delta \ell}{\delta \boldsymbol{\sigma} }+\boldsymbol{\pounds}_ {\boldsymbol{\sigma}  _\mu }  \frac{\delta \ell}{\delta \boldsymbol{\sigma_\mu} }=0,
\end{equation} 
where $\boldsymbol{\pounds}_ {\boldsymbol{\sigma}  _\mu }\delta \ell/ \delta \boldsymbol{\sigma} _\mu$ denotes the Lie derivative of $\delta \ell/ \delta \boldsymbol{\sigma} _\mu \in TX \otimes \Omega _{ex}^2(M)\otimes \Lambda ^{n+1}X$ with respect to $M$,  along the vector field $ \boldsymbol{\sigma} _\mu $.

In the 2D case, on can define a covariant stream section $ \psi : U \subset X \rightarrow L(TX, \mathcal{F} (M))$ for $ \boldsymbol{\sigma} $ and rewrite the equations as
\begin{equation}\label{2D_cov_vol} 
\operatorname{div}_X \frac{\delta \ell}{\delta \psi }  + \left\{ \frac{\delta \ell}{\delta \psi _\mu }, \psi _\mu \right\}  =0,
\end{equation} 
where $\{\,,\}$ denotes the Poisson bracket associated with the volume form on $M$, viewed as a symplectic form.

\paragraph{Covariant Clebsch variational principle and Clebsch variables.} Recall that the classical Clebsch variables for perfect fluids arise from the momentum map $ \mathbf{J} :T^* \mathcal{F} (M) \rightarrow \mathfrak{X}  (M) ^\ast = \Omega _{ex} ^2 (M)$, $ \mathbf{J} ( \lambda , \mu   )= \mathbf{d} \lambda \wedge \mathbf{d} \mu $ associated to the cotangent lifted action the right composition by $ \operatorname{Diff}( M)$ on $ \mathcal{F} (M)$, \cite{MaWe1983}.

In the covariant case, we take the bundle $Y= X \times \mathcal{F} (M)$, and we get the restricted dual jet bundle $ \Pi =TX \otimes T^* \mathcal{F} (M) \otimes \Lambda ^{n+1}X\ni ( \lambda , \omega )$. The covariant momentum map is $ \mathfrak{j} ( \lambda , \omega )= \partial _\mu \otimes \mathbf{d} \lambda \wedge \mathbf{d} \omega ^\mu \otimes d^{n+1}x$ and the covariant Clebsch variational principle reads
\[
\delta \int_ U \ell( \boldsymbol{\sigma}  )+ \left\langle \omega  , \mathbf{d}_X \lambda + \mathbf{d} \lambda \cdot \boldsymbol{\sigma}   \right\rangle =0,
\]
over sections $ \boldsymbol{\sigma} : U \subset X \rightarrow L(TX, \mathfrak{X} _{vol}(M))$ and $( \lambda , \boldsymbol{\omega}) :U \subset U \rightarrow \Pi $, where $ \mathbf{d} _X \lambda $ denotes the derivative of the map $ \lambda : U \subset X \rightarrow \mathcal{F} (M)$. The stationarity conditions are
\[
\frac{\delta \ell}{\delta \boldsymbol{\sigma}  }= \mathfrak{j} ( \lambda , \omega ), \quad \mathbf{d} _X \lambda + \mathbf{d} \lambda \cdot \boldsymbol{\sigma} =0, \quad  \operatorname{div}_X \omega + \operatorname{Tr}( \mathbf{d} \omega \cdot \boldsymbol{\sigma} )=0.
\]
In the hyperregular case, the first relation can be written as $ \boldsymbol{\sigma} = \frac{\delta h}{\delta \mathfrak{j} ( \lambda , \omega )}$. The last two equations can be written as a covariant canonical Hamiltonian system
\[
\mathbf{d} _X \lambda = \frac{\partial \mathsf{H}}{\partial \omega } , \quad  \operatorname{div}_X \omega = -\frac{\partial \mathsf{H}}{\partial \lambda  },
\]
for the collective Hamiltonian $\mathsf{H}= h \circ \mathfrak{j} $ on $ \Pi $, and yields the covariant EPDiff$_{vol}$ equations \eqref{cov_vol_vort}, together with the compatibility condition $ \mathbf{d} \lambda \cdot ( \mathbf{d} _X \boldsymbol{\sigma} -[ \boldsymbol{\sigma} , \boldsymbol{\sigma} ])=0$. The sections $( \lambda , \omega )$ are thus interpreted as \textit{covariant Clebsch variables} since they obey canonical covariant Hamilton's equations and yields the  noncanonical covariant Lie-Poisson equations for $h$, in a similar way with the classical Clebsch variables $(\lambda , \mu )$.

For example, with the $L ^2 $ Lagrangian density $\ell( \boldsymbol{\sigma} )=\sum_{ \mu =0}^n\frac{1}{2} \| \boldsymbol{\sigma} _\mu \|_{L ^2} ^2 $ we have  $ \frac{\delta \ell}{\delta \boldsymbol{\sigma} _\mu }= \mathbf{d} \boldsymbol{\sigma} _\mu ^\flat $ and the covariant Hamilton's equations for Clebsch variables become
\[
 \mathbf{d} _X \lambda + \mathbf{d} \lambda \cdot \mathbf{d} ^{-1} ( \mathbf{d} \lambda \wedge \omega ) =0, \quad  \operatorname{div}_X \omega + \operatorname{Tr}( \mathbf{d} \omega \cdot \mathbf{d} ^{-1} ( \mathbf{d} \lambda \wedge\omega  )  )=0,
\]
where $ \mathbf{d} ^{-1} $ denotes the inverse of the exterior differential seen as an isomorphism $ \mathbf{d} : \Omega _{vol} ^1 (M) \rightarrow \Omega ^2 _{ex}(M)$.
In the two-dimensional case, see \eqref{2D_cov_vol}, these equations are
\[
\partial _\mu \lambda +\{ \lambda , \{ \lambda , \omega _\mu \}\}=0\;\;\text{(no sum on $ \mu $)} , \qquad \partial_\mu \omega ^\mu +\{ \omega _\mu \{ \lambda , \omega _\mu \}\}=0\;\;\text{(sum on $ \mu $)}
\] 
and yield a \textit{covariant coupled double bracket formulation} of the two-dimensional EPDiff$_{vol}$ equations. They are reminiscent of the coupled double bracket formulation obtained in Proposition \ref{coupled_DB} in the context of adjoint actions, and extend to the covariant case the coupled double bracket formulation of perfect 2D fluids obtained in \cite[\S6]{GBRa2011}.

\section{The covariant Pontryagin variational principle}\label{Sec_5}

In this section we develop a general covariant variational principle that generalizes the covariant Clebsch variational principle and reproduces, as general case, the classical and covariant Hamilton-Pontryagin variational principles of \cite{YoMa2006} and \cite{VaYoMa2012}, respectively, together with variational principles arising in geometric optimal control problems, e.g. \cite{IbPeSa2010}.

Given a fiber bundle $\pi _{X,Y}:Y \rightarrow X$, we consider another fiber bundle $\pi _{Y,B} : B \rightarrow Y$ over $Y$ with local coordinates $( x ^\mu , y ^A , b ^\alpha )$, and define the bundles
\begin{equation}\label{def_ZP} 
\mathfrak{Z}:= J^1Y^\star \times _Y B\rightarrow Y \quad \text{and} \quad  \mathfrak{P}= \Pi \times _Y B\rightarrow Y,
\end{equation} 
referred to as the \textit{generalized Pontryagin bundle} and \textit{restricted generalized Pontryagin bundle}, respectively.

The line bundle $ \mu : J^1Y^\star \rightarrow \Pi $ induces naturally a line bundle $\mathfrak{m} :\mathfrak{Z} \rightarrow\mathfrak{P}$, locally given by $( x ^\mu , y ^A , \pi,p_A ^ \mu  ,b^ \alpha ) \rightarrow ( x ^\mu , y ^A ,p_A ^ \mu  ,b^ \alpha )$. The canonical forms $ \Theta \in \Omega ^{n+1}(J ^1 Y^\star)$ and $ \Omega  \in \Omega ^{n+1}(J^1Y^\star)$ induce the  forms $ \Theta _{\mathfrak{Z}}:= \pi _{J^1Y^\star, \mathfrak{Z}}^\ast \Theta \in \Omega ^{n+1}(\mathfrak{Z} )$ and $ \Omega  _{\mathfrak{Z}}:= \pi _{J^1Y^\star, \mathfrak{Z} }^\ast \Omega  \in \Omega ^{n+2}(\mathfrak{Z})$.

The variational principle that we define is associated to a smooth section $\mathsf{e}: \mathfrak{P} \rightarrow \mathfrak{Z} $ called a \textit{generalized energy section}, written locally as
\[
( x ^\mu , y ^A , p_A^\mu , b ^\alpha )\mapsto ( x ^\mu , y ^A , - e_{loc}( x ^\mu , y ^A, p_A ^\mu, b ^\alpha ) ,p_A^\mu , b ^\alpha ).
\]
By definition, a \textit{generalized energy density} is a smooth map $ \mathcal{E} : \mathfrak{Z} \rightarrow \Lambda ^{n+1}X$ covering the identity on $X$ and such that
\begin{equation}\label{cond_energy_dens} 
\mathbf{i} _X (\mathbf{d} \mathcal{E} + \Omega_ \mathfrak{Z} )=0 , \;\;\text{for all $X \in \mathfrak{X}^V  (\mathfrak{Z} ) $} ,
\end{equation} 
where $ \mathfrak{X}^V  (\mathfrak{Z} )= \{X \in \mathfrak{X}  (\mathfrak{Z} )\mid T \mathfrak{m}   \circ X=0\}$ is the space of all $ \mathfrak{m}  $-vertical vector fields on the generalized covariant Pontryagin bundle. From the condition \eqref{cond_energy_dens} we obtain that a generalized energy density is locally given by
\[
\mathcal{E} (x ^\mu , y ^A , p _A ^\mu , \pi, b ^\alpha  )= (\pi + e_{loc}(x ^\mu , y ^A , p _A ^\mu, b ^\alpha ))d^{n+1}x.
\]
This definition of generalized energy density is the natural extension of the definition \eqref{cond_Ham_dens} of Hamiltonian densities. Similarly to this case, there is a bijective correspondence between generalized energy densities and generalized energy sections. This correspondence being given by
\[
\operatorname{im} \mathsf{e}= \mathcal{\mathcal{E}} ^{-1} (0). 
\]

The Cartan $(n+1)$-form and $(n+2)$ forms associated to a generalized energy section $\mathsf{e}$ are defined by
\[
\Theta _\mathsf{e}:= \mathsf{e}^* \Theta_ \mathfrak{Z}  \in \Omega ^{n+1}( \mathfrak{P}  ) \quad \text{and} \quad  \Omega  _\mathsf{e}:= \mathsf{e}^* \Omega _ \mathfrak{Z} \in  \Omega ^{n+2}( \mathfrak{P} )
\]
and we have the relations
\[
\mathfrak{m}  ^\ast \Theta _\mathsf{e}= \Theta_ \mathcal{Z}  - \mathcal{E} \quad \text{and} \quad  \mathfrak{m}  ^\ast\Omega  _\mathsf{e}=\Omega_ \mathcal{Z}  + \mathbf{d} \mathcal{E}.
\]
One verifies the first equality by checking that both terms are locally given by
\[
-e_{loc}( x ^\mu , y ^A , p_A^ \mu , b ^\alpha )d^{n+1}x+ p_A ^\mu dy^A \wedge d^nx_ \mu .
\]

\begin{definition}[The covariant Pontryagin variational principle] Let $ \pi _{X,Y}: Y \rightarrow X$ be a fiber bundle and fix a fiber bundle $ \pi _{Y,B}: B \rightarrow Y$ and a generalized energy section $ \mathsf{e}: \mathfrak{P} \rightarrow \mathfrak{Z} $.
Let $U \subset X$ be an open subset whose closure $\bar U$ is compact, and let $ \psi : \bar U \subset X \rightarrow \mathfrak{P} $ be a local smooth section of $ \pi _{X ,\mathfrak{P} }: \mathfrak{P}   \rightarrow X$.

The covariant Pontryagin variational principle is
\begin{equation}\label{gen_def_cov_Clebsch} 
\delta \int_{ U} \psi ^\ast   \Theta _ {\mathsf{e}}   =0,
\end{equation} 
for all variations $\psi _ \varepsilon : \bar U \subset X\rightarrow  \mathfrak{P} $ of $ \psi $ (among smooth sections) such that $\psi _0 = \psi  $ and $ \phi _\varepsilon |_{ \partial U}= \phi |_{ \partial U}$, where $ \phi _\varepsilon$ is the section of $ \pi _{X,Y}: Y \rightarrow X$ induced from $ \psi _\varepsilon$.
\end{definition}

\medskip

Locally, the variational principle \eqref{gen_def_cov_Clebsch} reads
\[
\delta \int_U \left( \psi _A ^\mu \frac{\partial \psi ^A }{\partial x ^\mu }-e_{loc}( x ^\mu , \psi ^A , \psi _A ^\mu , \psi ^\alpha ) \right) d^{n+1}x=0,
\]
where we denoted locally the section as $ \psi ( x ^\mu )=( x ^\mu , \psi ^A  ( x ^\mu ), \psi _A ^\nu ( x ^\mu) , \psi ^\alpha  ( x ^\mu ))$

\begin{remark}[Formulation on $ \mathfrak{Z}$]{\rm The variational principle \eqref{gen_def_cov_Clebsch} can also be formulated on sections $\bar \psi :U \subset X \rightarrow \mathfrak{Z} $ of the generalized Pontryagin bundle as
\[
\delta \int_U \bar \psi ^\ast ( \Theta _{ \mathcal{Z} }- \mathcal{E} )=0,
\]
since we have $\bar\psi ^\ast ( \Theta _{ \mathcal{Z} }- \mathcal{E} )= \bar\psi ^\ast \mathfrak{m} ^\ast \Theta _\mathsf{e}= ( \mathfrak{m} \circ \bar \psi ) ^\ast \Theta _\mathsf{e}= \psi ^\ast \Theta _\mathsf{e}$, where we defined the section $ \psi := \mathfrak{m}  \circ \bar \psi$ of $\mathfrak{P} $. Of course, the associated stationarity conditions only involve the projection $ \psi := \mathfrak{m} \circ \bar \psi $ of $\bar \psi $ and the section $\bar \psi $ is reconstructed from $ \psi $ by imposing the zero energy constraint
\[
\mathcal{E} \circ \bar \psi =0.
\]
In coordinates, the section $ \bar \psi $ is thus given by
\[
\bar \psi ( x ^\mu )=\left(  x ^\mu , \psi ^A  ( x ^\mu ), \psi _A ^\nu ( x ^\mu) , -e_{loc}( x ^\mu , \psi ^A ( x ^\mu), \psi _A ^\mu( x ^\mu) , \psi ^\alpha ( x ^\mu)) ,\psi ^\alpha  ( x ^\mu )\right) .
\]
}
\end{remark} 

\medskip

In the same way as in Theorem \ref{3_statements}, we can prove the following result.

\begin{theorem}\label{}   The following statements for a smooth section $ \psi : \bar U \subset X \rightarrow \mathfrak{P}$ are equivalent:
\begin{itemize}
\item[\rm (1)] $ \psi $ is a critical point of the covariant Pontryagin principle \eqref{gen_def_cov_Clebsch}.
\item[\rm (2)] $ \psi ^\ast ( \mathbf{i} _V \Omega _\mathsf{e})=0$, for all $ \pi _{X, \mathfrak{P}  }$-vertical vector fields on $\mathfrak{P}  $.
\item[\rm (3)] $ \psi $ satisfies the following equations in local coordinates
\begin{equation}\label{local_GP} 
\frac{\partial \psi ^A }{\partial x ^\mu }= \frac{\partial e_{loc}}{\partial \psi ^\mu _A }, \quad   \frac{\partial \psi _A^\mu  }{\partial x ^\mu }=- \frac{\partial e_{loc}}{\partial \psi ^A }, \quad \frac{\partial e_{loc}}{\partial \psi ^\alpha }=0. 
\end{equation} 
\end{itemize}
\end{theorem}

\medskip

We now show that this covariant principle unifies several known variational principles in field theories and classical mechanics, including the covariant Clebsch variational principle developed in this paper and the Hamilton-Pontryagin principle.

\paragraph{Recovering the covariant Hamilton-Pontryagin principle.} Let $ \mathcal{L} : J^1Y \rightarrow \Lambda  ^{n+1}X$ be a Lagrangian density and consider the fiber bundle $B:= J^1Y\rightarrow Y$. The generalized Pontryagin bundle is thus given by $ \mathfrak{P} = J^1Y^\star \times _Y J^1Y$. Define the generalized energy density $ \mathcal{E} : J^1Y^\star \times _Y J^1Y\rightarrow \Lambda ^{n+1}X$ by
\[
\mathcal{E} (z, \gamma )= \left\langle z, \gamma \right\rangle - \mathcal{L} ( \gamma ).
\]
Since it is locally given by $ \mathcal{E} ( x ^\mu , y ^A , p_A ^\mu , v_ \mu ^A )= \left( \pi+ p_A ^\mu v_ \mu ^A - L( x ^\mu, y ^A , v_ \mu ^A )\right)  d^{n+1}x$, it verifies the condition \eqref{cond_energy_dens}. In this case, \eqref{gen_def_cov_Clebsch} recovers the covariant Hamilton-Pontryagin variational principle developed in \cite{VaYoMa2012}. Since $e_{loc}= p_A ^\mu v_ \mu ^A - L( x ^\mu, y ^A , v_ \mu ^A )$, the stationarity conditions \eqref{local_GP} read
\[
\frac{\partial \psi ^A }{\partial x ^\mu }= \psi _\mu ^A , \quad   \frac{\partial \psi _A^\mu  }{\partial x ^\mu }=\frac{\partial L}{\partial \psi ^A } , \quad \psi _A ^\mu - \frac{\partial L}{\partial \psi _\mu ^A }=0 
\]
and consistently recover the Euler-Lagrange equations in implicit form.

\paragraph{Recovering the covariant Clebsch variational principle.} Let $ \mathcal{G} \subset  \mathcal{A}ut(Y)$ be a Lie group acting on $Y$ by bundle automorphisms covering the identity. Let $ \mathcal{J} :J ^1 Y^\star \rightarrow L( \mathfrak{g}  , \Lambda ^n Y)$ be the covariant momentum map and consider the bundle $B:=L(TX, \mathfrak{g}  ) \rightarrow Y$. Given an Ehresmann connection $ \gamma $, we consider the energy section defined by
\[
\mathsf{e}( \omega , \sigma ):=\left( \mathsf{s} ^\gamma ( \omega )-j( \omega , \sigma )+ \pi _{X,Y} ^\ast \ell( \sigma ), \sigma \right),
\]
see \eqref{def_section_e}. Then the covariant Pontryagin principle \eqref{gen_def_cov_Clebsch} recovers the covariant Clebsch variational principle of Definition \ref{Def_CCVP}. Since $e_{loc}= p _A ^\mu ( \sigma ^A _\mu - \gamma _\mu ^A )-l( \sigma  ^\alpha _\mu )$, one verifies that the stationarity conditions \eqref{local_GP} agree with \eqref{stationary_cond_local}. 

\paragraph{Recovering the covariant Hamilton phase space principle.} Given a Hamiltonian density $ \mathcal{H} :J ^1 Y ^\star \rightarrow \Lambda ^{n+1}X$, we consider the case when the bundle $ B \rightarrow Y$ is absent, i.e., $ \mathfrak{Z} = J ^1 Y^\star$, and we take $ \mathcal{E} := \mathcal{H} $, so that $ \mathsf{e}=\mathsf{h}$. In this case the covariant Pontryagin principle \eqref{gen_def_cov_Clebsch} recovers the covariant Hamilton phase space principle recalled in \eqref{covariant_phase_space}. The stationarity conditions \eqref{local_GP} yield the covariant Hamilton equations \eqref{covHE}.

\paragraph{Recovering classical Pontryagin principles in optimal control and mechanics.} In order to particularize the variational principle \eqref{gen_def_cov_Clebsch} to classical mechanics, we choose $X:= \mathbb{R}  $, $Y:= \mathbb{R}  \times Q \rightarrow \mathbb{R}$, and the bundle $B:= \mathbb{R}  \times C \rightarrow \mathbb{R}  \times Q, (t, c _q ) \mapsto (t,q)$, where $C \rightarrow Q$ is a fiber bundle over $Q$. In this case, the generalized Pontryagin bundle and restricted Pontryagin bundle defined in \eqref{def_ZP} are $ \mathfrak{Z} =T^*( \mathbb{R}  \times Q) \times _{( \mathbb{R}  \times Q)} \times ( \mathbb{R}  \times C)$ and $ \mathfrak{P} = (T^*Q \times \mathbb{R}  )\times  _{( \mathbb{R}  \times Q)} \times ( \mathbb{R}  \times C)$. A generalized energy section reads $ \mathsf{e}( \alpha _q , t, c _q )=( \alpha _q , t, - e(t, \alpha _q , c _q ), c _q )$ and a local section of $ \mathfrak{P} \rightarrow X$ reads $ \psi (t)=( \alpha (t),t,c(t))$. With these notations, the variational principle \eqref{gen_def_cov_Clebsch} becomes
\begin{equation}\label{classical_Pontryagin} 
\delta \int_{0}^{T} \psi ^\ast \Theta _\mathsf{e}=\delta \int_{0}^{T} \left(  \left\langle \alpha (t),\dot q(t) \right\rangle - e( t, \alpha (t) , c(t)) \right) dt=0.
\end{equation} 
By choosing $e(t, \alpha _q  , c _q  ):= \left\langle \alpha _q ,\Gamma (c _q )\right\rangle -L(c _q )$, where $L:C \rightarrow \mathbb{R}  $ and $ \Gamma $ is a fiber bundle map $ \Gamma :C \rightarrow TQ$ covering the identity on $Q$, \eqref{classical_Pontryagin} recovers the variational principle associated to geometric nonlinear optimal control problems via the Pontryagin maximum principle, see e.g. \cite{IbPeSa2010}. In this case $C$ is the control bundle, $L$ is the cost function of the problem, $e$ is the Pontryagin Hamiltonian, and $ \dot q(t)= \Gamma (c(t))$ is the differential equation constraint.

When $C=TQ$ and $e(t, \alpha _q , v _q )= \left\langle \alpha _q , v _q \right\rangle-L( v _q ) $, (i.e., $ \Gamma = id_{TQ}$) then \eqref{classical_Pontryagin} recovers the Hamilton-Pontryagin variational principle (\cite{YoMa2006}); when $C= Q \times \mathfrak{g}  $ and $e(t, \alpha _q , \xi )= \left\langle \alpha _q , \xi _Q (q) \right\rangle - \ell( \xi )$, (i.e. $ \Gamma (q, \xi )= \xi _Q(q)$), then \eqref{classical_Pontryagin} recovers the Clebsch variational principle (\cite{GBRa2011}); finally, when $C$ is absent and $e(t, \alpha _q )=H( \alpha _q )$, then \eqref{classical_Pontryagin} recovers the classical Hamilton phase space variational principle.

\paragraph{Acknowledgment.} We thank M. Bruveris, C. Campos, D. Holm, D. Meier, T. Ratiu, and C. Tronci, for helpful discussions during the course of this work.

{\footnotesize

\bibliographystyle{new}

\end{document}